\documentclass[review,3p]{elsarticle} 

\usepackage{subcaption}
\usepackage{caption}
\usepackage{graphicx}
\usepackage{amsmath}   
\usepackage{xcolor}
\usepackage{placeins} 
\usepackage[colorlinks]{hyperref}
\definecolor{mh}{rgb}{0.59, 0., 0.09} 

\journal{Nuclear Instruments and Methods in Physics Research Section A}
\begin{document}

\begin{frontmatter}

\begin{center}
\vspace*{-2cm}
\noindent \small This work has been submitted to \textit{Nuclear Instruments and Methods in Physics Research Section A}. Copyright may be transferred without notice, after which this version may no longer be available.
\end{center}

\title{Quality Control of Mass-Produced GEM Detectors \\ for the CMS GE1/1 Muon Upgrade\tnoteref{t1}}


\tnotetext[t1]{Manuscript prepared by S.\ D.\ Butalla (FIT), M.\ Hohlmann (FIT), J.\ Merlin (UoS), R.\ Venditti (Bari) on behalf of the CMS Muon GEM Collaboration.}
\cortext[cor1]{Corresponding authors.\\ \indent Emails: \href{mailto:stephen.butalla@cern.ch}{\textbf{stephen.butalla@cern.ch}}, \href{mailto:jeremie.alexandre.merlin@cern.ch} {\textbf{jeremie.alexandre.merlin@cern.ch}},
\href{mailto: rosamaria.venditti@cern.ch}{\textbf{rosamaria.venditti@cern.ch}}
}
\fntext[fn1]{Now at Rice University, Houston, USA} 
\fntext[fn2]{Now at UC Davis, Davis, USA} 
\fntext[fn3]{Now at INFN Sezione di Genova, Genova, Italy} 

\author[n]{M.~Abbas} 
\author[s]{M.~Abbrescia}
\author[h,j]{H.~Abdalla}
\author[h,k]{A.~Abdelalim}
\author[h,i]{S.~AbuZeid}
\author[d]{A.~Agapitos}
\author[ad]{A.~Ahmad}
\author[q]{A.~Ahmed}
\author[ad]{W.~Ahmed}
\author[w]{C.~Aim\`e}

% ...
\address[d]{Peking University, Beijing, China} %
% ...
\address[h]{Academy of Scientific Research and Technology - ENHEP, Cairo, Egypt} %
\address[i]{Ain Shams University, Cairo, Egypt}
\address[j]{Cairo University, Cairo, Egypt}
\address[k]{Helwan University, also at Zewail City of Science and Technology, Cairo, Egypt}
% ...
\address[n]{Karlsruhe Institute of Technology, Karlsruhe, Germany}
% ...
\address[q]{Delhi University, Delhi, India}
% ...
\address[s]{Politecnico di Bari, Universit\`{a} di Bari and INFN Sezione di Bari, Bari, Italy}%
% ...
\address[w]{Universit\`{a} di Pavia and INFN Sezione di Pavia, Pavia, Italy} %
% ...
\address[ad]{National Center for Physics, Islamabad, Pakistan}
% ...

\begin{abstract}
The series of upgrades to the Large Hadron Collider, culminating in the High Luminosity Large Hadron Collider, will enable a significant expansion of the physics program of the CMS experiment. However, the accelerator upgrades will also make the experimental conditions more challenging, with implications for detector operations, triggering, and data analysis. The luminosity of the proton-proton collisions is expected to exceed $2-3\times10^{34}$~cm$^{-2}$s$^{-1}$ for Run 3 (starting in 2022), and it will be at least $5\times10^{34}$~cm$^{-2}$s$^{-1}$ when the High Luminosity Large Hadron Collider is completed for Run 4. These conditions will affect muon triggering, identification, and measurement, which are critical capabilities of the experiment. To address these challenges, additional muon detectors are being installed in the CMS endcaps, based on Gas Electron Multiplier technology. For this purpose, 161 large triple-Gas Electron Multiplier detectors have been constructed and tested. Installation of these devices began in 2019 with the GE1/1 station and will be followed by two additional stations, GE2/1 and ME0, to be installed in 2023 and 2026, respectively. The assembly and quality control of the GE1/1 detectors were distributed across several production sites around the world. We motivate and discuss the quality control procedures that were developed to standardize the performance of the detectors, and we present the final results of the production. Out of 161 detectors produced, 156 detectors passed all tests, and 144 detectors are now installed in the CMS experiment. The various visual inspections, gas tightness tests, intrinsic noise rate characterizations, and effective gas gain and response uniformity tests allowed the project to achieve this high success rate.

\end{abstract}


\end{frontmatter}

\section{Introduction}
The muon system \cite{CMS:1997iti} of the CMS experiment installed at the Large Hadron Collider (LHC) at CERN currently exploits three different gaseous detector technologies commensurate with the background rates in each detector region: 
Resistive Plate Chambers (RPCs) \cite{Santonico:1981sc} in the barrel and endcaps, Cathode Strip Chambers (CSCs) \cite{Hauser:1996mc} in the endcaps, and Drift Tubes (DTs) \cite{Abbiendi:2009zza} in the barrel.
The system provided information for muon triggering, reconstruction, and identification with excellent performance during Run 1 \cite{Chatrchyan:2013sba} and Run 2 \cite{Sirunyan:2018fpa} data-taking.
After the upgrade of the LHC injector chain during the second long shutdown (LS2) of the LHC, the instantaneous luminosity of the colliding beams will regularly exceed $2\times10^{34}$~cm$^{-2}$s$^{-1}$, which is twice the original design luminosity \cite{CMSCollaboration:2015zni}. Under these conditions, each proton-proton bunch crossing will experience up to 100 pileup (PU) interactions, meaning that during each bunch crossing at the interaction point, up to 50 proton-proton collisions will occur. This increased pileup, along with an increase of the integrated radiation dose, consequently leads to a degradation of the current detector performance. 
In this scenario, the muon reconstruction in the forward region of the CMS muon system will become particularly challenging, especially for the Level-1 (L1) Trigger. The smaller track bending in the magnetic field surrounding the endcap detectors relative to the track bending in the barrel, combined with multiple scattering of muons in the yoke, leads to muon momentum mis-measurements by the L1 trigger. This factor contributes to an increase of the overall muon L1 trigger rate in the forward region and, together with the general increase in background rates, will lead to a situation where the L1 rates in the pseudorapidity region $ \lvert \eta  \rvert \geq 1.6$ can no longer be controlled by the CSC detectors alone. This effect will be further exacerbated in the High Luminosity LHC (HL-LHC) scenario, where 200 PU interactions per bunch-crossing are expected.

\begin{figure}[ht] 
\centering
\includegraphics[width=\columnwidth]{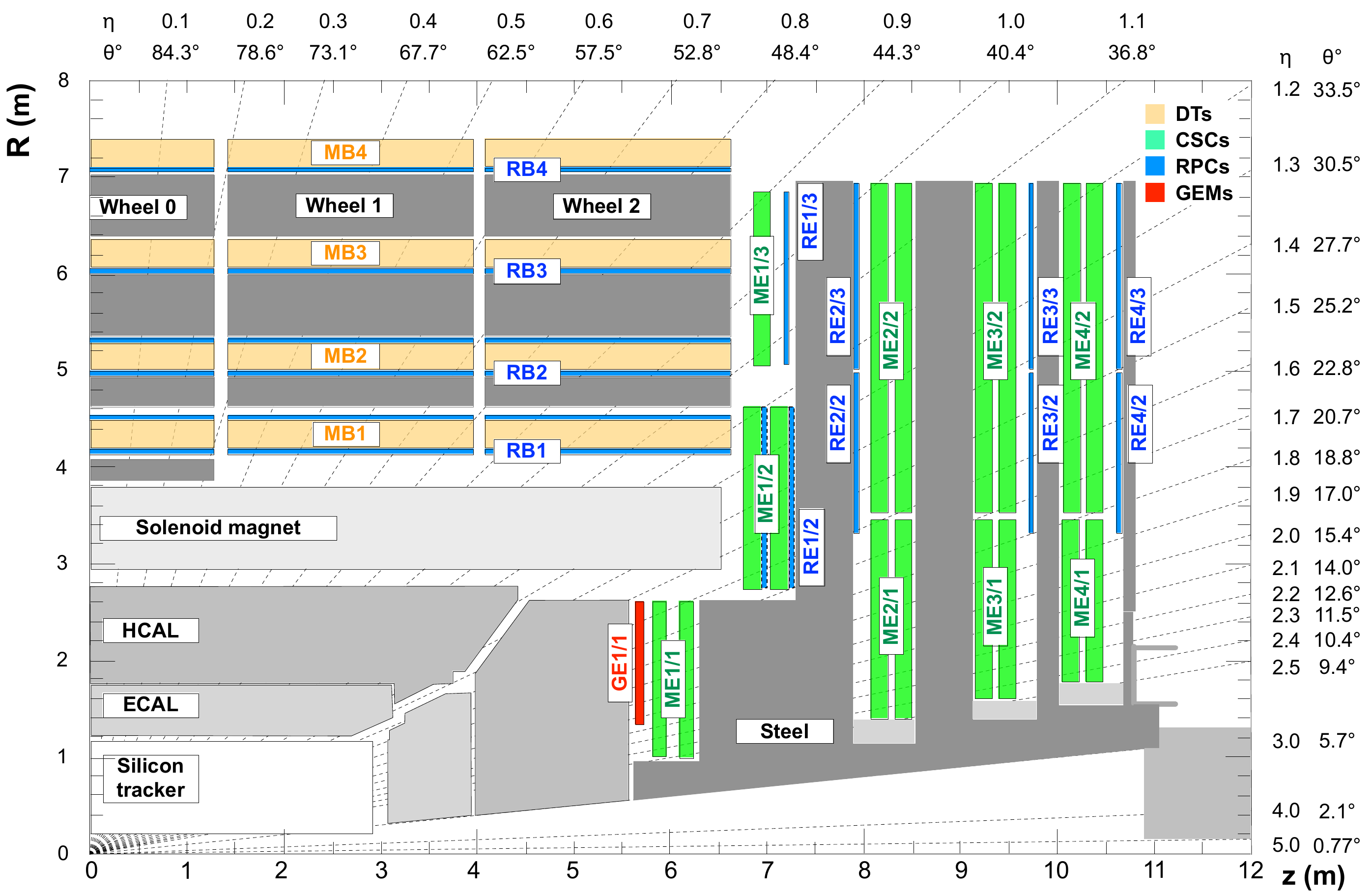}
\caption{\label{fig:MuSys}A quadrant of the CMS detector, showing the different types of chambers used in the muon system \cite{CMSCollaboration:2015zni}. The original system consisted of the DT chambers (yellow) in the barrel region, together with the RPCs (light blue) and CSCs (green) in the endcap regions. The new GEM detector station GE1/1 is indicated in red.}
\end{figure}

As a consequence, a general upgrade \cite{bib:MuTDR} of the present muon system has been carefully designed and planned by the CMS collaboration. Here we focus on the first new muon endcap station (GE1/1)~\cite{Colaleo:2015vsq}, which is based on Gas Electron Multiplier (GEM) detector technology~\cite{Sauli:1997qp} and was installed in front of the first CSC station (ME1/1) before Run 3 (Fig.~\ref{fig:MuSys}). Simulation studies  \cite{Colaleo:2015vsq} have shown  that this new station will improve the muon L1 trigger performance in the forward region. It is expected to keep the L1 trigger rates under control in the high-radiation environment of Run 3.

This paper is organized as follows: An overview of the GE1/1 station is presented in Section \ref{sect:GE11_struct}, while the quality control procedures and detector performance results are discussed in Section \ref{sec:qcProcess}. The selection of the detector working point is described in Section \ref{sec:workingPoint}, and Section \ref{sec:summary} provides a summary and conclusions of the GE1/1 construction and quality control process, and perspectives for the next two GEM stations to be built.

\section{Overview of the GE1/1 station}
\label{sect:GE11_struct}
The GE1/1 station provides full coverage in the azimuthal coordinate, $\phi$, while instrumenting the pseudorapidity region \mbox{$1.55<|\eta|<2.18$}, as shown in Fig.~\ref{fig:MuSys}.
A triple-GEM chamber (also simply referred to as ``a detector''), consists of three GEM foils separated by spacers and held between an anode readout board and a cathode drift board.
The choice of the triple-GEM technology is driven by the excellent performance of these detectors: triple-GEM detectors exhibit a rate capability well above the required $\sim$10 kHz/cm$ ^{2}$, a time resolution of $\sim$8 ns or better, and a spatial resolution of $\sim$200 $\mu$m, which make them well suited for the purposes of the CMS Muon System Upgrade \cite{Colaleo:2015vsq}.

In the GE1/1 station, a pair of triple-GEM chambers is combined to form a ``superchamber'' (SC) that provides two measurements of the muon hit position per muon track. 
As each superchamber covers a $\sim$10$^{\circ}$ sector, 36 superchambers are installed in each endcap. Short and long chambers alternate in $\phi$ to maximize the pseudorapidity coverage. A total of 144 (plus 17 spare) triple-GEM detectors were constructed and instrumented with electronics for installation.

\begin{figure}[ht] 
\centering
\includegraphics[trim={0.4cm 6cm 0 6cm}, clip, width=0.77\columnwidth]{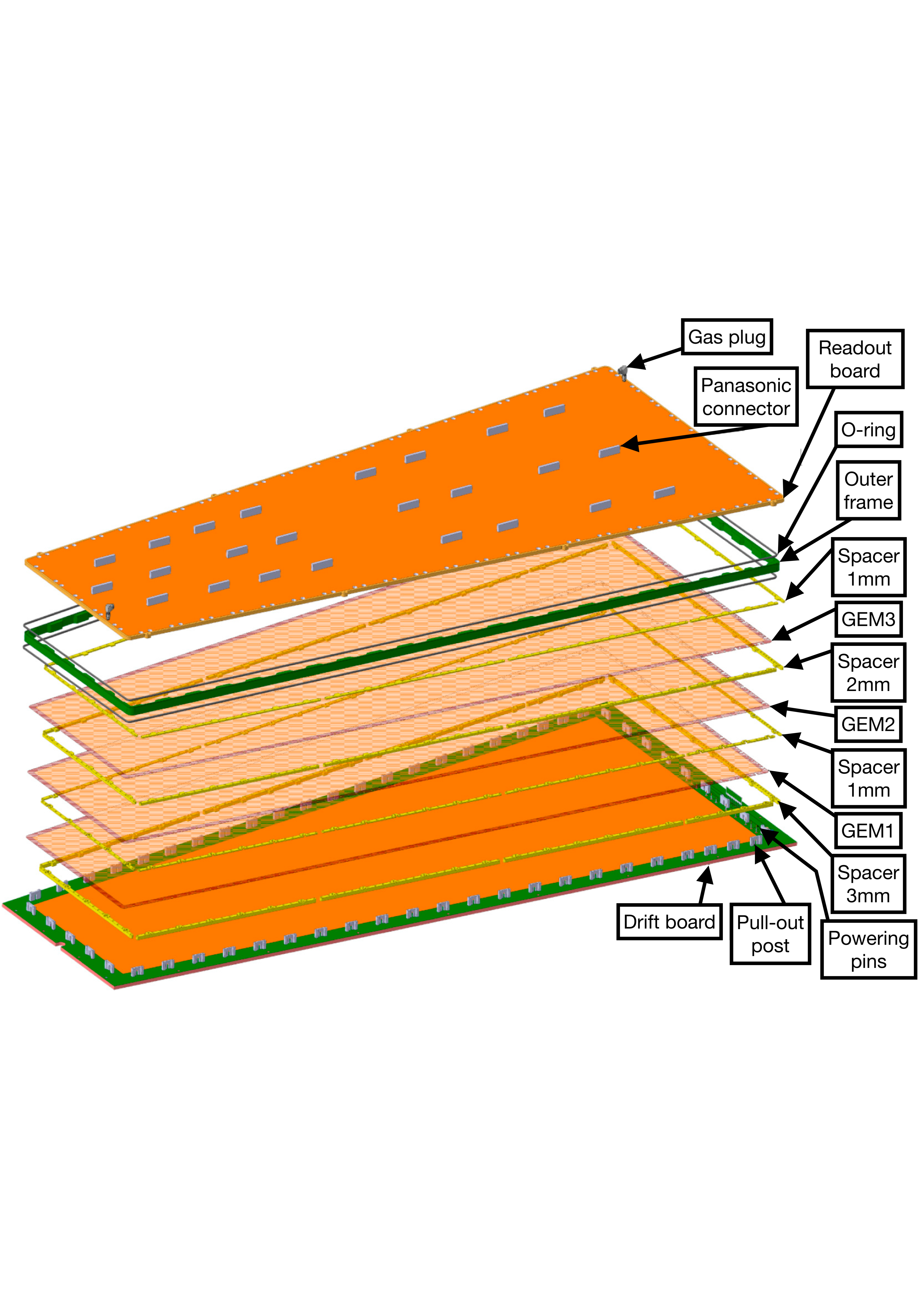}
\caption{\label{fig:View}Exploded view of the mechanical design of a triple-GEM detector.}
\end{figure}

An exploded view of a GE1/1 detector is shown in Fig.~\ref{fig:View}. A detailed description of the chamber design can be found in Ref.~\cite{Colaleo:2015vsq}.
Three GEM foils, i.e.\ copper-clad polyimide foils perforated with a high density of microscopic holes (140 $\mu$m pitch and 70 $\mu$m hole diameter), are stacked between an unsegmented drift cathode and a segmented readout anode..

\begin{figure}
     \centering
     \begin{subfigure}[t]{0.48\textwidth}
         \centering
         \includegraphics[width=0.9\textwidth]{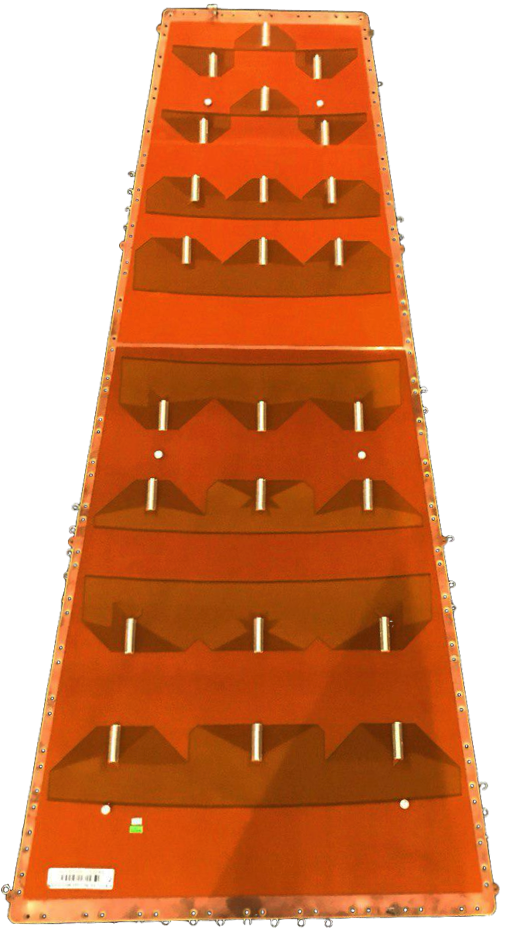}
         \caption{\label{fig:ge11ROB}The readout board of the GE1/1.}
     \end{subfigure}
     \hfill
     \begin{subfigure}[t]{0.5\textwidth}
         \centering
         \includegraphics[width=0.9\textwidth]{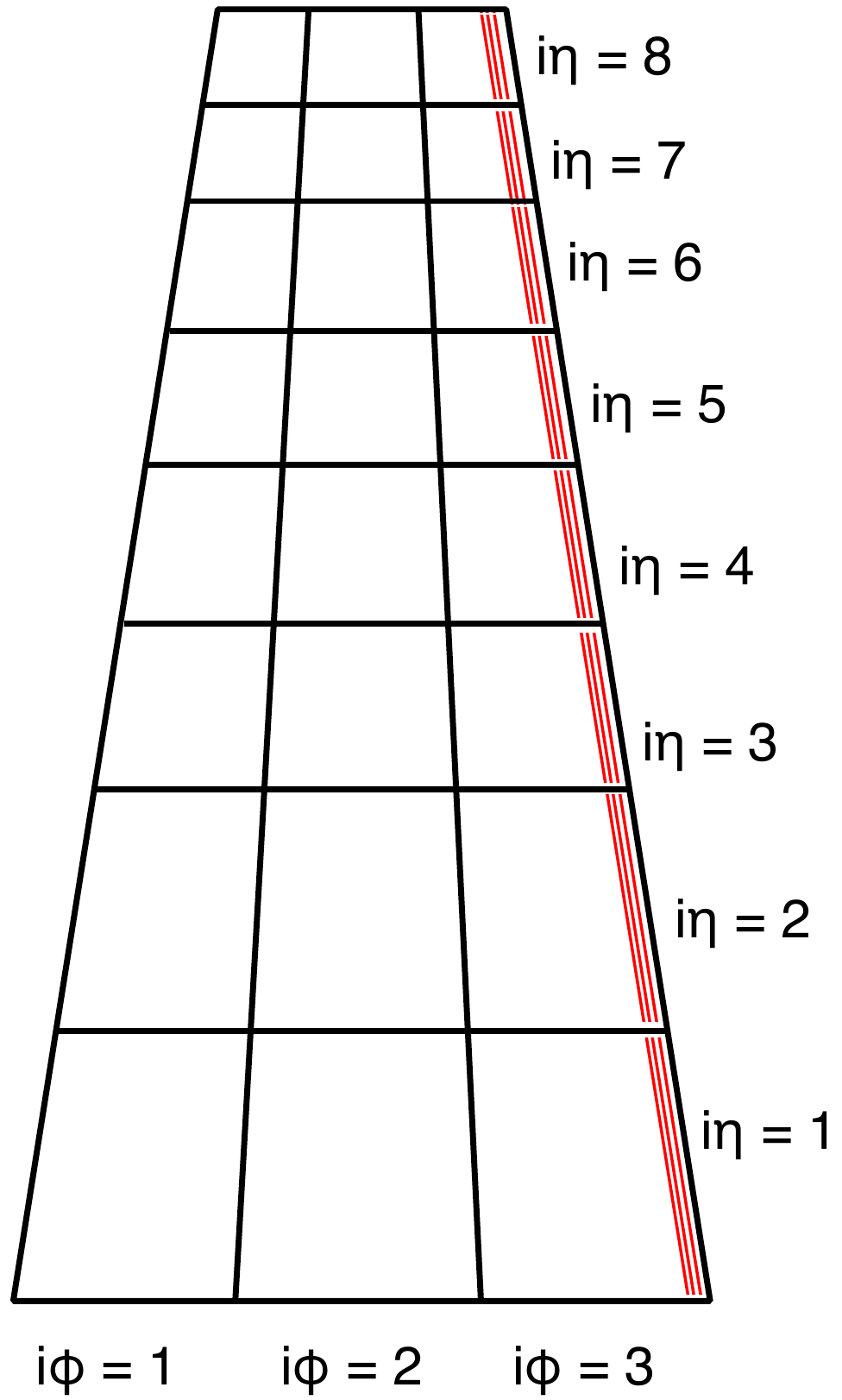}
         \caption{\label{fig:ge11schematic}The local coordinate system, showing the local pseudorapidity coordinate, denoted by $i\eta$, and the local azimuthal coordinate, denoted by $i\phi$. Some readout strips (red lines) are pictured to indicate their orientation on the readout board. Note: readout strips not drawn to scale.}
     \end{subfigure}
        \caption{\label{fig:GE11mapping}The readout board and the partitioning scheme of the readout sectors.}
\end{figure}

The induced electrical signal produced by the particles that hit the detector is picked up by the anode readout board, which is segmented into 384 radial readout strips with a pitch of 461 $\mu$rad. These readout strips are grouped into three main sectors along the azimuthal coordinate and eight sectors along the $\eta$ coordinate, thus resulting in a total of 24 readout sectors. Each of these 24 readout sectors are assigned a two-coordinate designation, $(i\eta,\, i\phi)$, with $i\eta=8$ at the narrow end (closest to the beamline), and $i\eta=1$ at the wide end, and from $i\phi=1$ to $i\phi=3$ from left to right (see Fig.~\ref{fig:GE11mapping}).

While operating in CMS, each sector is read out by one VFAT3 chip, which is an Application Specific Integrated Circuit (ASIC), adapted for gaseous detector readout~\cite{Aspell:2008zz}. The voltage is provided to the GEM foils through the drift board, a single, contiguous drift cathode that routes a total of seven potentials (drift electrode plus two sides of each of the three GEM foils) to the various GEM electrodes and to the drift cathode.

\begin{figure}[htp!]
\centering
    \begin{subfigure}{0.5\textwidth}
        \centering
        \resizebox{7cm}{5.6cm}{\includegraphics{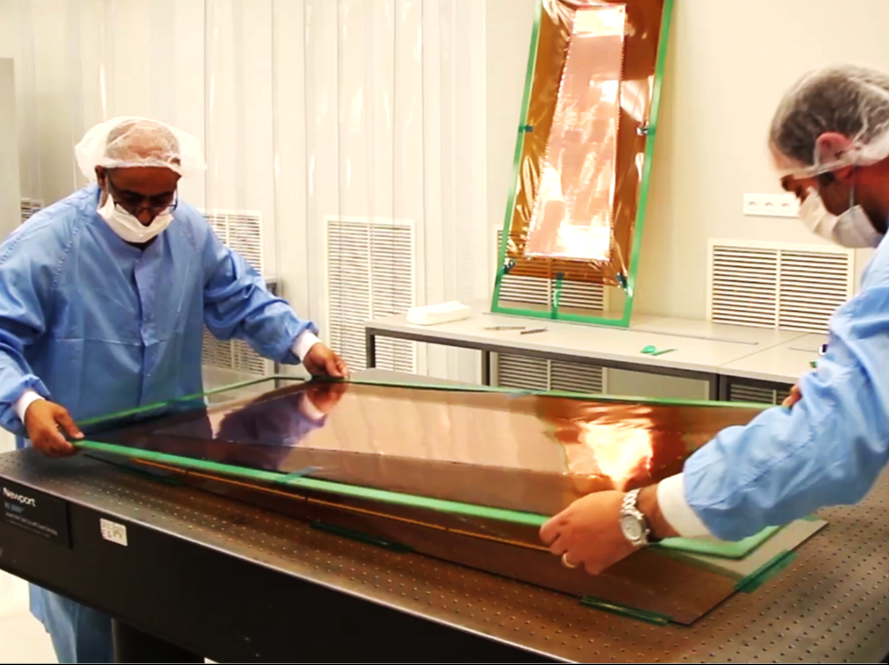}}
        \caption{\label{fig:placingGEM}Placing the first GEM foil in the stack.} 
    \end{subfigure}\hspace*{\fill}
    \begin{subfigure}{0.5\textwidth}
        \centering
        \resizebox{7cm}{5.6cm}{\includegraphics{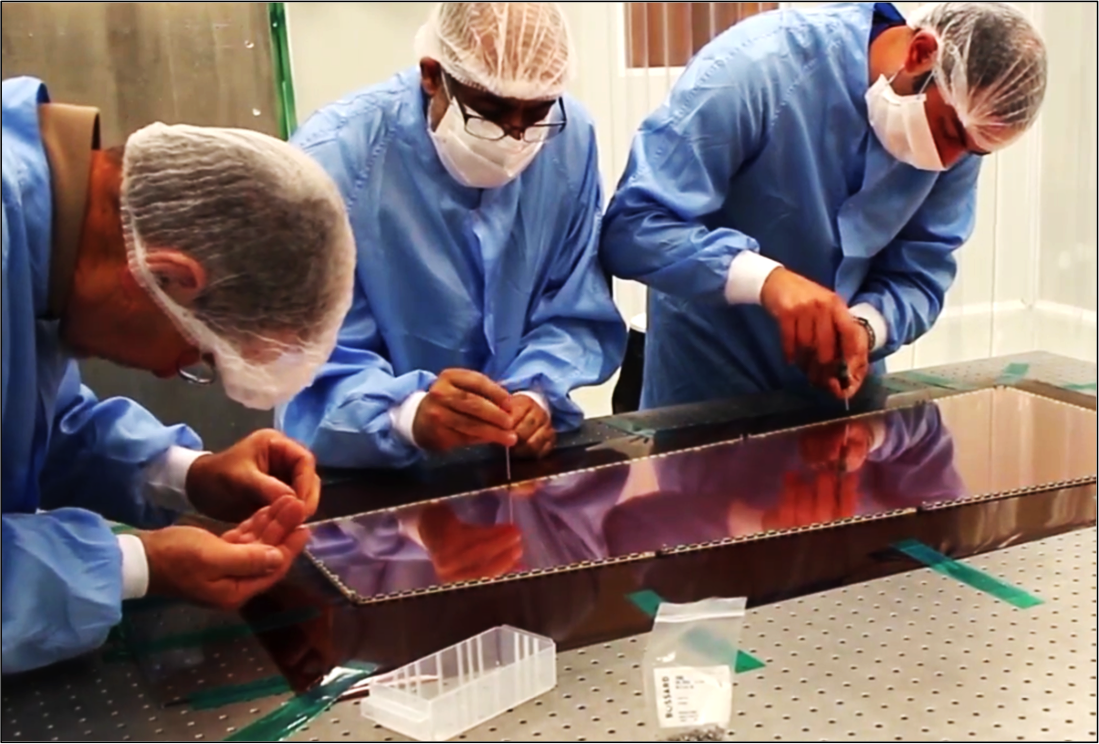}}
        \caption{\label{fig:internalFrames}Securing the GEM stack.}
    \end{subfigure}
    \begin{subfigure}{0.5\textwidth}
        \centering
        \resizebox{7cm}{4cm}{\includegraphics{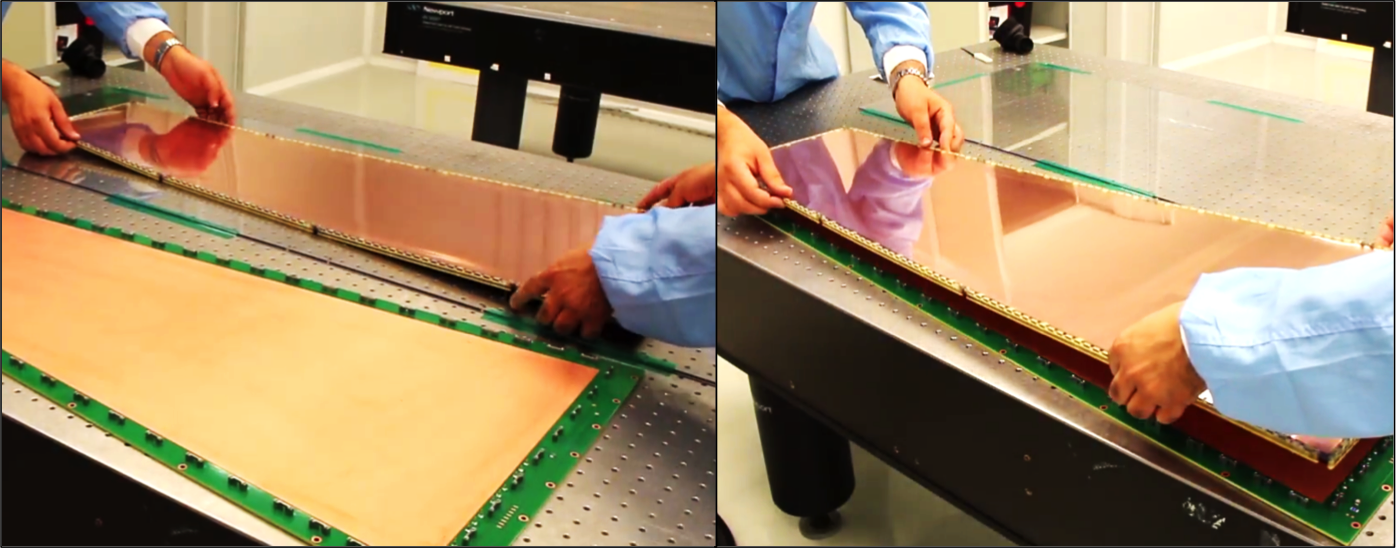}}
        \caption{\label{fig:stackTransfer}Transferring the stack to the drift board.} 
    \end{subfigure}
    \begin{subfigure}{0.5\textwidth}
        \centering
        \resizebox{7cm}{4cm}{\includegraphics{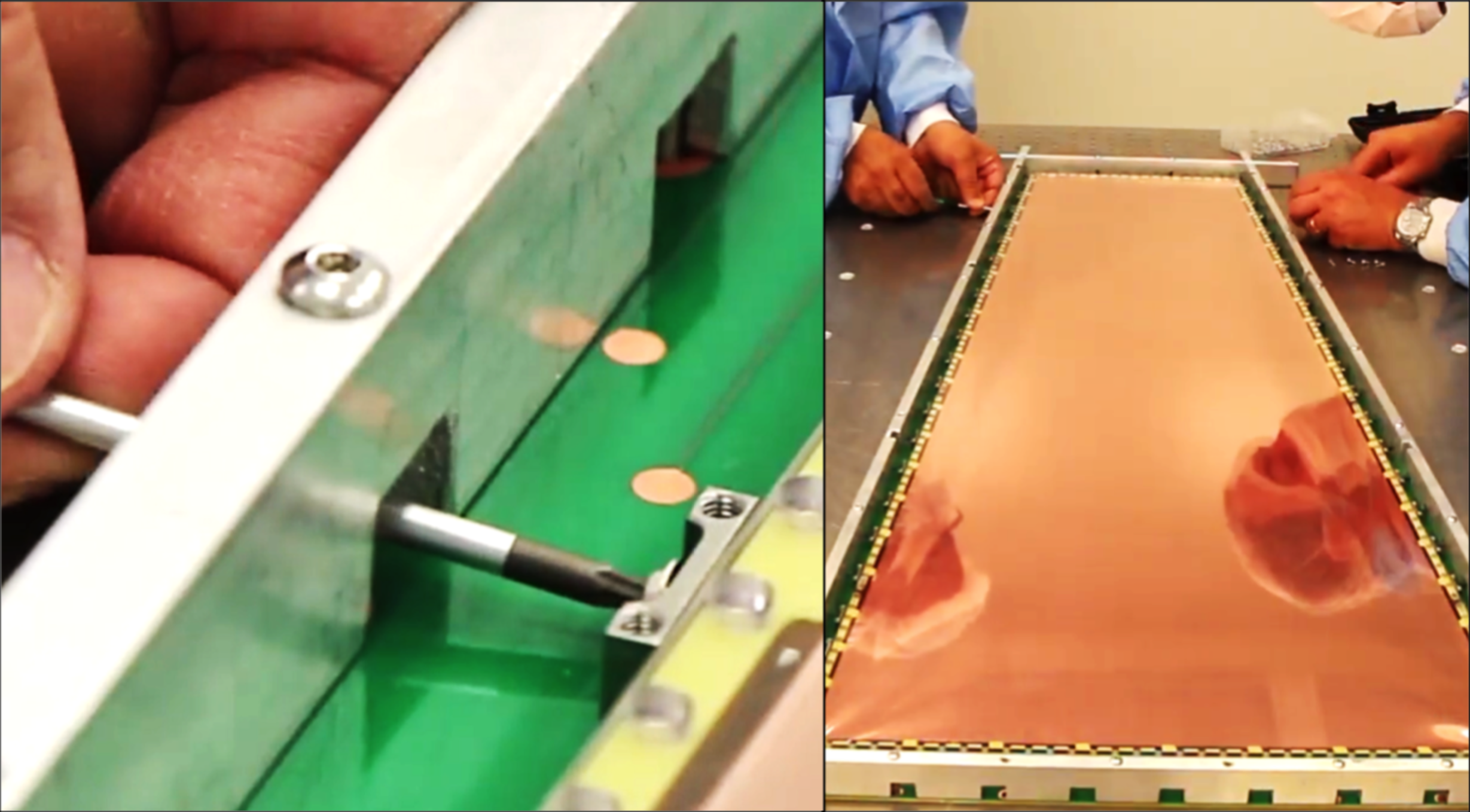}}
        \caption{\label{fig:stretching}Stretching the GEM foils.} 
    \end{subfigure}

    \begin{subfigure}{0.5\textwidth}
        \centering
        \resizebox{7cm}{7.2cm}{\includegraphics{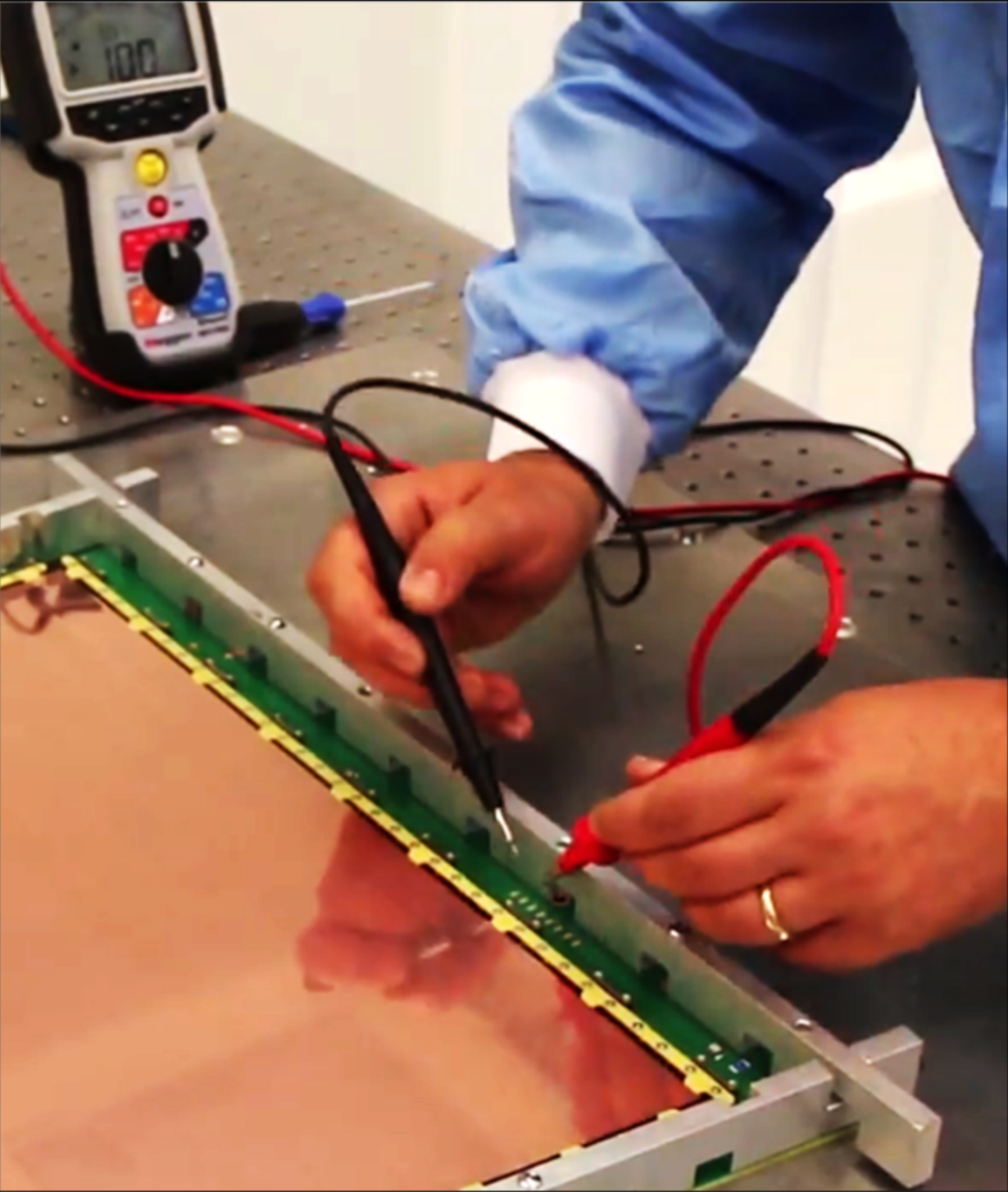}}
        \caption{\label{fig:preClosingMeggering}Resistance testing of all GEM foils and gaps.}
        \end{subfigure}\hspace*{\fill}
    \begin{subfigure}{0.5\textwidth}
        \centering
        \resizebox{7cm}{7.2cm}{\includegraphics{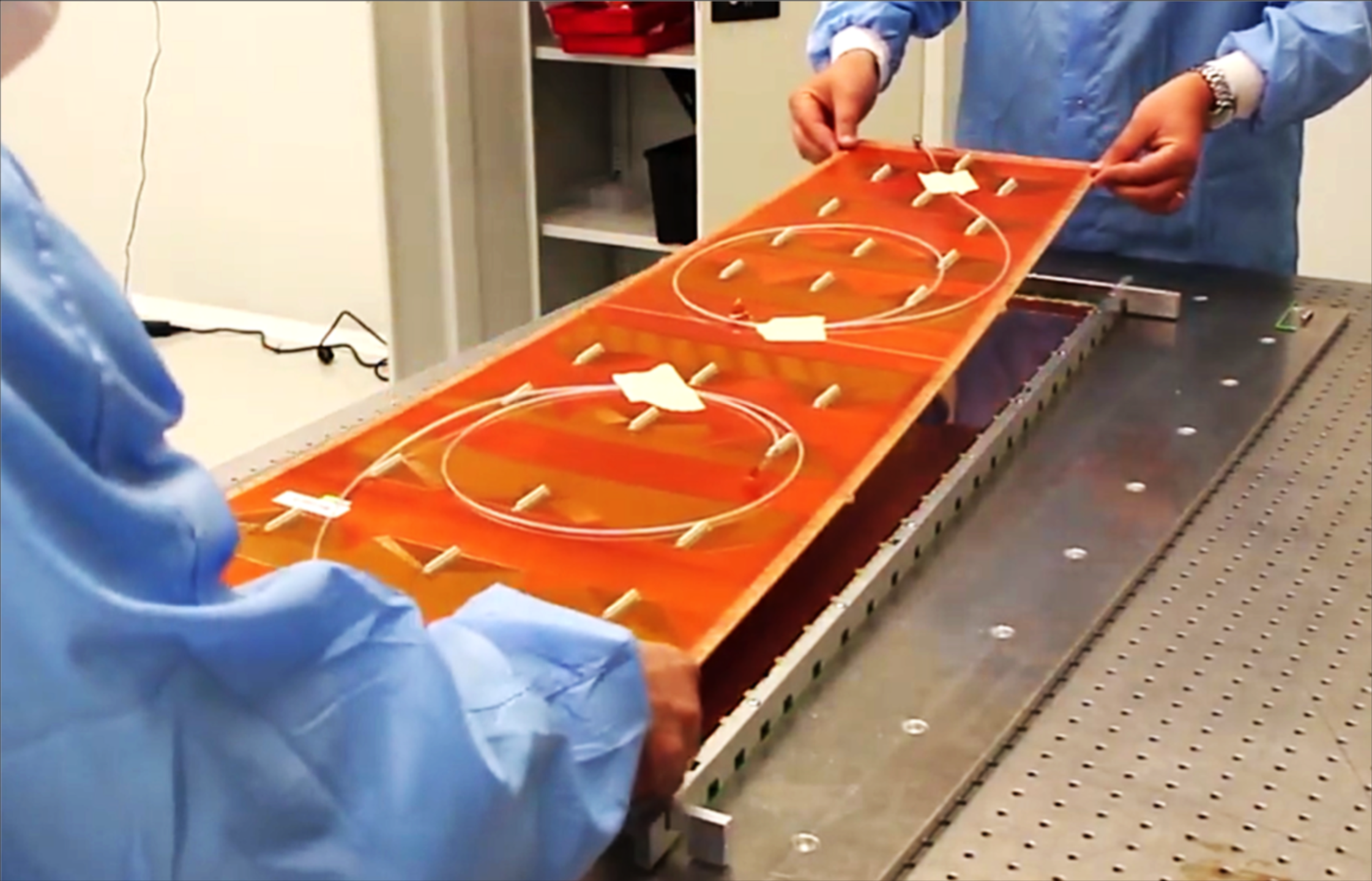}}
        \caption{\label{fig:closingChamber}Closing the chamber with the readout board.} 
        \end{subfigure}
\caption{\label{fig:GEMassembly}Various steps of the assembly process of a CMS GE1/1 GEM detector.}
\end{figure}

Assembly and gas sealing of the detector are entirely mechanical. A brief overview of the construction process is shown in Fig.~\ref{fig:GEMassembly}. The stack of three GEM foils is first completed outside of the drift board, as shown in Figs.~\ref{fig:placingGEM} and \ref{fig:internalFrames}. Figure~\ref{fig:stackTransfer} displays the transferring of the GEM stack to the drift board. The GEM foils are then tensioned by the stretching screws (Fig.~\ref{fig:stretching}), after which the resistances of the GEM foils and gaps are tested (Fig.~\ref{fig:preClosingMeggering}). Finally, the chamber is closed by fixing the readout board to the drift board, as shown in Fig.~\ref{fig:closingChamber}. For a comprehensive and detailed description of the assembly techniques, see Ref.~\cite{CMS-Muon:2018wsu}. No glue is applied during assembly, which makes it possible to open a detector again for repairs if needed, and to avoid possible gas contamination from glues.
The three GEM foils are sandwiched at their edges between four thin frames. Square stainless steel nuts are embedded into the frames every few centimeters to host stainless steel screws that are inserted into small ``pull-out" posts, which are located within the gas volume. 
When the pull-out screws are tightened manually, the GEM foils in the stack are tensioned against the posts. A large outer glass-epoxy frame is placed around the tensioned GEM stack and around the pull-outs providing the enclosure of the gas volume. The standard gas mixture for operating and testing this triple-GEM detector is Ar/CO\textsubscript{2} (70:30). 

\section{Quality Control Process and Performance of GE1/1 detectors}\label{sec:qcProcess}
Several institutions from six countries participated in the assembly and quality control process of the GE1/1 detectors. These institutions are located in Switzerland  (CERN),  Italy (INFN Bari and Frascati), Belgium (U.\ Ghent),  India (U.\ Delhi, BARC, Punjab U., Saha Inst.\ of Nuclear Physics), Pakistan (National Centre for Physics), and USA (Florida Inst.\ of Technology). An additional satellite site in Germany (RWTH Aachen U.) performed quality control tests on a subset of the chambers produced in Belgium. This distributed production model was adopted by the collaboration to ensure timely mass production with built-in redundancy.

Before becoming a production site, each institution had to follow a two-year certification program organized by the GEM Collaboration. This program is designed to ensure that all GE1/1 chambers are produced in suitable conditions using the same infrastructure, and that the same procedures for construction and quality control testing are performed. All steps of the certification program are reviewed by the collaboration, culminating in a final certification after an on-site visit. The entire process is supported by a set of technical documents that describe the composition of the test setups and provide guidance for the local management of the production \cite{QCManual:2016}.

To ensure proper and robust operation of the chambers within their performance specifications once installed in CMS, a detailed Quality Control (QC) procedure \cite{CMS-Muon:2018wsu, QCManual:2016, AssemblyManual:2017} is employed in all the production sites, which are equipped with the same laboratory infrastructure and instrumentation. This protocol aims to carefully assess the detector performance at each step. If measurements are found that are incompatible with predefined standards, the detector undergoes further adjustment until it fulfills those requirements. The results of all QC tests are systematically reviewed by the production community and the final validation of the chambers is submitted to the GEM collaboration every two months during dedicated reviews.

The first group of tests is performed on the components before the assembly. 
The second set of tests are completed on a single chamber while the third set of tests are performed after superchamber assembly. Single-chamber assembly and QC tests are done at the different production sites, while the assembly and the QC tests on the SC are only performed centrally at CERN. The detailed step-by-step description of the QC process performed at the production sites and the results obtained for single chambers are the focus of this paper. The five main steps that comprise this component and single-chamber QC are listed in Table \ref{tab:QC_steps}, and are discussed in detail below.

\begin{table*}[ht] 
\centering
\caption{\label{tab:QC_steps}Quality control steps for GE1/1 chamber components and assembly.}\vspace{2mm}
\begin{tabular}{|c|l|}\hline
QC Step &  QC procedure\\ \hline
1  & Initial inspection of the chamber components \\ \hline
2a & Electrical cleaning of the GEM foils and resistance check \\ \hline
2b & Long-term monitoring of the GEM foil leakage current \\ \hline
3  & Leak test of closed detector volume \\ \hline
4  & Linearity test of high voltage  divider and intrinsic noise rate measurement \\ \hline
5a & Effective gas gain measurement \\ \hline
5b & Response uniformity measurement \\ \hline
\end{tabular}
\end{table*}

\subsection{QC Step 1 -- Initial inspection of the chamber components}
The quality control procedure starts with the initial inspection of the components that will form the main structure of the chamber, i.e.\ the printed circuit boards (PCBs) that host the drift and readout electrodes, the internal frames, and the external frame with its embedded Viton O-rings. While the mechanical dimensions of all components are carefully verified using a precision caliper, an additional systematic visual inspection of the materials helps to identify potential defects, such as mechanical degradation, scratches and cracks, and the presence of any residues of chemicals used during manufacturing or other contaminants. Components outside the specifications or showing defects are discarded from the production materials or returned to the factory for rectification.

Particular attention is paid to the planarity of the drift and readout PCBs, which directly impacts the electric field uniformity, and consequently the overall detector performance. To determine the planarity, the boards are placed on a flat marble table, and then gaps between PCBs and the table are measured with precision calipers. Boards with deviations from flatness above 3~mm from the center are rejected. The readout boards undergo an additional electrical test to ensure the integrity of all readout paths. The setup for this test, based on an ARDUINO microprocessor, automatically measures  the continuity of the signals between the internal strips and the external connectors and also identifies any short circuits between neighbouring strips.

\subsection{QC Step 2a -- Electrical cleaning of the GEM foils and resistance check}
\begin{figure}[!htp]
\centering
\includegraphics[width=0.88\textwidth]{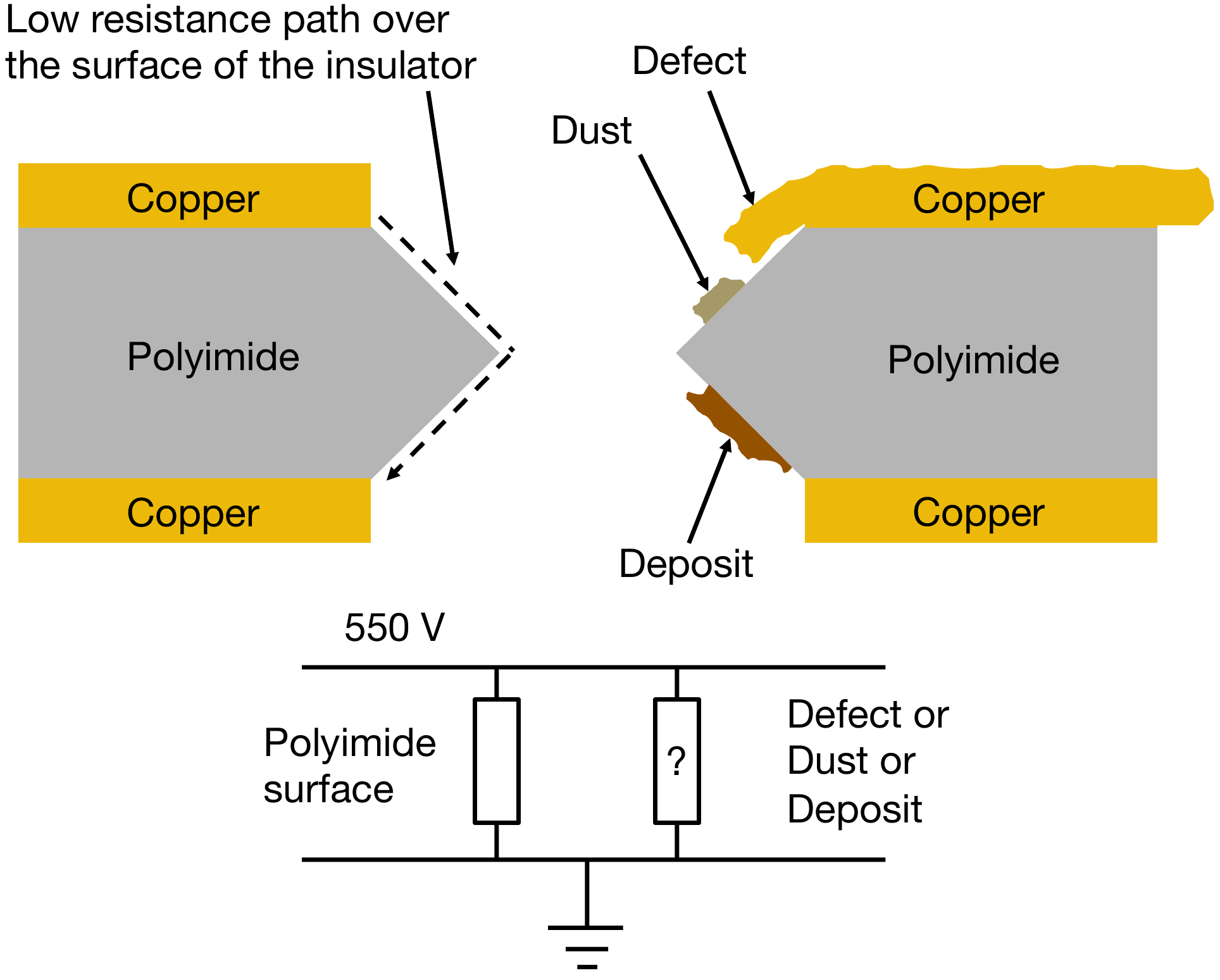}
\caption{\label{fig:QC2principle}Schematic representation of the cross-section of a GEM foil at the location of a hole. As shown on the right of the figure, defects in the copper, deposits, and dust can affect the resistance of the GEM foil.}
\end{figure}

The GEM foil is the core component of the amplification structure of the detector. Because of its microscopic nature, the qualification of the GEM foils is not done optically, but electrically. The presence of dust, chemical contaminants, or mechanical defects creates a low resistance bridge between the two electrodes of the GEM foil as shown in Fig.~\ref{fig:QC2principle}. For healthy foils, the total resistance should be greater than 20 G$\Omega$ in environments with relative humidity below 50\%. In the extreme scenario of a short-circuit between the GEM electrodes, the measured resistance of the foil is equal to 10 M$\Omega$---the value of the protection resistors that are soldered directly onto the GEM foils---or lower in case of multiple short-circuits. 

\begin{figure}[!htp]
\centering
\includegraphics[width=\textwidth]{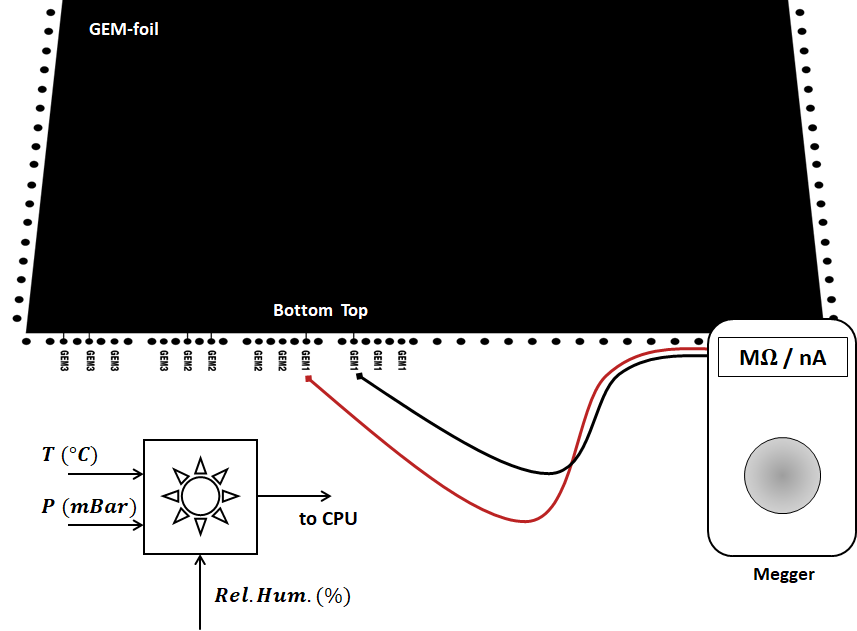}
\caption{\label{fig:QC2fast}Resistance test setup of a single GEM foil with a Megger Giga-Ohmmeter.}
\end{figure}

\begin{figure}[!hbp]
\centering
\includegraphics[width=0.9 \columnwidth]{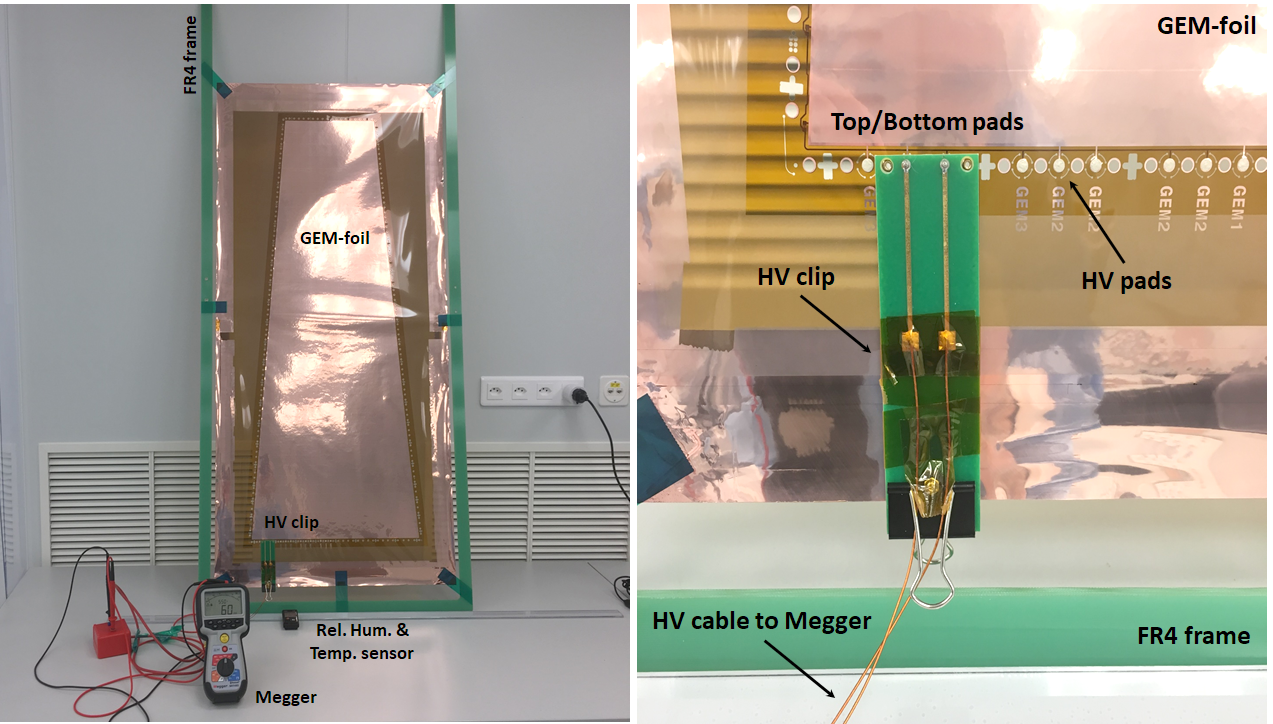}
\caption{\label{fig:QC2_short_setup}Typical setup for the electrical resistance test of a GE1/1 GEM foil.}
\end{figure}

The resistance across the foil is measured in a cleanroom environment (ISO 7/class 10,000 or better) using a handheld Giga-Ohm insulation meter (Megger MIT485). The meter applies 550 V across the foil and measures the corresponding leakage current to derive the foil resistance (see Fig.~\ref{fig:QC2fast} for a schematic representation of this test, and Fig.~\ref{fig:QC2_short_setup} for a photo of the procedure being performed).

\begin{figure}[!ht]
\centering
\includegraphics[width=0.8\textwidth]{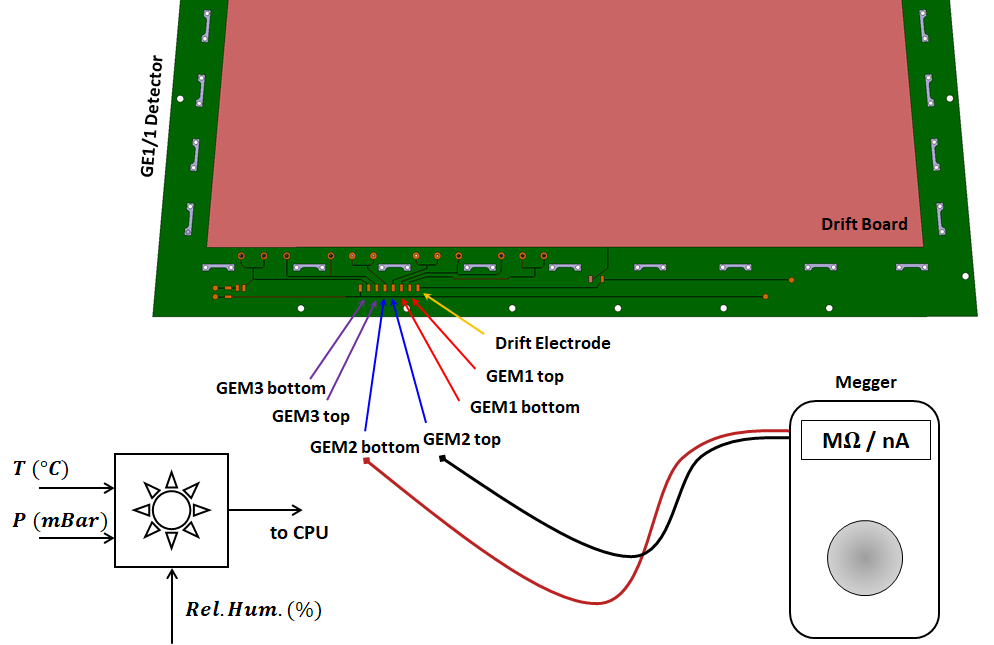}
\caption{\label{fig:QC2assembly}Resistance test setup of a GEM stack during assembly.}
\end{figure}

The same procedure is repeated multiple times during the various steps of the detector assembly to detect any foils that might have developed short-circuits because of the manipulations required for testing, transport, and assembly. In this case, the voltage is applied on dedicated pads present on the assembly table or on the drift PCB on which the GEM stack is mounted, as shown in Fig.~\ref{fig:QC2assembly}. The voltage is applied to the active area of the GEM foil via spring-loaded pins.

\FloatBarrier

The same technique is also used as needed during production, testing, or operation of a chamber, to clean the GEM foils in case of contamination, defects, or dust particles trapped inside the GEM holes. The current injected by the Giga-Ohmmeter flows primarily through contaminated areas with low impedance, resulting in the evaporation or burning of the impurity. In dry environments, it is also possible to clean the inside surfaces of a damaged hole responsible for a short-circuit between the GEM electrodes.

\subsection{QC Step 2b -- Long-term monitoring of the GEM foil leakage current}

\subsubsection{Motivations}
While the impedance check helps to identify major problems with a GEM foil, it might not be sufficiently sensitive to minor issues or hidden defects that can affect the long-term stability of the chamber. Specifically, some GEM foils could reach their nominal voltages with high impedances, while simultaneously experiencing discharges at high frequency (e.g., several per hour). 
These discharges can seriously and permanently damage the GEM foils, and they can also prevent stable operation of the detector, once in continuous run mode. Consequently, it is essential to monitor the leakage current and discharge rate of all the GEM foils for several hours before the detectors are assembled.

\subsubsection{Selection criteria}
\begin{figure}[!ht]
\centering
\includegraphics[width=\columnwidth,trim={0.2cm 0.4cm 2cm 1.4cm}, clip]{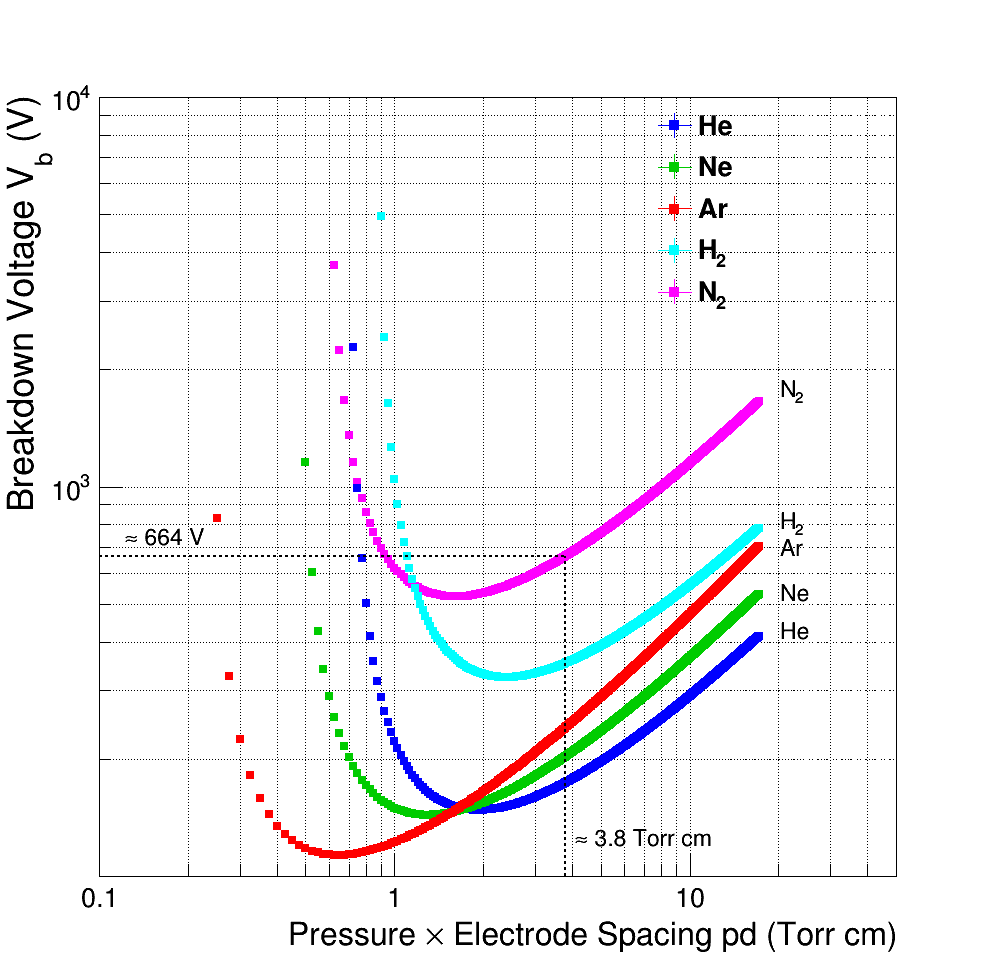}
\caption{\label{fig:Paschen_Curve}Paschen curves for electrical breakdown voltages of parallel plates in different gas mixtures \cite{Lieberman_2005} at standard temperature. For GEM foils surrounded by gas at atmospheric pressure and room temperature, the pressure $\times$ spacing value is \mbox{3.8 Torr $\times$ cm.}}
\end{figure}

The total leakage current of a GEM foil mainly depends on the resistivity of the biconical holes of the polyimide layer between the two copper layers of the foil. This resistivity can vary with the ambient humidity or because of contaminants. For a precise measurement of the leakage current, the foil needs to be operated at the maximum stable voltage, i.e.\ just below the breakdown voltage of the gas. As the Paschen curves in Fig.~\ref{fig:Paschen_Curve} indicate that the breakdown voltage in pure nitrogen is close to 660~V for a GEM foil where the two copper layers are separated by 50~$\mu$m, an appropriate voltage for testing GEM foil stability is 600~V in pure nitrogen. In such conditions, and with a relative gas humidity below 10\%, we require that the leakage current of a GE1/1-sized foil not exceed 1~nA, with a maximum discharge rate below 3 discharges per 5 hours.

\subsubsection{Test setup and procedure}
Before performing this quality control step, both sides of the foils are mechanically cleaned using an anti-static silicone roller to remove dust and contaminants. The GEM foils are enclosed in a plexiglass box filled with pure nitrogen flowing at a rate of 30-40~L/hr. The box, which can accommodate up to 8 foils, is placed in a cleanroom and covered to prevent ambient light from reaching the foils. This is done because electrons can be released from the GEM copper as a consequence of a photoelectric effect induced by UV photons emitted from fluorescent lighting. We observed that when the foils operate at the maximum voltage of 600~V, these photo-electrons can get sufficiently amplified by the GEM to generate a leakage current of the order of 2--3~nA, which can bias the test results for the intrinsic leakage current measurement.

\begin{figure}[!htp]
\centering
\includegraphics[width=\columnwidth]{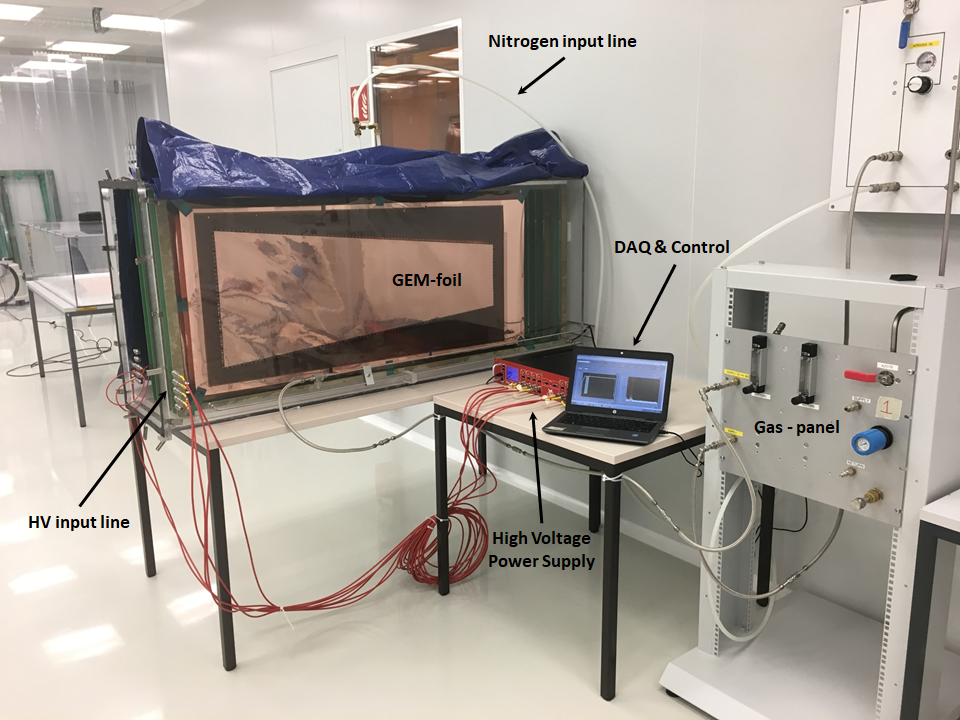}
\caption{\label{fig:QC2_long_setup}Test setup for the long-term measurement of leakage currents in the GE1/1 GEM foils.}
\end{figure}

The  foils are powered by an 8-channel CAEN R1471HETD power supply, which has a current monitoring resolution down to 50~pA. 
The power supply is controlled by a LabView interface that is programmed to run the test automatically.
A picture of the QC2b test setup is shown in Fig.~\ref{fig:QC2_long_setup}.
\newpage

The test procedure consists of three main steps: the initial ramp-up phase, a power cycling phase, and the final stability phase.
During the ramp-up phase, the voltage is increased from 0 to 500~V in steps of 100~V, and then from 500~V to 600~V in smaller steps, typically in increments of 50~V. After a desired voltage is reached, it is maintained for 15 seconds before moving to the next step.
During the cycling phase, the voltage is quickly ramped up and down between 600~V and 100~V within ten seconds in order to stress the foil and to lift potential solid contaminants off the foil via the oscillating electric field.
During the stability phase, the foil is maintained at the maximum voltage of 600~V for 6--8~hours. This final step is designed to verify the long-term behavior and the stability of the foil during continuous operation.
At any phase of the test, if the leakage current exceeds a limit of 20~$\mu$A (a typical magnitude of a discharge current), the system trips and automatically restarts at the same phase. If the system records more than three trips in the same phase, it automatically goes back to the previous phase. Three phase failures result in aborting the test and automatic rejection of the foil from chamber production.

\subsubsection{Typical results}
\begin{figure}[!htp]
\centering
\includegraphics[width=0.62\columnwidth]{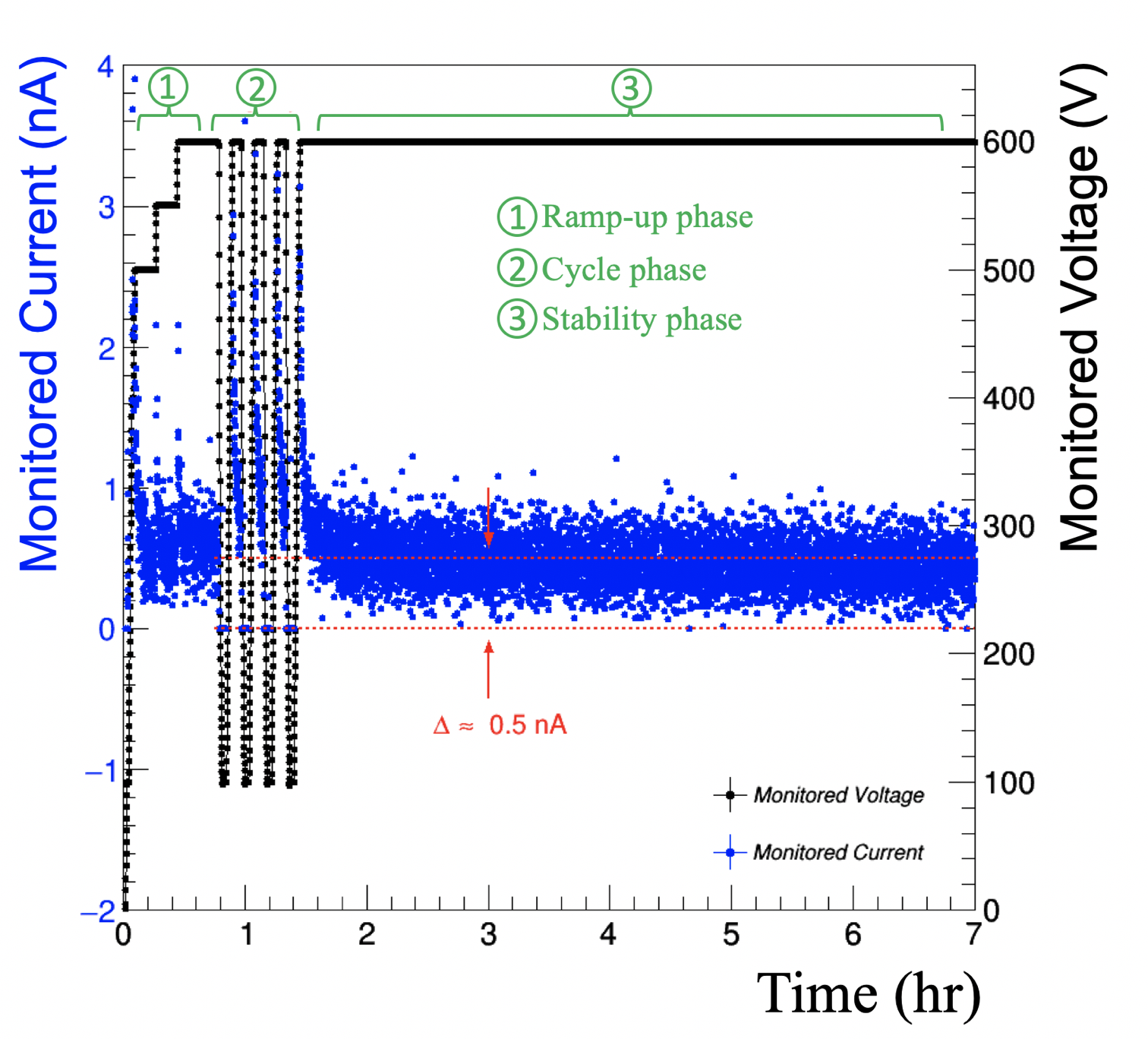}
\caption{\label{fig:QC2_accpected_foil}Typical leakage current behavior in a validated GEM foil.}
\end{figure}

Figure \ref{fig:QC2_accpected_foil} shows typical QC2b data for a ``good" GEM foil, that is fully validated and approved for detector assembly. All three phases are completed and it is demonstrated that the foil can safely operate long-term at 600~V without high voltage (HV) trips and with an average leakage current below 0.5~nA.

\begin{figure}[!ht]
\centering
\includegraphics[width=0.62\columnwidth]{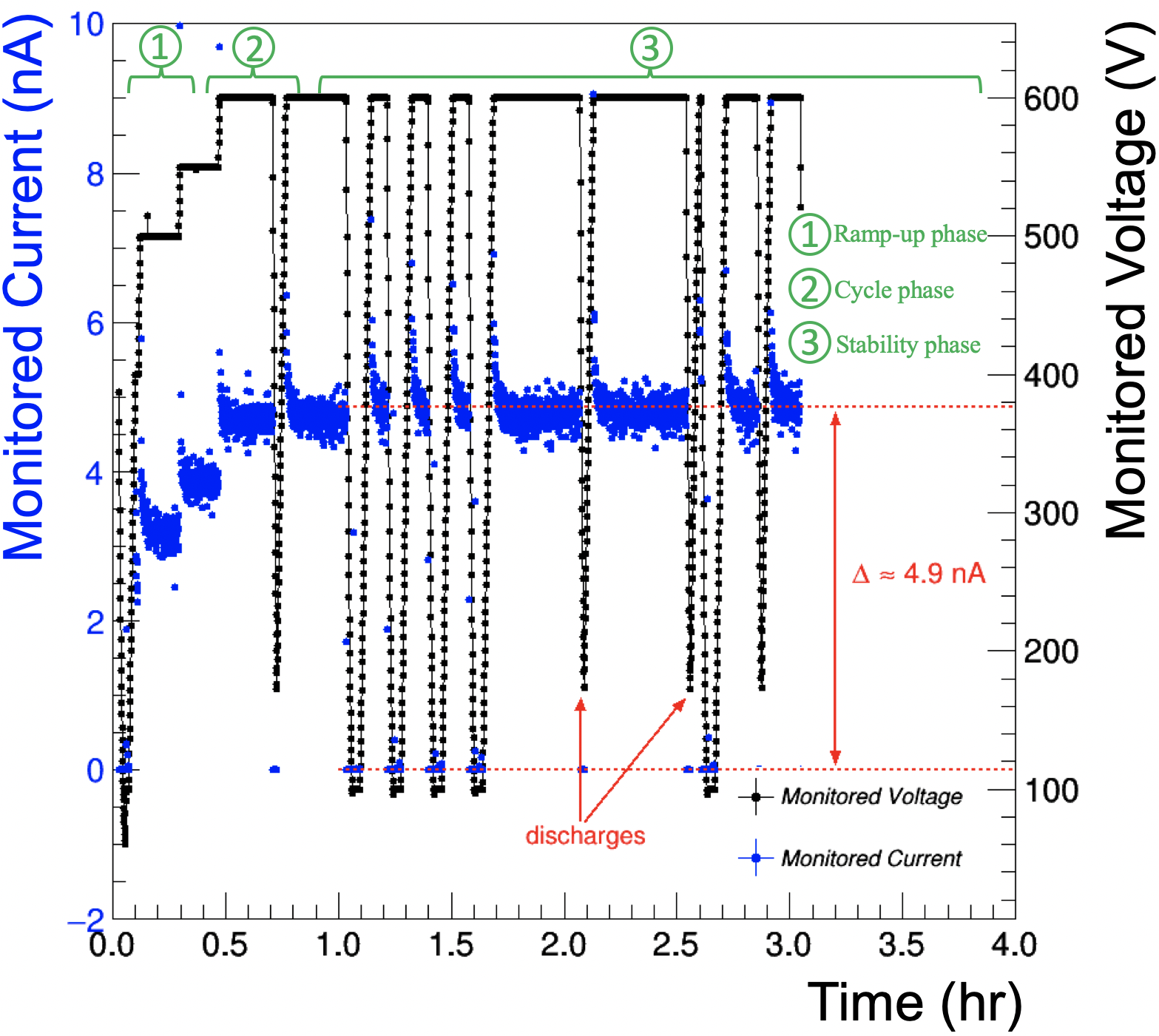}
\caption{\label{fig:QC2_rejected_foil}Typical leakage current behavior in a rejected GEM foil.}
\end{figure}

Conversely, Fig.~\ref{fig:QC2_rejected_foil} shows the typical QC2b data of a ``bad" GEM foil that suffers from instabilities. The plot shows multiple trips that force the system to go back from the stability phase to the cycling phase. The repetition of the trips and the measurement of a high leakage current prematurely abort the test. Such a foil would require an additional, in-depth cleaning at the manufacturer's facility before going back through the entire QC protocol from the beginning.

Over the course of the GE1/1 detector mass production, on average, about one to two foils per batch of 15 foils were rejected by the QC2 tests and sent back to the manufacturer. After additional, in-depth cleaning at the manufacturer’s facility, those foils were tested again and most of them could be validated for assembly. Overall, only one foil out of 485 had to be completely rejected without any possibility for recovery.

\subsection{QC Step 3 -- Leak test of closed detector volume}

\subsubsection{Motivations}
Detector gas-tightness is crucial to avoid gas leaks and to reduce the possibility of contamination by external pollutants.
The presence of undesired gas molecules, e.g.\ those found in air, particularly oxygen and other electronegative elements, can deteriorate the charge amplification and electron transfer process, which will affect the detector performance during operation. Additionally, dust particles could enter the gas volume and jeopardize the foil integrity. Such effects can have a significant impact on the HV stability and the long-term behavior of the chamber.
Given the design of the GE1/1 detectors, where a large number of screws are employed, the leak test is the first verification performed after the assembly.

\subsubsection{Selection criteria}
To maintain high gas purity, the maximum acceptable leak rate for a single detector is set to 1\% of the total incoming gas flow rate. For GE1/1, this rate is initially set to 2.5 L/hr with a maximum internal over-pressure of 25 mbar. Directly measuring a leak rate as little as 0.025 L/hr is challenging and would require a specific and expensive tooling, which is not necessarily available at all production sites. The leak rate, however, can also be derived from the measurement of the internal pressure loss when the chamber input and output are both closed. Such a measurement is much easier and more reliable with standard laboratory equipment.

Assuming that the detector volume remains constant during the test, the time evolution of the internal pressure can be expressed as a simple exponential:

\begin{equation}
\label{eqn:pressure}
P_{\textrm{int}} = P_0e^{-t/\tau}
\end{equation}

\noindent where $P_{0}$ is the initial over-pressure and $\tau$ is the time constant of the system that depends on the leak point dimensions. From the ideal gas law, $PV = nRT$, with $P$ being the absolute pressure inside the detector volume $V$, we can then derive the quantity of gas $n$ lost after a given time $t$:

\begin{equation}\label{eqn:pvExpanded} n = \dfrac{V}{RT}\big( P_{\textrm{atm}} + P_{\textrm{int}}(t)\big). \end{equation}

\noindent Substituting $P_{\textrm{int}}$ from equation \eqref{eqn:pressure} in equation \eqref{eqn:pvExpanded} and differentiating with respect to time, we find

\begin{equation} \dfrac{dn}{dt}=\dfrac{VP_0}{RT}\bigg(\dfrac{-1}{\tau}\bigg)e^{-t/\tau}.
\end{equation}

\begin{figure}[bp!]
\centering
\includegraphics[width=0.7\columnwidth]{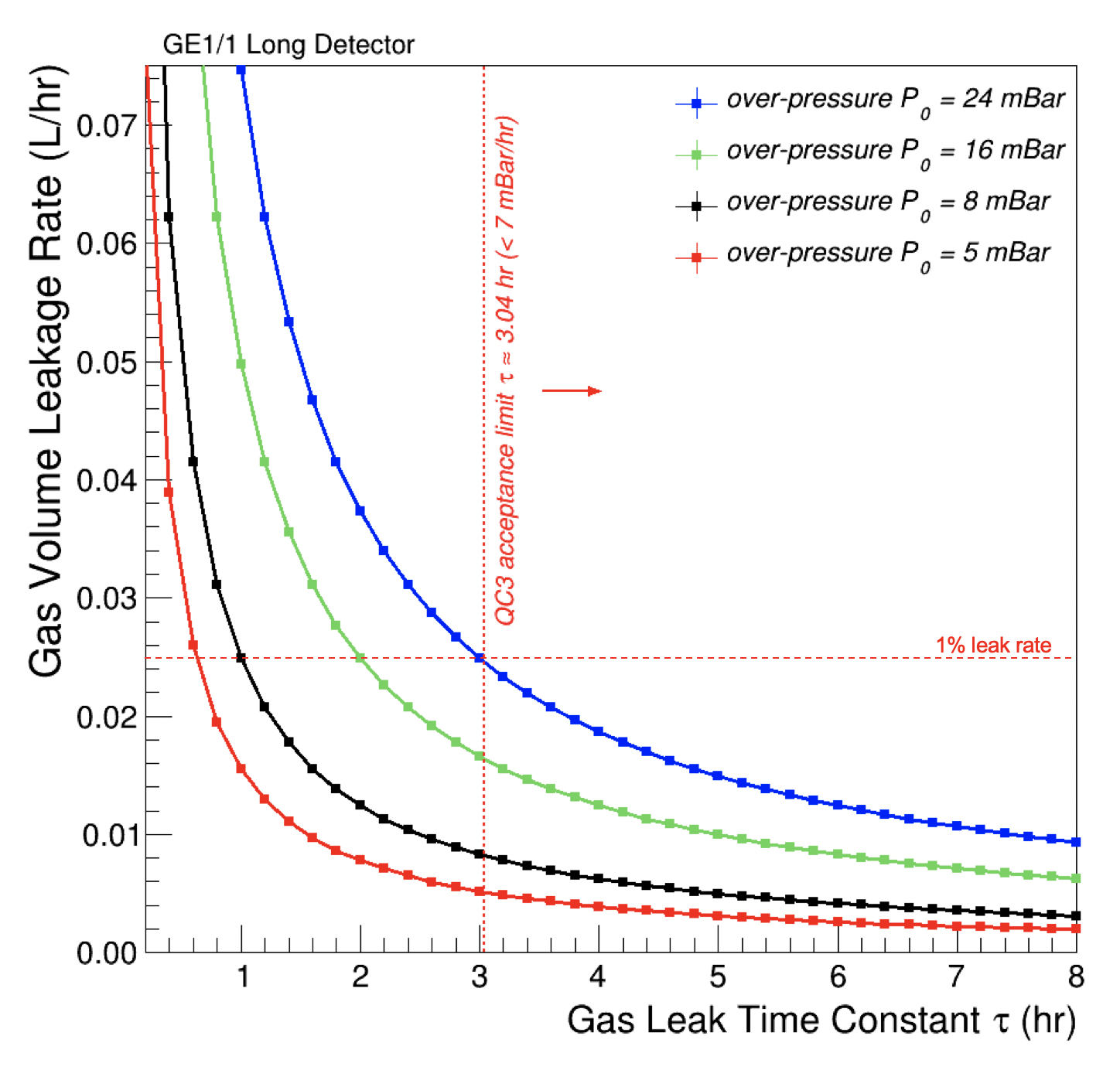}
\caption{\label{fig:QC3Leakage1} Calculations of the gas leak rate as a function of the time constant measured during QC Step 3 for the GE1/1 chambers.}
\end{figure}

\begin{figure}[!ht]
\centering
\includegraphics[width=0.6\columnwidth]{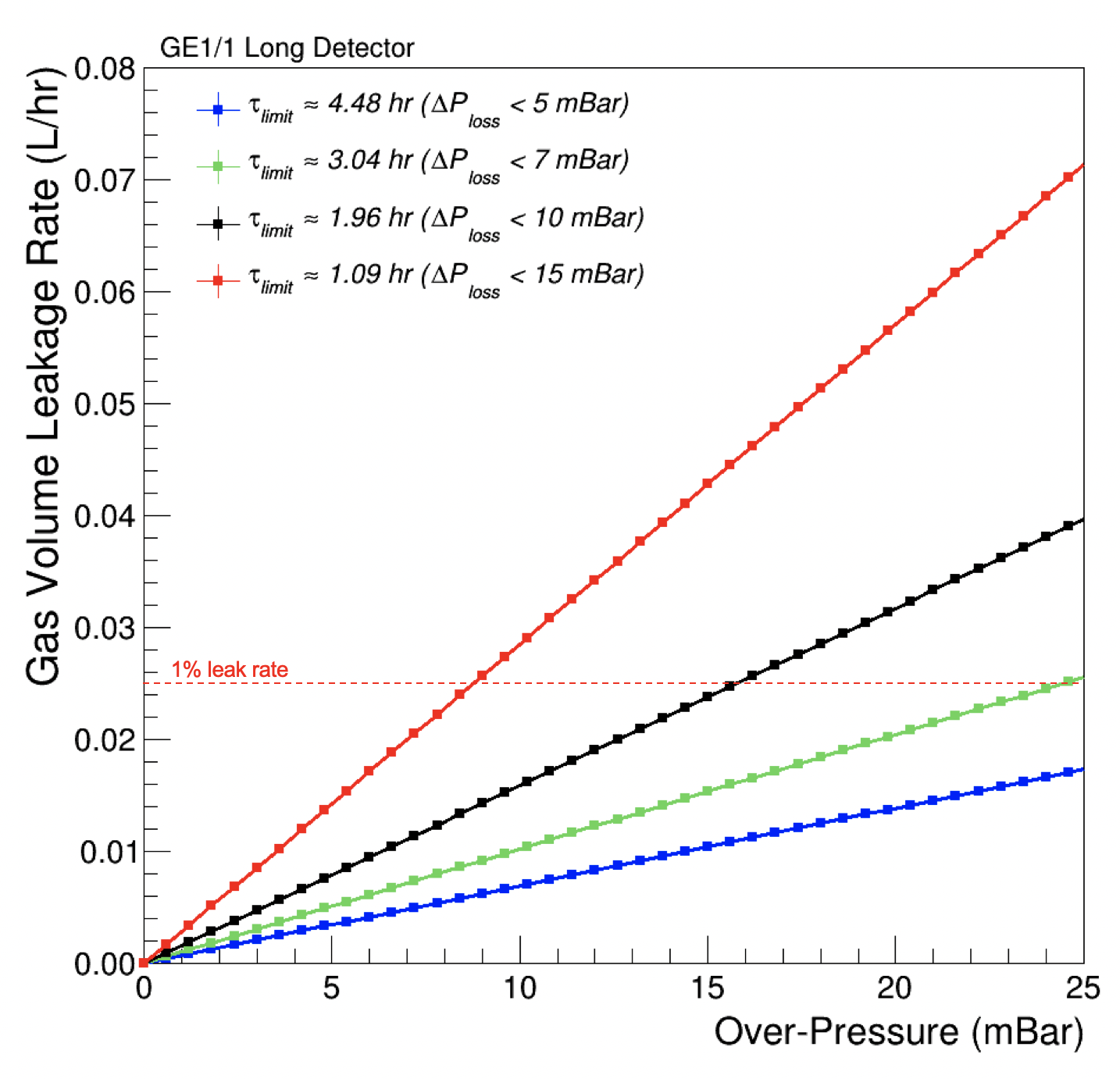}
\caption{\label{fig:QC3Leakage2}Calculations of the gas leak rate in GE1/1 chambers as a function of the internal over-pressure.}
\end{figure}

Consequently, we can calculate the detector leak rate from the measurement of the pressure-loss time constant $\tau$. Figure \ref{fig:QC3Leakage1} shows the gas leak rate as a function of the time constant for different values of internal over-pressure $P_0$ during operation. Figure \ref{fig:QC3Leakage2} shows the gas leak rate as a function of the internal over-pressure for different time constants $\tau$. To ensure that the leak rate will remain below 1\% of the total incoming flow rate, the time constant for each GE1/1 chamber should be greater than 3.04 hours. 

\subsubsection{Test setup and procedure}
\begin{figure}[tp!]
\centering
\includegraphics[width=0.9\columnwidth]{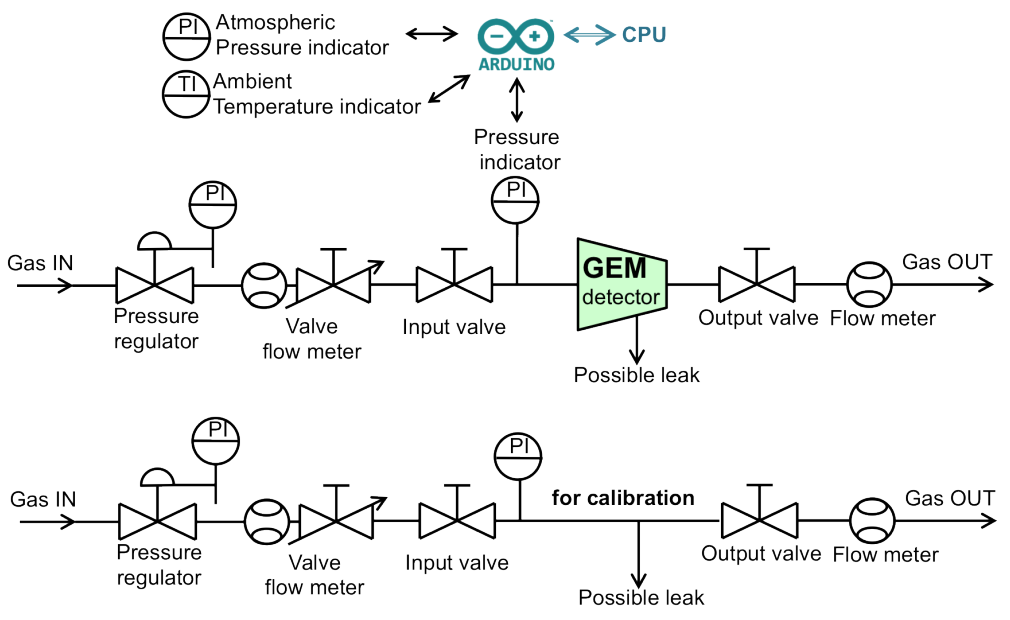}
\caption{\label{fig:arduinoDiagram}Diagram of the Arduino DAQ and the gas system for the leak test (top diagram) and for calibration (bottom diagram).}
\end{figure}

The leak test aims to  quantify the gas leak rate of a GE1/1 detector by monitoring the drop of the internal over-pressure as a function of time. 
A pressure sensor connected to an Arduino-Mega micro-controller allows the monitoring of the over-pressure in the detector and the gas system. The ambient temperature and pressure are also monitored via sensors connected to the Arduino board during this quality control procedure. The top schematic in Fig.~\ref{fig:arduinoDiagram} shows a basic diagram of the data acquisition (DAQ) and gas system setup for this test.
Before the leak test is conducted, the DAQ system is calibrated to ensure accurate measurement of the leak rate (see the bottom diagram of Fig.~\ref{fig:arduinoDiagram}). After calibration of the system, a chamber is connected to the gas system with its outflow gas line plugged. The chamber is then pressurized to about 25 mbar, and the over-pressure is monitored every minute over the course of one hour.

\subsubsection{Typical results}

\begin{figure}[!htp]
\centering
\includegraphics[width=0.6\columnwidth]{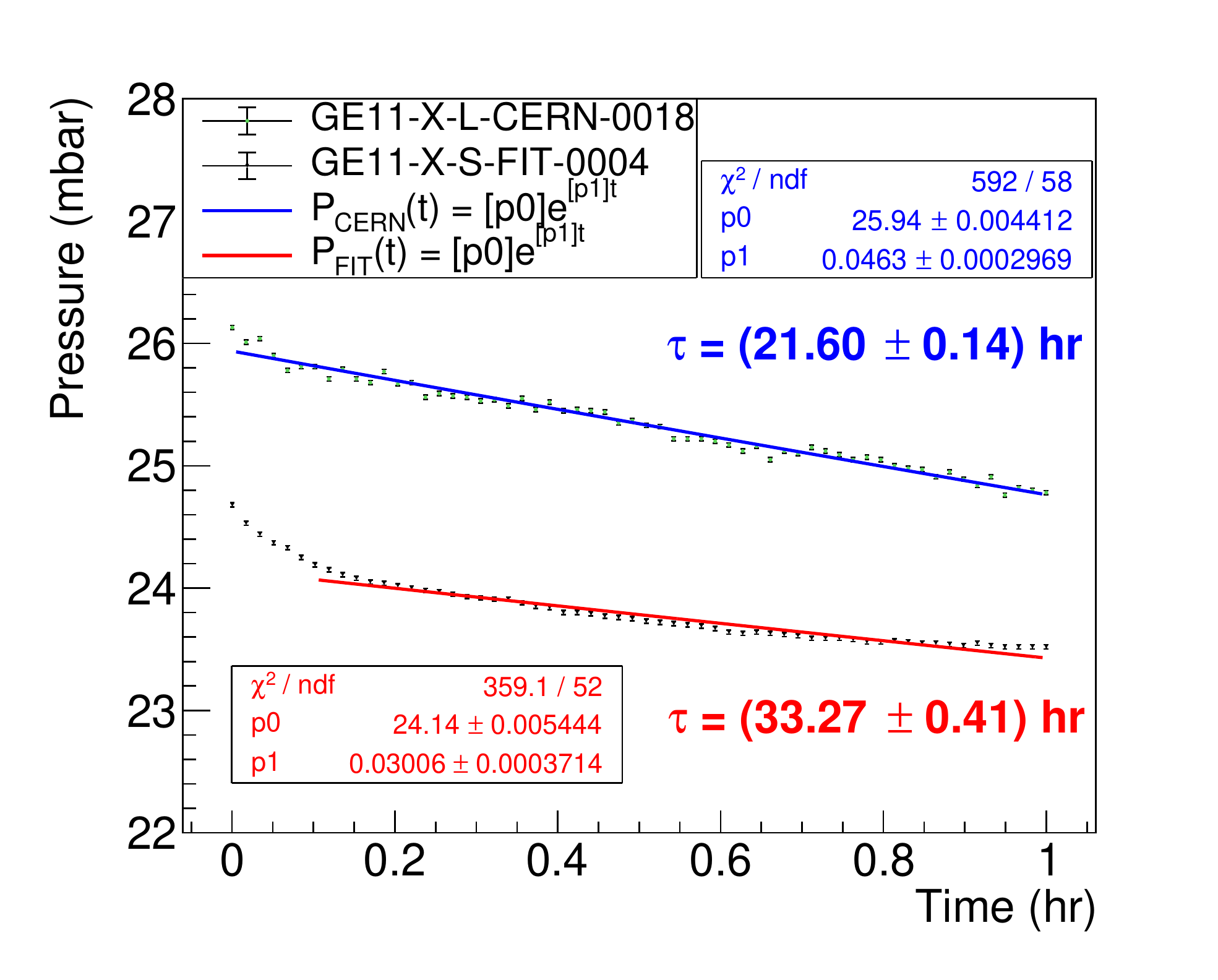}
\caption{\label{fig:QC3CERNandFIT}Examples of the measured gas leak rate for closed GE1/1 chambers over one hour (serial numbers: GE1/1-X-L-CERN-0018 and GE11-X-S-FIT-0004). Both chambers display time constants that are well above the acceptance criterion for a gas-tight chamber ($\tau\geq 3.04$ hr).}
\end{figure}

\begin{figure}[bp!]
\centering
\includegraphics[width=0.6\columnwidth]{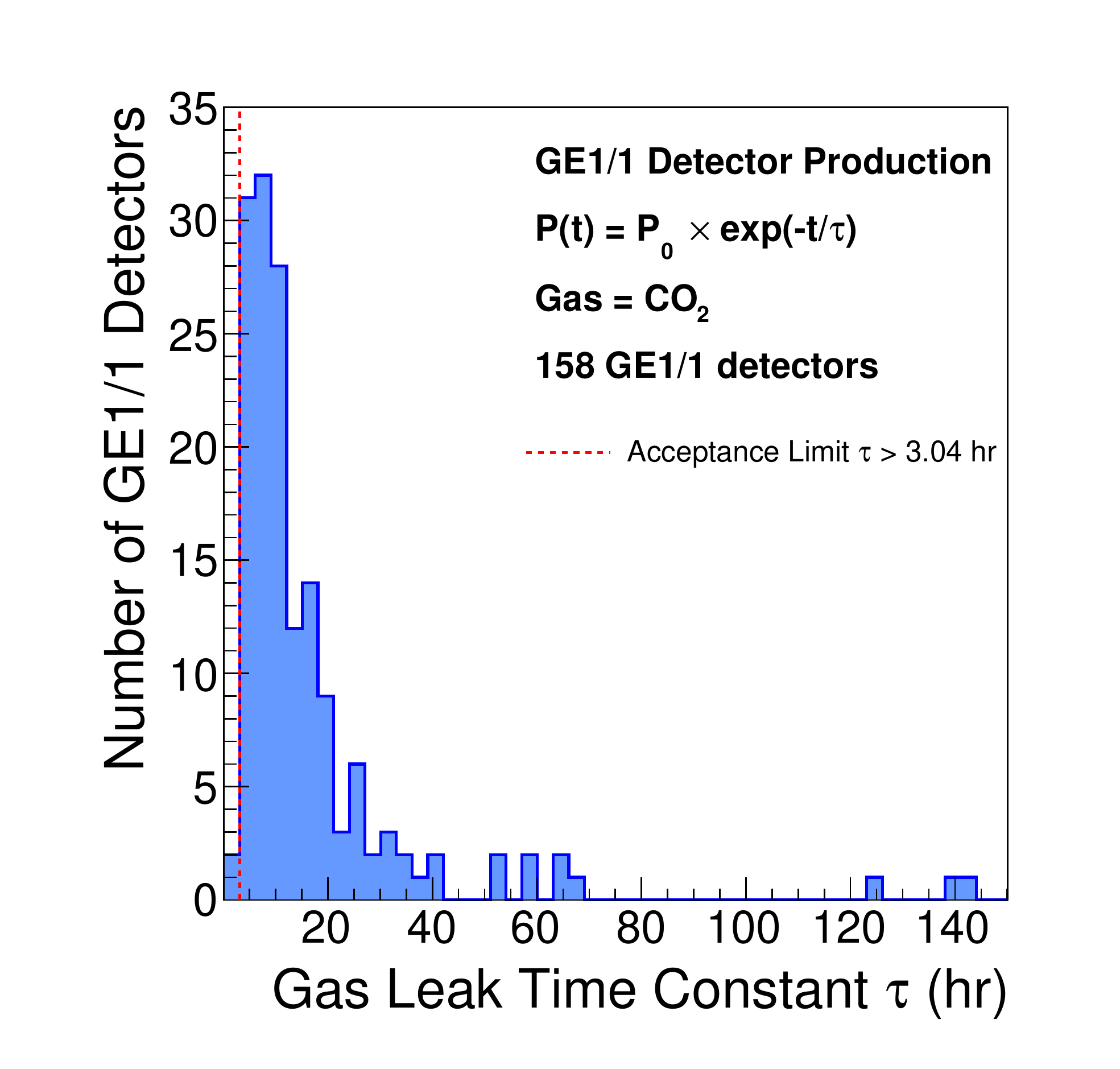}
\caption{\label{fig:QC34gas}Gas leak time constants of all the GE1/1 detectors.  The threshold for acceptance is also shown. Results for 158 GE1/1 detectors out of 161 produced are shown. Out of these, 156 detectors are accepted and two are rejected.}
\end{figure}

The pressure drop is modeled by the function $P(t)=P_0\ e^{-t/\tau}$ discussed above, where $P_0$ is a constant representing the initial over-pressure of the detector (set to about 25 mbar at all the production sites and for each detector) and the time constant $\tau$ quantifies how fast the over-pressure inside the detector decreases as a function of time.
\newpage
Results of the leak tests for two different chambers measured at two different sites (serial numbers: GE1/1-X-L-CERN-0018 and GE1/1-X-S-FIT-0004) are displayed in Fig.~\ref{fig:QC3CERNandFIT}. The expected exponential behavior is exhibited by GE1/1-X-L-CERN-0018 (green data points with a blue fitted exponential curve), and shows a time constant of $\tau=(21.6\pm0.14)$ hr, which is well above the acceptance criterion for a gas-tight chamber ($\tau\geq 3.04$ hr). In some cases, we observe chambers that initially do not follow an exponential leak profile. This can be attributed due to the bulging of the readout and drift PCBs after the initial pressurization of the chamber to 25~mbar. Once the PCBs return from the deformed shape to their planar shape, the gas volume returns to its nominal, constant shape, and  the gas escaping from the closed chamber exhibits the theoretically predicted exponential leak behavior. Consequently, the time constant determined from the fit for GE1/1-X-S-FIT-0004 (black data points with a red fitted exponential curve) has the first few points excluded. Note that this chamber exhibits a time constant of $\tau=(33.27\pm0.41)$ hr, which is also well above the acceptance criterion. Out of the 161 fully assembled chambers, 158 chambers could be tested in this QC step; three chambers were eliminated due to failures in earlier QC steps and were not tested for leak tightness. As shown in Fig.~\ref{fig:QC34gas}, all 158 GE1/1 detectors are found to be gas-tight except two built with a pre-series internal frame. These two detectors with excessive leak rates are not installed in CMS. 

\newpage

\subsection{QC Step 4 -- Linearity test of the HV divider}\label{sect:QC4}
\subsubsection{Motivation}\label{sect:QC4Mot}
The goal of this test is to check the on-detector circuitry that distributes the high voltage to the detector electrodes. The proper functioning of this circuitry is important for the operational stability of the GE1/1 detector with respect to the applied HV, both for guaranteeing a stable effective gas gain and for avoiding discharges that could damage the surface of the GEM foils or the readout electronics. This need is particularly crucial for GE1/1 detectors because of the large size of the GEM foils. Even with the foils being segmented into 40 HV sectors (with the sensitive sector areas ranging from dozens to a hundred cm$^2$), the foils still have large total capacitances---on the order of tens of nF---and store considerable charge.

\begin{figure}[bp!]
\centering
\includegraphics[width=0.85\columnwidth]{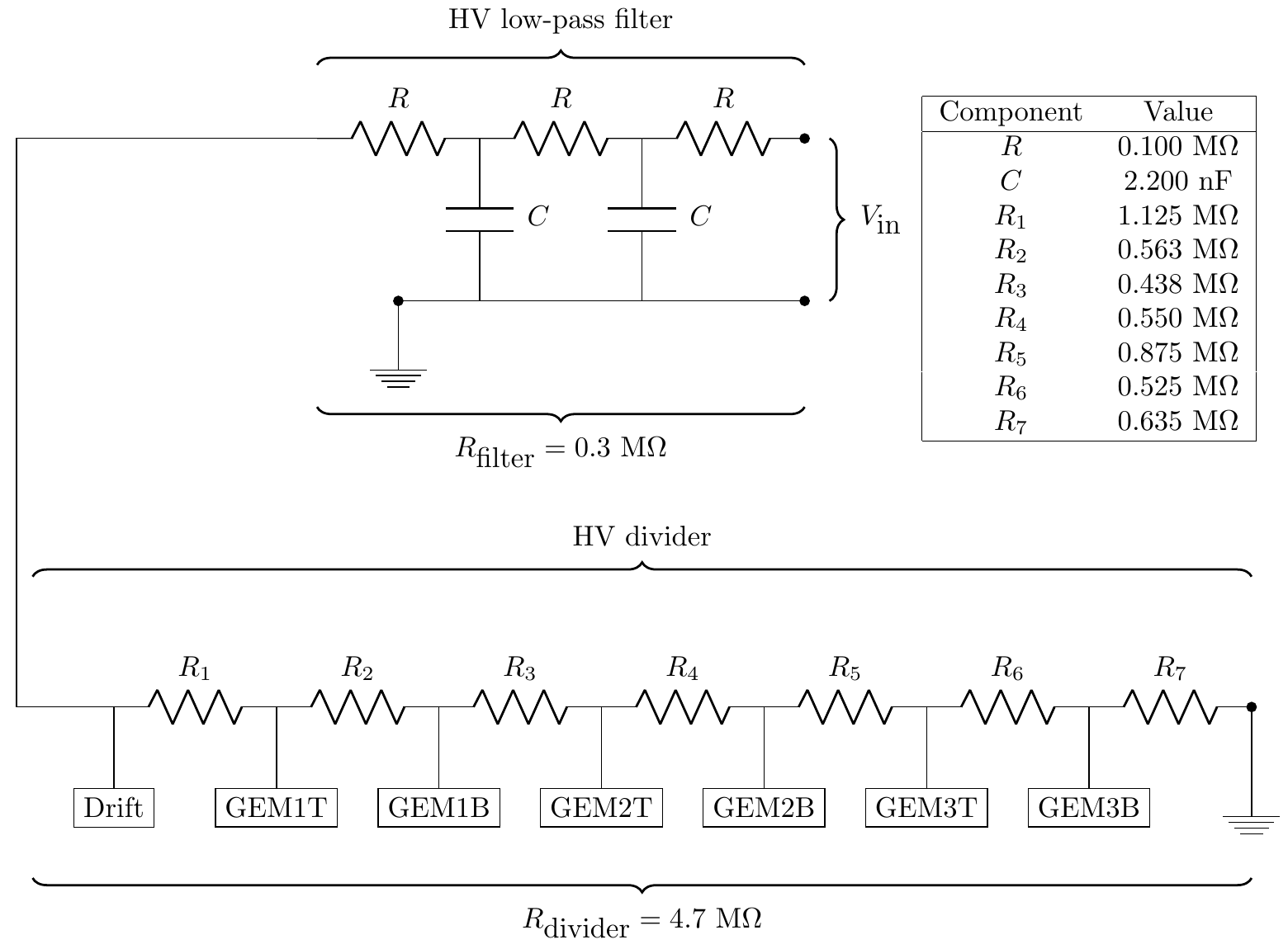}
\caption{\label{fig:QC4circuit}The circuit diagram for the HV filter (top) and the ceramic HV divider (bottom).}
\end{figure}

During the detector production and test phase, a simple ceramic voltage divider, directly soldered onto the drift board, is used to divide the voltage from a single high voltage input line~\cite{Colaleo:2015vsq}. This ceramic divider comprises a small network of ohmic resistors (see the bottom of Fig.~\ref{fig:QC4circuit} for the circuit diagram) with resistances $R_{i}$ (see the top right of Fig.~\ref{fig:QC4circuit} for the values of the resistors). This HV divider has an equivalent resistance of $R_{\mathrm{divider}}=4.7$ M$\Omega$. The current $I_{d}$ through this network produces voltage drops  $V_{i} = R_{i} \times I_{d}$ across the various resistors that are used to create the appropriate electric potentials needed to power the different electrodes $i$ of the detector. A custom low-pass protection filter (with equivalent resistance $R_{\mathrm{filter}}=0.3$ M$\Omega$) connects the detector to the power supply to eliminate the high frequency noise (see the top of Fig.~\ref{fig:QC4circuit} for the circuit diagram). The total equivalent resistance of the ceramic divider and low-pass filter is 5.0 M$\Omega$. 
The HV stability is quantified by monitoring two quantities: the deviation of the measured total resistance of the powering circuit with respect to the one predicted by Ohm's law, and the noise induced on the bottom of the third GEM foil in a gas with high ionization energy.

\subsubsection{Selection criterion}
For each HV step, the current through the powering circuit is recorded, as shown in Fig.~\ref{fig:QC4example}.
The measured resistance of the powering circuit $R_{m}$ is then obtained by computing the average of all the values recorded in each step. This value is compared with the nominal resistance $R_{n}$ of the powering circuit measured with a standard multimeter (at a few volts) before soldering the divider onto the detector, and the resistance deviation is computed.
The deviation of the resistance of the powering circuit with respect to Ohm's law, labeled $D_{R}$ in the following discussion, is defined in Eq.~\eqref{eqn:ResDev} as the relative deviation of the measured resistance, $R_{m}$, with respect to the nominal value, $R_{n}$:
\begin{equation}\label{eqn:ResDev}
D_{R} = \frac{R_{m} - R_{n}}{ R_{n}}.    
\end{equation}
The value of the resistance deviation includes the tolerances on the resistors in the HV divider and HV filters used for biasing the GE1/1 detector. Ten resistors are employed in total in this powering circuit, each of them rated with a 1\% tolerance. Summing all of the involved tolerances in quadrature shows that a deviation $D_{R}$ of up to 3\% can be tolerated. Consequently, GE1/1 detectors exhibiting resistance deviation greater than this value are rejected or undergo further investigation.\\

\subsubsection{Test Setup}

A commercial power supply able to provide up to 1 mA current with 1 $\mu$A monitoring resolution is used to ramp the HV applied to the filter and divider (connected to the detector) up to 5 kV in steps of 100 V, and also used to monitor the current. The detector is flushed with pure CO$_2$ gas with a flow rate of 5 L/h. 

\subsubsection{Results}

\begin{figure}[!ht]
\centering
\includegraphics[width=0.7\columnwidth]{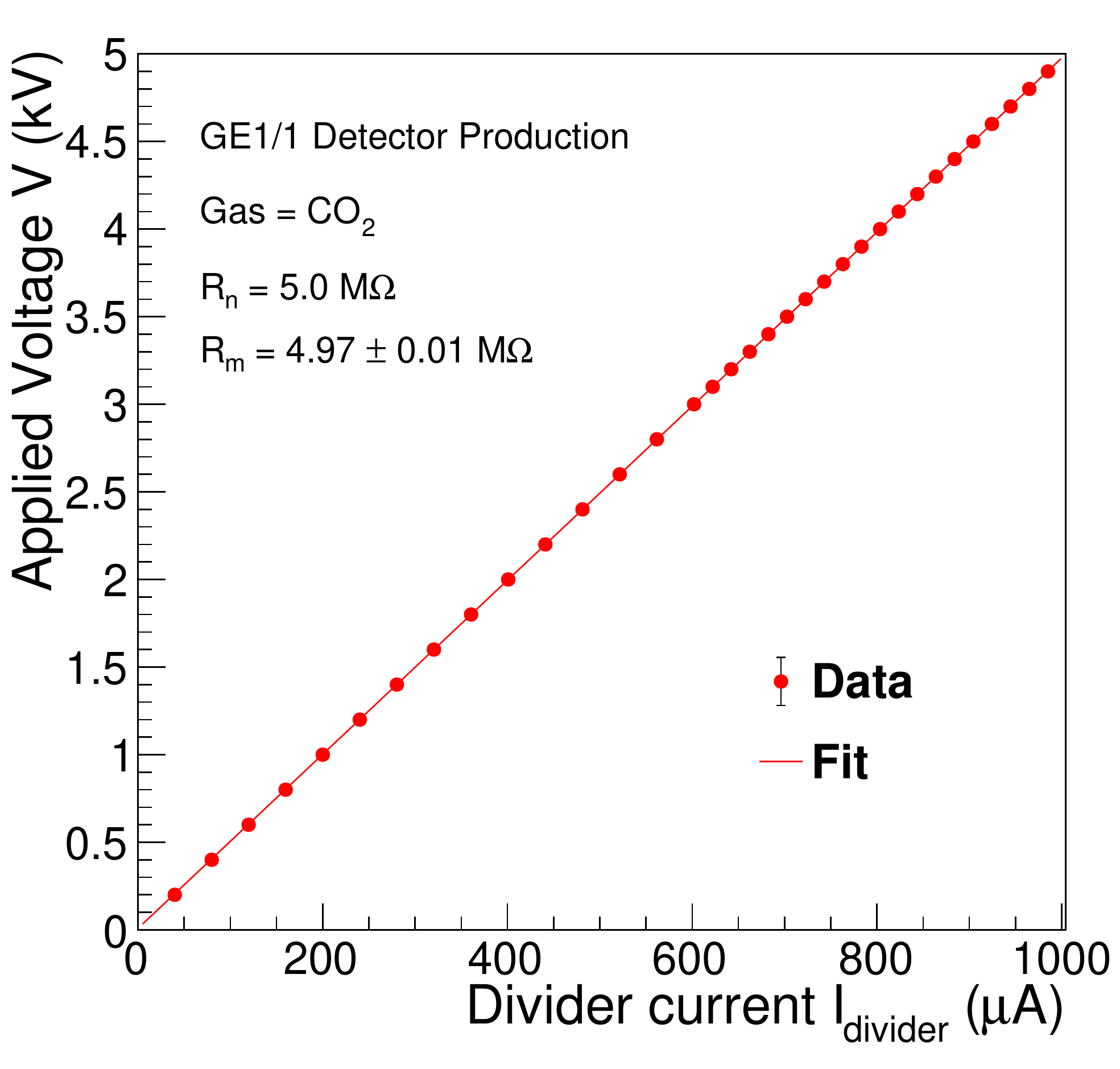}
\caption{\label{fig:QC4example}High voltage vs.\ divider current curve for a GE1/1 detector (GE11-X-S-CERN-0012), measured in pure CO\textsubscript{2}. Error bars are smaller than the markers. A linear fit of the measured data points is shown. The slope from the fit is taken as the measured resistance $R_m$.}
\end{figure}

\begin{figure}[!ht]
\centering
\includegraphics[width=0.6\columnwidth]{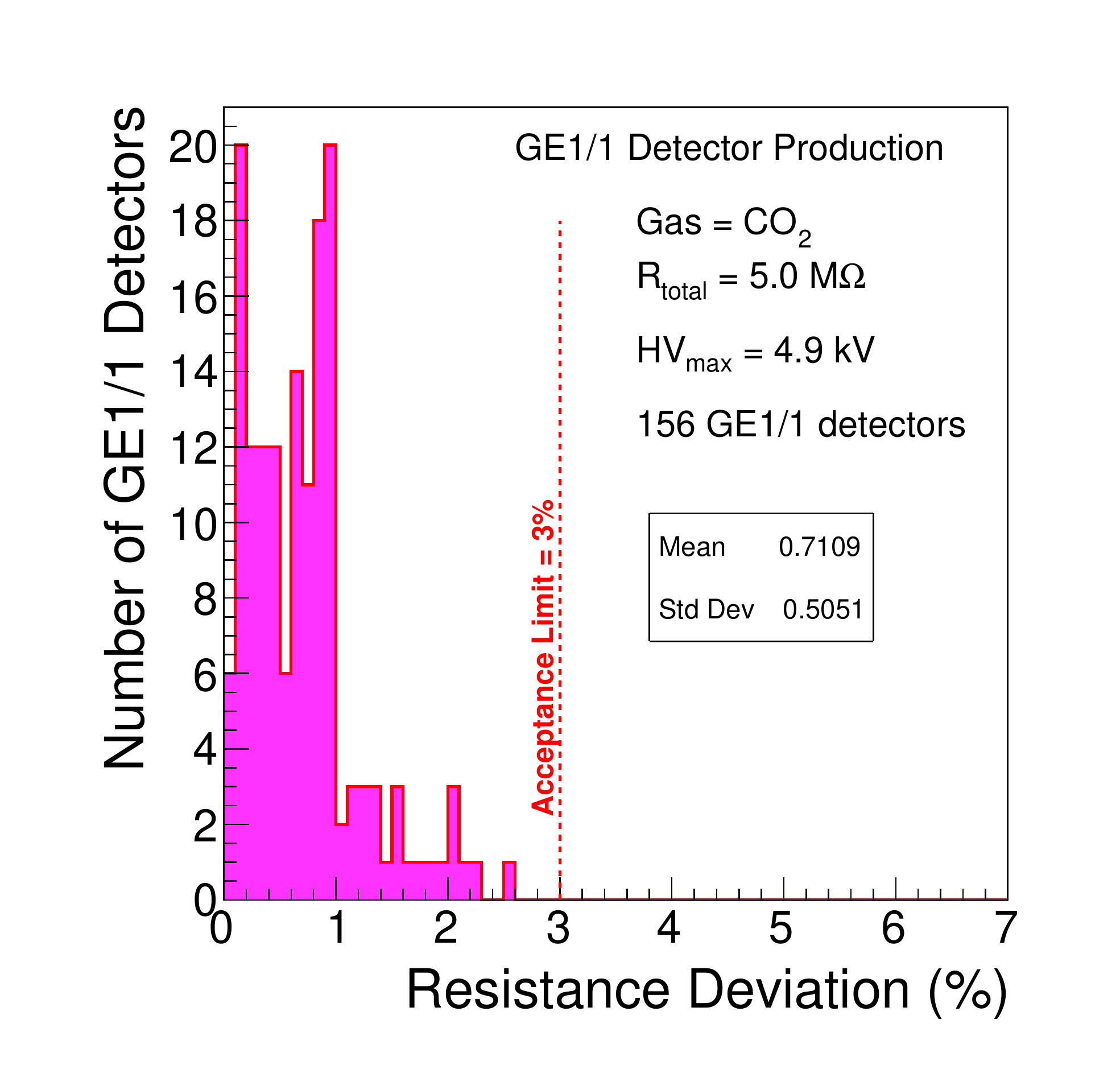}
\caption{\label{fig:QC4resistance}Deviation of the measured resistance, $D_{R}$, for the high voltage powering circuit connected to the detector, with respect to its nominal value for all 156 GE1/1 detectors (the two missing detectors did not pass the previous QC test). The 3\% threshold for accepting a detector is also shown.}
\end{figure}

To characterize the biasing circuit, the slope from the linear fit on the points of a plot of the applied voltage as a function of divider current is obtained (see Fig.~\ref{fig:QC4example}), and then the resistance deviation is computed using equation~\ref{eqn:ResDev}.
The distribution of the resistance deviations for all of the GE1/1 detectors is displayed in Fig.~\ref{fig:QC4resistance}.
The HV biasing circuits for all detectors were found to have a resistance deviation $D_{R}$ better than the 3\% tolerance.

\subsection{QC Step 4 -- Intrinsic noise rate measurement}
\subsubsection{Motivation}
Micro-pattern gaseous detectors are known to suffer from spurious discharges, particularly when operating at very high counting rates or at exceptionally high gas gains \cite{Peskov:1216915}. These glow discharges, which are not induced by the direct detection of any ionizing particles, can produce spurious signals and must therefore be reduced to an acceptably low rate. The test described in this section is designed to quantify the intrinsic noise rate of the detectors due to those spurious signals. To perform this measurement, the detectors are put into a safe mode by flushing them with a gas that has a high ionization energy, such as pure CO\textsubscript{2}, which suppresses signals from cosmic rays. The rate of spurious signals as a function of the applied high voltage is then observed. This procedure serves to disentangle effects related to the detection of actual radiation from those induced solely by the application of high voltage.

\begin{figure}[!h]
\centering
\includegraphics[width=0.7\columnwidth]{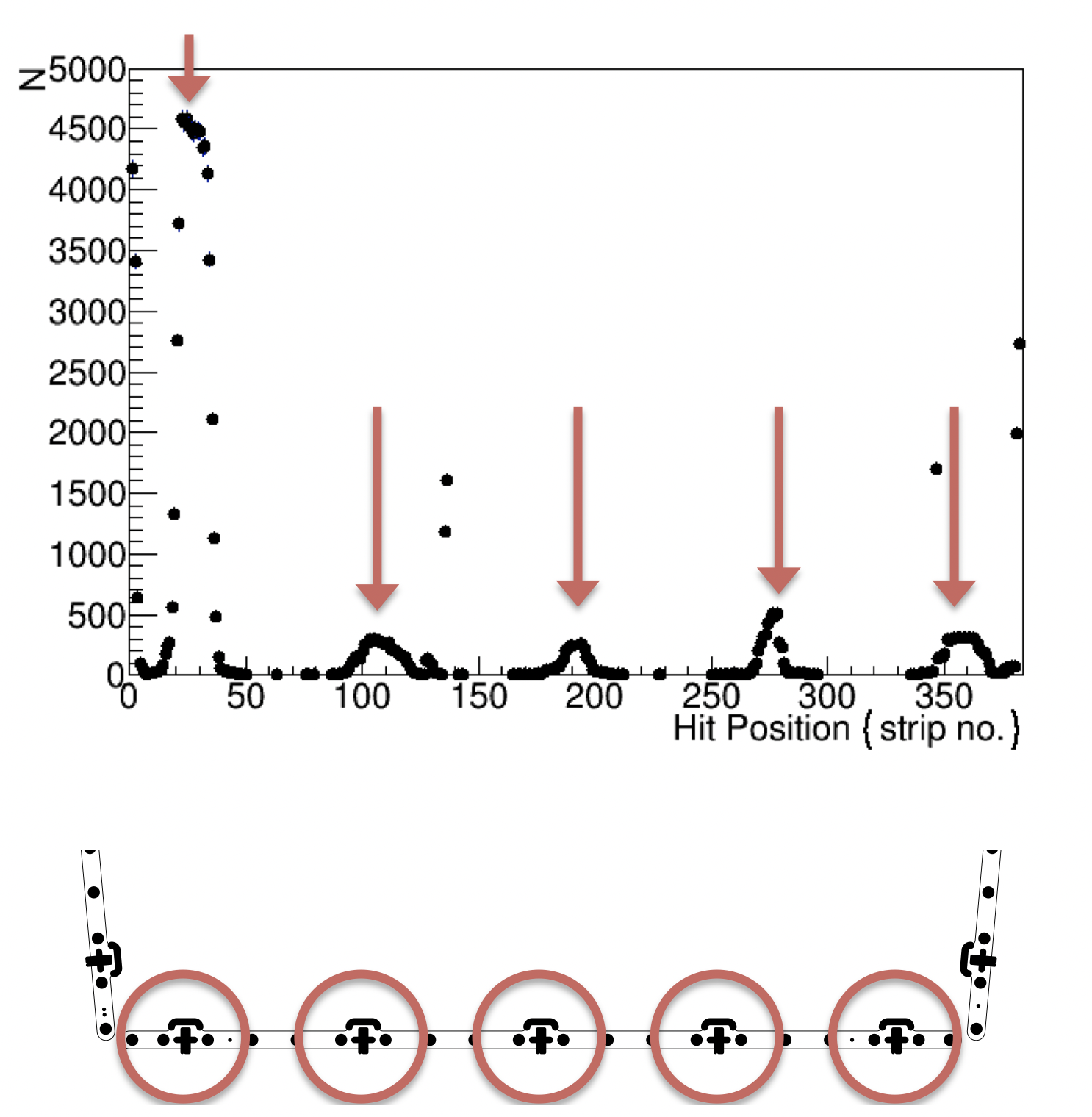}
\caption{\label{fig:SR-Investigation}Number of intrinsic noise signals recorded by each strip in the eighth $i\eta$ partition (next to the short base of the trapezoid) for a prototype GE1/1 detector (serial number GE1/1-VII-L-CERN-0001). The layout in the bottom part of the picture shows the location of the stretching screws along the internal frame in this partition that the intrinsic noise signals correlate to.}
\end{figure}

 Figure \ref{fig:SR-Investigation} shows the number of spurious signals recorded by each strip in the eta sector next to the narrow base of the trapezoid for a prototype of the GE1/1 detector. In this reference measurement, the detector is operated with  Ar/CO\textsubscript{2} (70:30) with an average effective gas gain of only $\sim$300 to reduce the detection efficiency for cosmic ray muons to a few percent. It is found that the discharges that create the spurious signals here are induced mainly in strips near the screws used to stretch the GEM foils. Similar results are found for the eta partition next to the long base of the trapezoid while measurements performed in the remaining sectors showed no substantial contribution to the spurious signal rate. We conclude that the origin of the spurious signals in the GE1/1 detectors under these conditions is a coronal discharge from the active area of the GEM along the internal frame (which holds the GEM foils inside the gas volume) to ground through the anode strips where the signal is read out.

\subsubsection{Selection criteria}
The rate of intrinsic noise signals produced by coronal discharges across the full 3500 -- 4000 cm$^2$ surface area of a normally operating GE1/1 detector is found to not exceed $\mathcal{O}(10^{-2})$ Hz/cm\textsuperscript{2}.
Consequently, the detector is accepted if the intrinsic noise rate for the full detector does not exceed 100 Hz. This value is negligible when compared to the 4.5 kHz/cm\textsuperscript{2} background rate \cite{Colaleo:2015vsq} expected in the CMS experiment during standard operating conditions.

\subsubsection{Test Setup}
\begin{figure}[!ht]
\centering
\includegraphics[width=0.75\columnwidth]{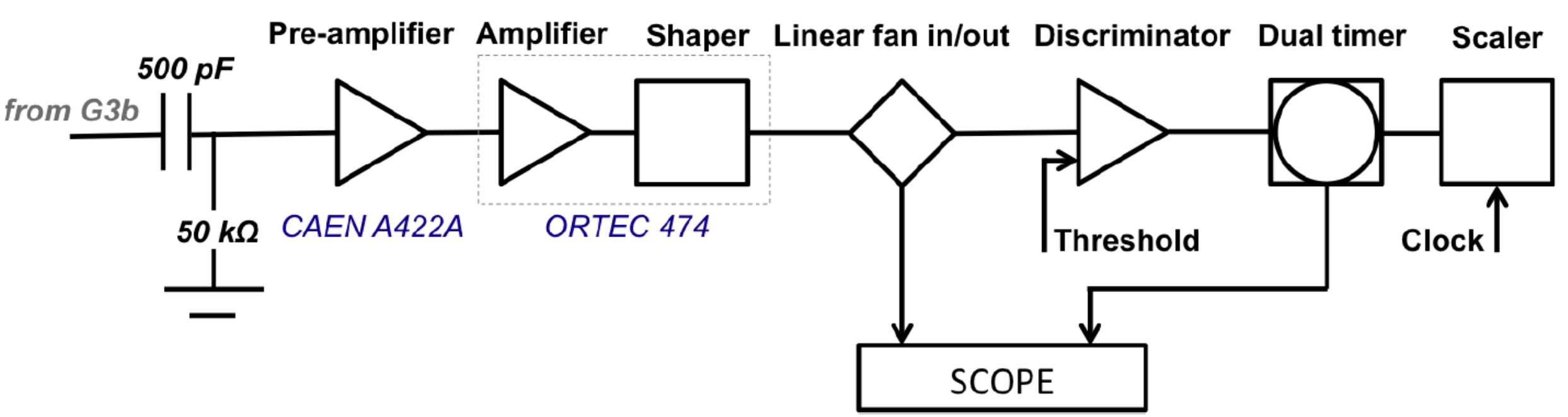}
\caption{\label{fig:QC4SSRSetup}The signal processing chain for the measurement of the rate of converted photoelectrons in the drift gap. Since the pre-amplifier and amplifier are among the most critical components of this processing chain, we indicate the specific models in the schematic. The measurement is performed on the contiguous bottom electrode of GEM3 (G3b) of a GE1/1 detector operated with pure CO\textsubscript{2}.}
\end{figure}

A grounding plate is attached to the backside of the detector readout board and connected to the ground pad of the drift PCB to reduce environmental RF noise pick-up. The readout sectors are grounded via Panasonic-to-LEMO adapters with 50~$\Omega$ termination. To measure the intrinsic noise rate, signals are read from the contiguous bottom of the third GEM foil through a decoupling CR differentiator circuit soldered onto the drift board of the GE1/1 detector.   
The signals are then processed with a readout chain (Fig.~\ref{fig:QC4SSRSetup}) that comprises a charge-sensitive preamplifier, an amplifier, and a discriminator with the threshold set to suppress the environmental noise (140 mV). The resulting digital pulses go through a dual timer and then to a counting unit for the rate measurement. A linear fan-in fan-out module is placed between the shaper and the discriminator to allow simultaneous monitoring of the intrinsic noise signals on the scope. 

\subsubsection{Results}
\begin{figure}[bp!]
\centering
\includegraphics[width=0.65\columnwidth]{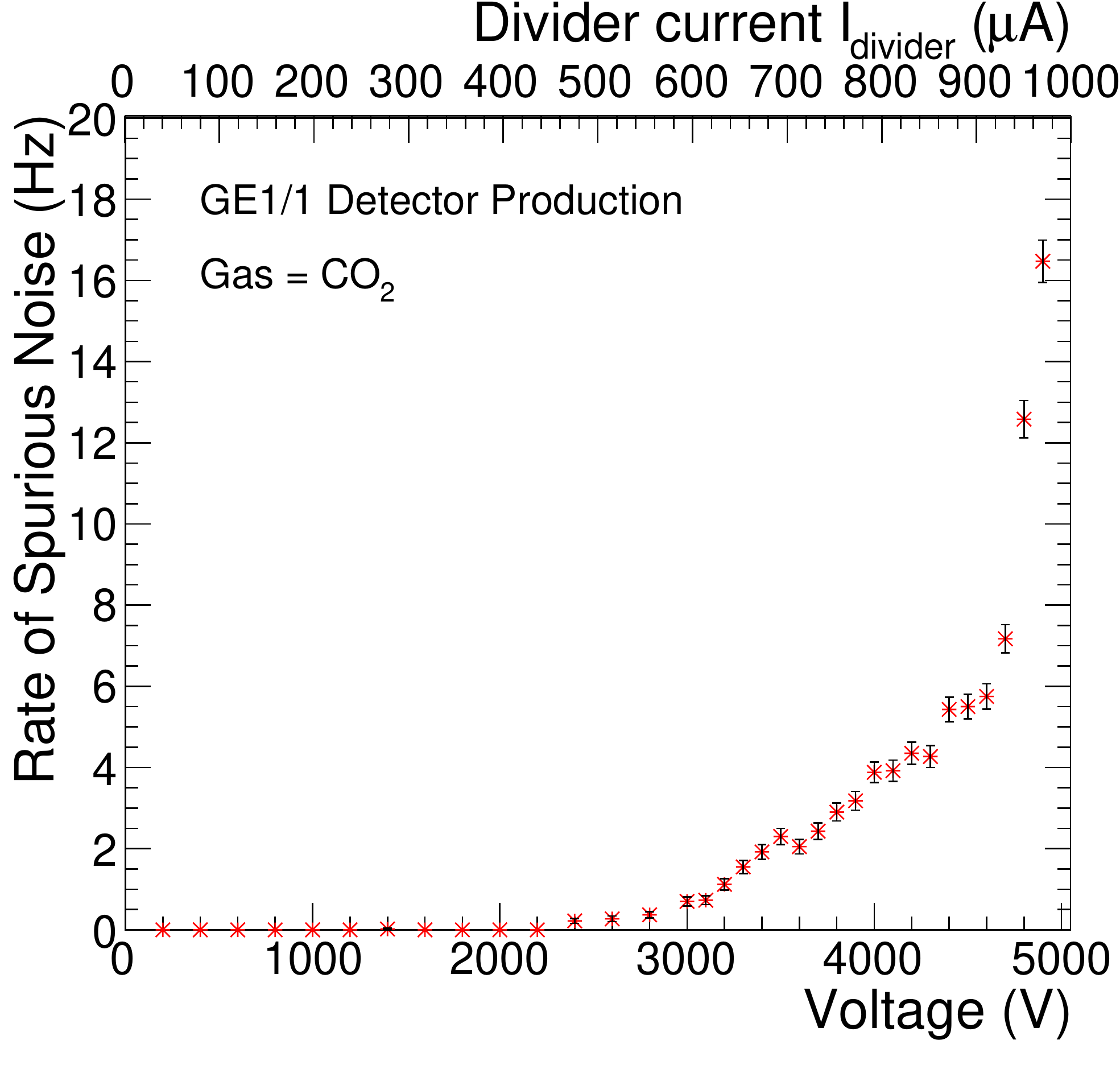}
\caption{\label{fig:QC4SSRexample}Measured intrinsic noise rate of a GE1/1 detector (GE11-X-S-CERN-0012) when operated in pure CO\textsubscript{2} and at a signal threshold of 140 mV.}
\end{figure}

\begin{figure}[!ht]
\centering
\includegraphics[width=0.65\columnwidth]{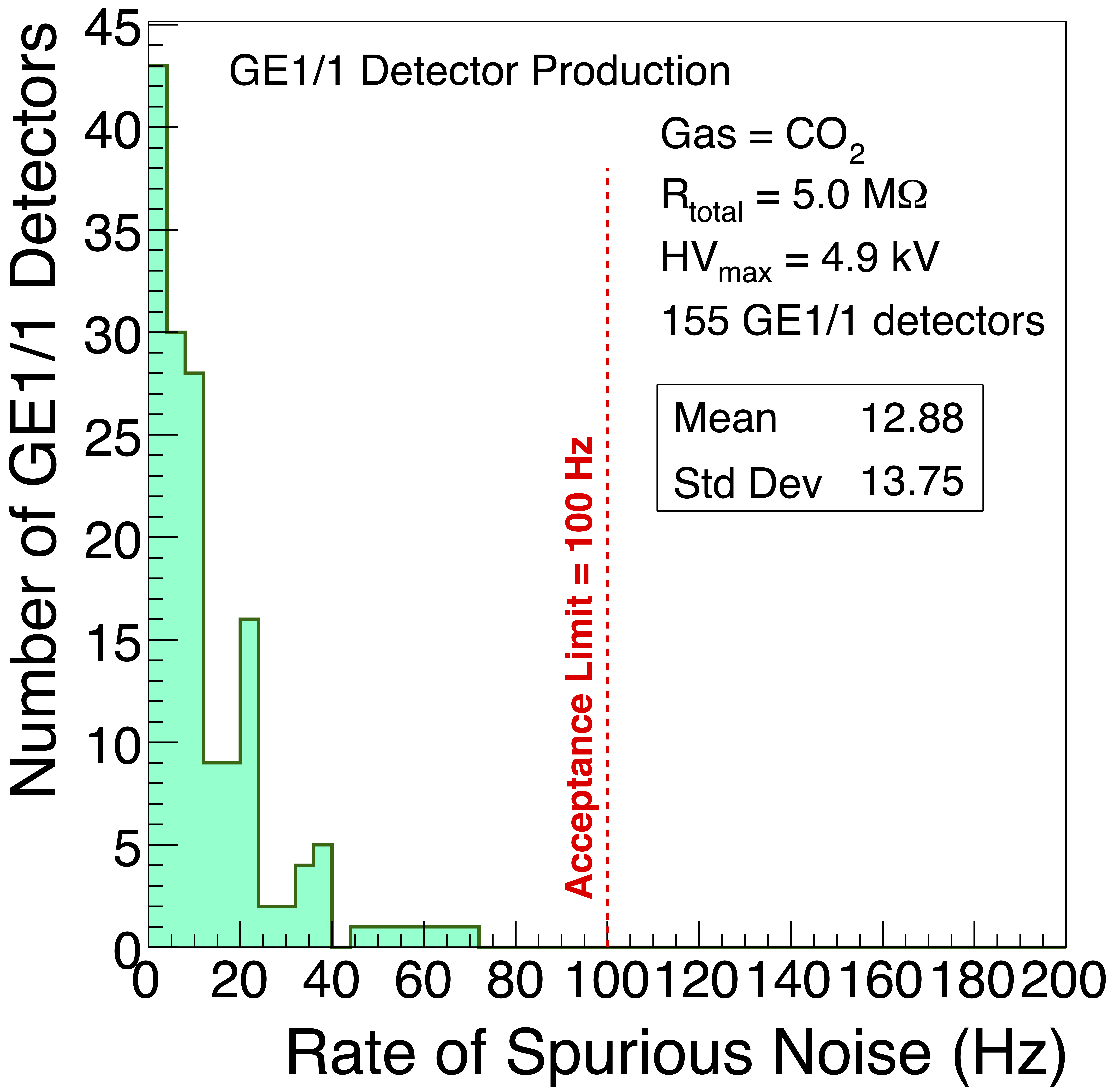}
\caption{\label{fig:QC4SSR}Intrinsic noise rates for 155 GE1/1 detectors operated with pure CO\textsubscript{2} and a signal threshold of 140 mV out of 156 that passed the previous QC test. (Results for the one missing detector have been lost.)}
\end{figure}

The rate of intrinsic noise signals induced on the bottom of the third GEM foil is measured as a function of applied HV for each detector as shown in Fig.~\ref{fig:QC4SSRexample}. The maximum measured value of the intrinsic noise rate at 4.9 kV is taken as a final result.
All GE1/1 detectors show a noise rate that is well below the 100 Hz acceptance threshold, as displayed in Fig.~\ref{fig:QC4SSR}.

\subsection{QC Step 5 -- Effective gas gain measurement}\label{effgasgain}

\subsubsection{Motivations}

The effective gas gain is a crucial parameter for operating a gaseous detector in proportional mode. It is a combination of two main components: the amplification factor given by each GEM foil and the transparency factor which depends on the fraction of the electrons that get lost because they recombine with the gas ions or attach to gas molecules or get absorbed by one of the GEM electrodes. It governs the detection performance such as efficiency and time resolution, which are key parameters for successfully reconstructing detector hits associated with charged particles that traverse the active detector volume.

The effective gas gain is determined experimentally by comparing the primary current induced in the drift gap by an incident, ionizing radiation quantum, with the output current provided by the amplification structure of the detector, i.e.\ the three GEM foils, with the signal induced on the readout electrode. The effective gas gain primarily depends on the operating gas and the electric field configuration inside the detector. For a triple-GEM detector, the effective gas gain is also a function of the electron multiplication factor in the GEM holes and the electron transparency of the GEM foils.

\subsubsection{Selection criteria}
To successfully detect muon hits used by the CMS trigger and offline reconstruction, a time resolution below 8 ns and at least a 97\% detection efficiency are necessary. Previous test beam campaigns showed that such performance can be obtained using GE1/1 detectors operating with an effective gas gain around $2 \times 10^{4}$. The gain performances of all detectors should fall within $\pm$37\% of this nominal effective gas gain value to ensure a detection system with sufficiently uniform gain \cite{Abbaneo:2665948}. The typical HV range for obtaining such gain values is 3290-3390 V applied on the drift electrode for GE1/1 detectors. Previous studies \cite{DischPaper} show that this operational regime is far from the breakdown voltage that would trigger discharges towards the readout electrode with non-negligible probability.

\subsubsection{Test setup and procedure}\label{QC5asetup}
\begin{figure}[ht]
\centering
\includegraphics[width=0.8\columnwidth]{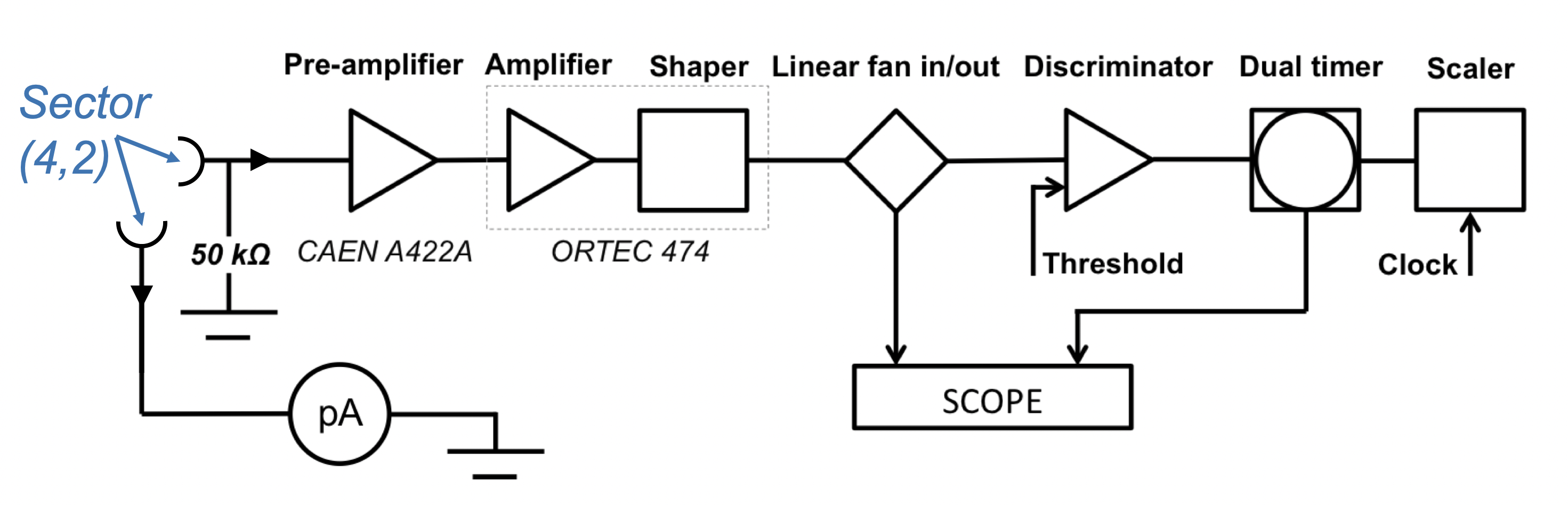}
\caption{\label{fig:QC5Setup}Signal processing chain for the measurement of the rate of converted photoelectrons in the drift gap. The most critical modules are highlighted and their specifications are available on the manufacturer's website. The measurement is performed in the central readout sector ($i\eta=$4, $i\phi=$2) of a GE1/1 detector; see the CMS local coordinate system (Fig.~\ref{fig:GE11mapping}).}
\end{figure}

The effective gas gain of a GE1/1 detector is measured in the central readout sector 
as a function of the current through the HV divider. To perform the test, we use an X-ray generator 
with an emission cone size of 120 degrees that makes it possible to irradiate the entirety of the chamber simultaneously. The X-ray generator uses an electron gun with electrons incident on a silver target. The X-rays emitted from the silver consist mainly of K$_{\alpha}$ and K$_{\beta}$ emission lines, centered on energies of 22.5 and 25 keV, on top of a Bremsstrahlung continuum.
The X-ray photons are absorbed by the copper atoms that constitute the drift electrode which in turn emit 8 keV photons by fluorescence. These 8 keV photons are then converted to electrons via the photoelectric effect in the gas atoms in the detector volume -- in particular in the drift gap. 
The primary current induced in the drift gap is given by the product of the rate $R$ of electrons converted from the photoelectric effect from the incident photons, multiplied by the elementary charge $e$ and the number of primary electrons $N_{p}$ produced by the incident photoelectron in the Ar/CO$_2$ (70:30) gas mixture. To measure the rate of incident photoelectrons, the readout electrode in the sector under test is connected to the signal processing chain in Fig.~\ref{fig:QC5Setup}.   

The current $I_{RO}$ induced on the readout electrode is measured separately with a pico-ammeter connected to the sector under test (Fig.~\ref{fig:QC5Setup}). For each measurement, the thermal noise current and noise rate are also measured and treated as pedestals to be subtracted from the readout current and the photoelectron rate, respectively, that are measured when irradiating the detector with the X-ray beam.
The effective gas gain $G$ is given by:
\begin{equation}
G = \frac{I_{RO}}{R \cdot e \cdot N_{p}}.
\end{equation}

For determining the experimental error on the gain values, Poissonian uncertainty is assumed on the number of recorded photoelectrons for the determination of $R$. For the determination of the readout current, a sample of 300 measurements is taken and the standard deviation of the mean is taken as the uncertainty on $I_{RO}$. 

\subsubsection{Measuring the number of primary electrons}
A dedicated calibration procedure is performed to measure the mean number of primary electrons $N_{p}$ created by an incident X-ray photon in a GE1/1 detector filled with the Ar/CO$_2$ (70:30) gas mixture. The energy spectrum of the resulting photoelectrons is recorded by a GE1/1 detector with a multi channel analyzer for several detector operating points, each corresponding to a particular value of the effective gas gain. 
The peak position of each energy spectrum, which is extracted with a fit procedure assuming a Gaussian model for the peaks and a polynomial distribution for the underlying continuum, is taken as the measured value of the amplified charge recorded by the GE1/1 detector for an incident photoelectron from copper fluorescence with an energy of 8~keV. By assuming that the effective gas gain of the operating detector is precisely known, the number of primary electrons can be obtained. The mean number of primary electron induced by an 8 keV X-ray photon in an Ar/CO$_2$ (70:30) gas mixture in the GE1/1 detector is found to be $N_{p} = 346 \pm 3$ from this measurement. 

As a cross-check, this procedure is performed with two more X-ray sources ($^{55}$Fe, $^{109}$Cd) with the same detector under the same conditions. These same X-ray sources are then used to measure the effective gain calibration curve of the detector. It is found that the resulting respective gain curves, obtained with the same detector under irradiation by different sources, agree with each other within the uncertainties.

\subsubsection{Typical results}
\begin{figure}[!hbp]
\centering
\includegraphics[width=0.7\columnwidth]{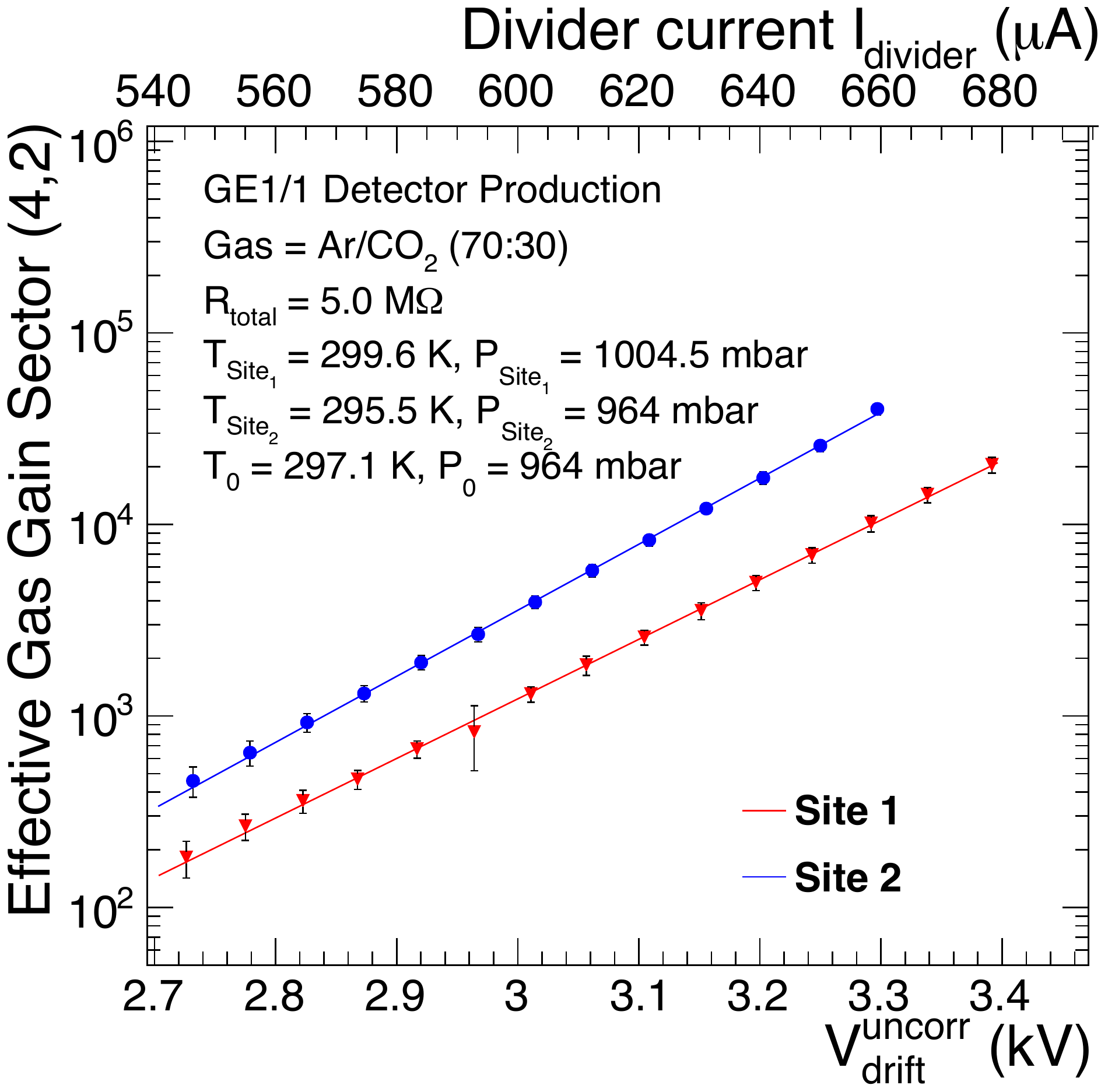}
\caption{\label{fig:gainbefore}
The effective gas gain of two GE1/1 detectors built and tested at CERN (GE11-X-S-CERN-0012, red curve) and in Italy (GE11-X-S-BARI-0011, blue curve) as a function of drift voltage and HV divider current before pressure and temperature corrections.}
\end{figure}

\begin{figure}[!htp]
\centering
\includegraphics[width=0.7 \columnwidth]{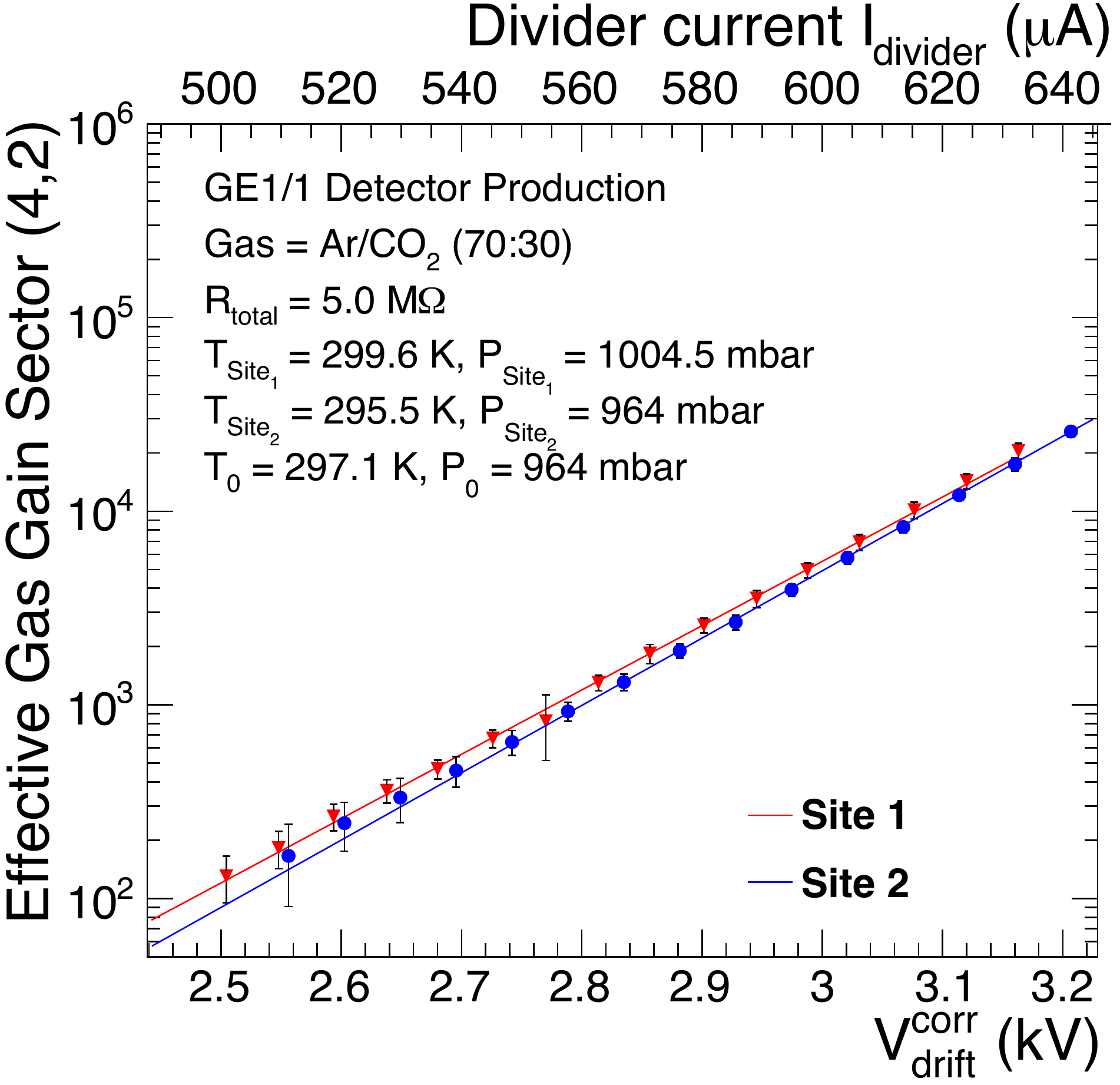}
\caption{\label{fig:gainafter}
The effective gas gain of two GE1/1 detectors built and tested at CERN (GE11-X-S-CERN-0012, red curve) and in Italy (GE11-X-S-BARI-0011, blue curve) after pressure and temperature corrections (see equ.~\ref{eqPTcorr}) are applied to drift voltage and HV divider current.}
\end{figure}

Examples of typical effective gas gain curves for GE1/1 production detectors, measured as a function of the HV applied to the drift electrode and the current in the HV divider, are shown in Fig.~\ref{fig:gainbefore}. The blue and red curves refer to GE1/1 detectors produced and tested at different production sites. The difference in the measured effective gain is due to the different environmental conditions at the two production sites.
Given the dependence of the effective gas gain on the exponential of the Townsend coefficient, which in turn depends on the ratio between environmental temperature and pressure, the gain curves are normalized to a predefined reference pressure and temperature. The reference values correspond to the average pressure and temperature at the location of the GE1/1 station in the CMS apparatus observed over one year, with $P_{0}=964$ mbar and $T_{0}=297$ K.
Rather than correcting the gain values, the value of the applied drift voltage $V_{\textrm{drift}}$ (or the current running through the divider), for a given production site $X$, is corrected by the relation
\begin{equation}\label{eqPTcorr}
V_{\textrm{drift}}^{\textrm{corr}} = V_{\textrm{drift}} \cdot \frac{T_{X}}{P_{X}} \cdot
\frac{P_{0}}{T_{0}}
\end{equation}
 The impact of the normalization procedure is clearly evident in Fig.~\ref{fig:gainafter}, where the two curves are shown after applying the pressure and temperature correction. While this particular correction was used during the QC process described here, we note that an empirical correction was developed subsequently to improve the description of the gain dependence on environmental temperature and pressure \cite{Mocellin:822893}.

\newpage

\subsection{QC Step 5b - Response Uniformity measurement}
\subsubsection{Motivations}
The gain measurement performed in a central readout sector of each GE1/1 detector sets the absolute gain scale of the detector as a function of HV. To find the gain across the entire detector and to quantify its variation, the relative response is measured for all readout strips.
\subsubsection{Test Setup}
\begin{figure}[!ht]
\centering
\includegraphics[width=0.8\columnwidth]{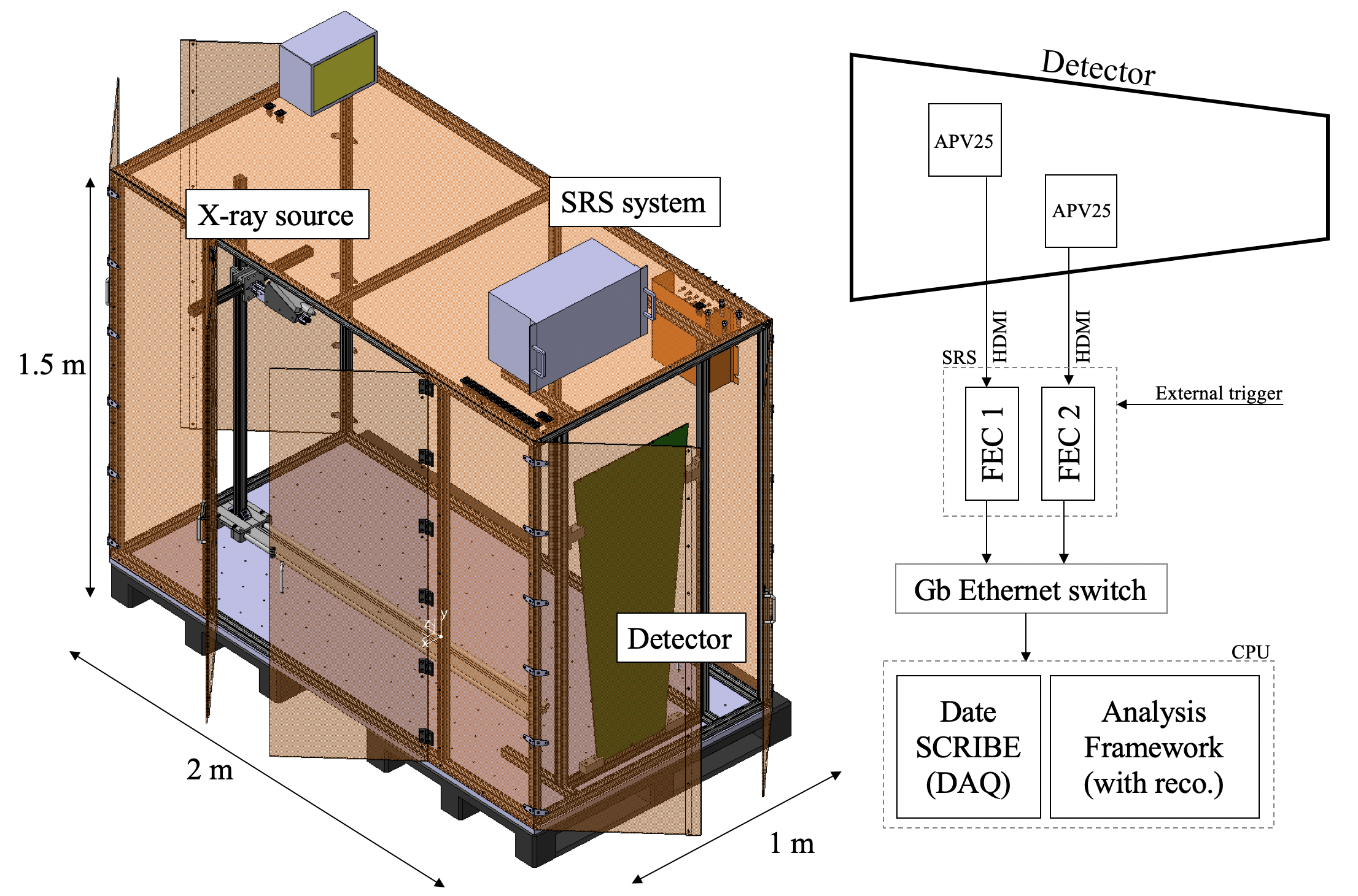}
\caption{\label{fig:gainUniformSchematic} Schematic overview of the X-ray station for the QC5 response uniformity test (left). Typical data flow in the SRS DAQ from the front-end APV25 to the analysis framework (right).}
\end{figure}

The drift electrode is irradiated with a wide X-ray beam from the X-ray generator described in Sec.~\ref{QC5asetup}. The charge induced on the readout strips is amplified by analog pipeline voltage 25 (APV25) analog readout chips~\cite{French:2001xb}, digitized by analog-to-digital converters (ADCs), and recorded by front-end concentrator cards (FECs), which are components of the RD51 Scalable Readout System (SRS)~\cite{Martoiu:2013}. Figure \ref{fig:gainUniformSchematic} gives a schematic of this test setup.

During data taking, one would ideally want to trigger on the signal from the contiguous bottom electrode of the third GEM foil, which is the mirror image of the signals collected on all the readout strips. However, because of the large capacitance of this electrode relative to the anode strips, the electronic noise is typically too high to reliably trigger the APV system, in particular for the lower range of the energy distribution of the interacting particles. Consequently, the APV25 is operated in random trigger mode. Given the rate from the X-ray gun, which is on the order of several MHz over the entire chamber, and the size of the APV25 acquisition time window (up to 750 $\mu$s), it is basically guaranteed that at least one event is recorded for each random trigger.

\begin{figure}[tp!]
\centering
\includegraphics[width=0.6\columnwidth]{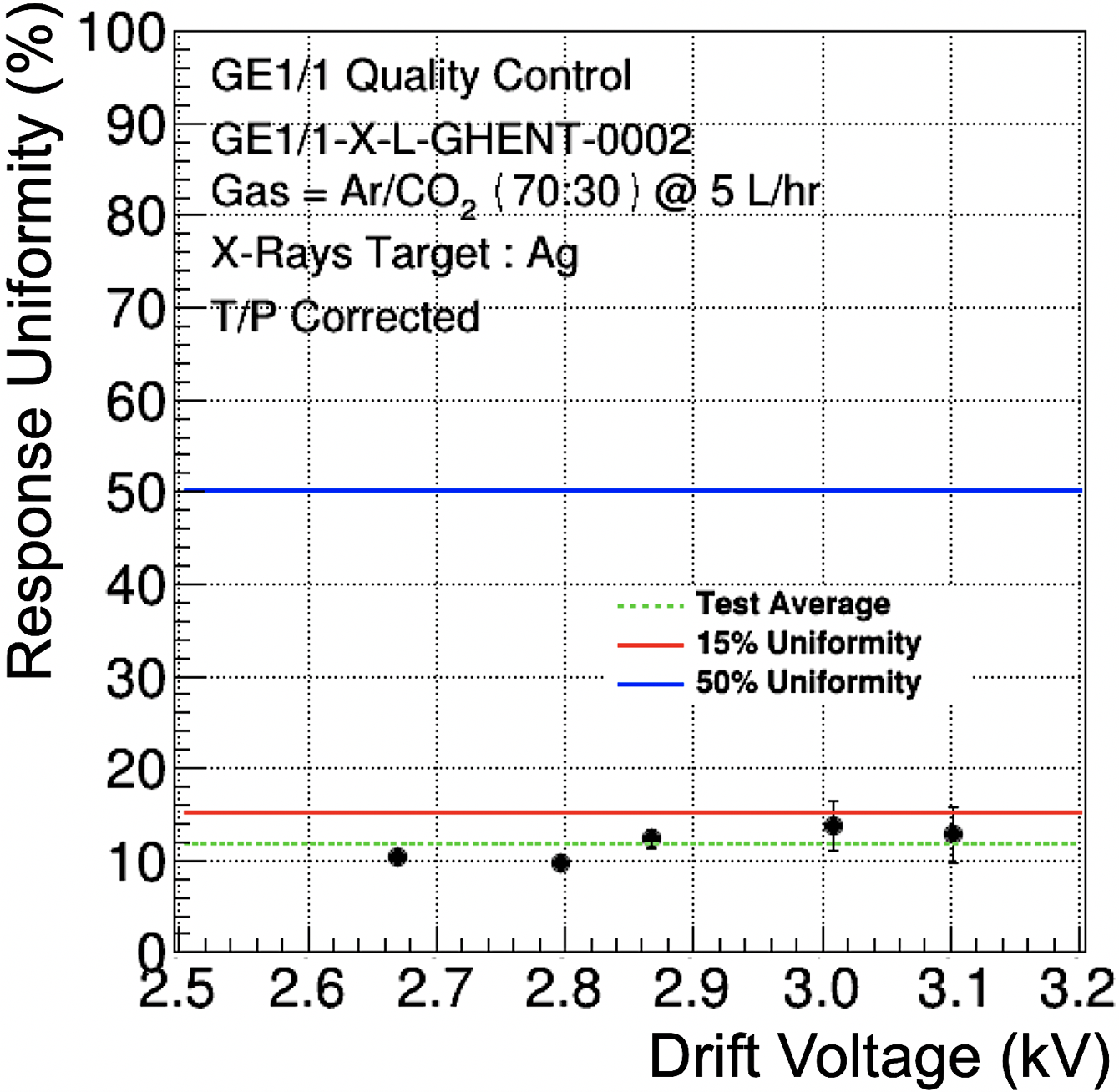}
\caption{\label{fig:QC5UniformityVsHV}Gas gain uniformity as a function of the HV applied to a GE1/1 chamber. The results shown are for GE1/1-X-L-GHENT-0002.}
\end{figure}

\begin{figure}[!ht]
\centering
\includegraphics[width=0.6\columnwidth]{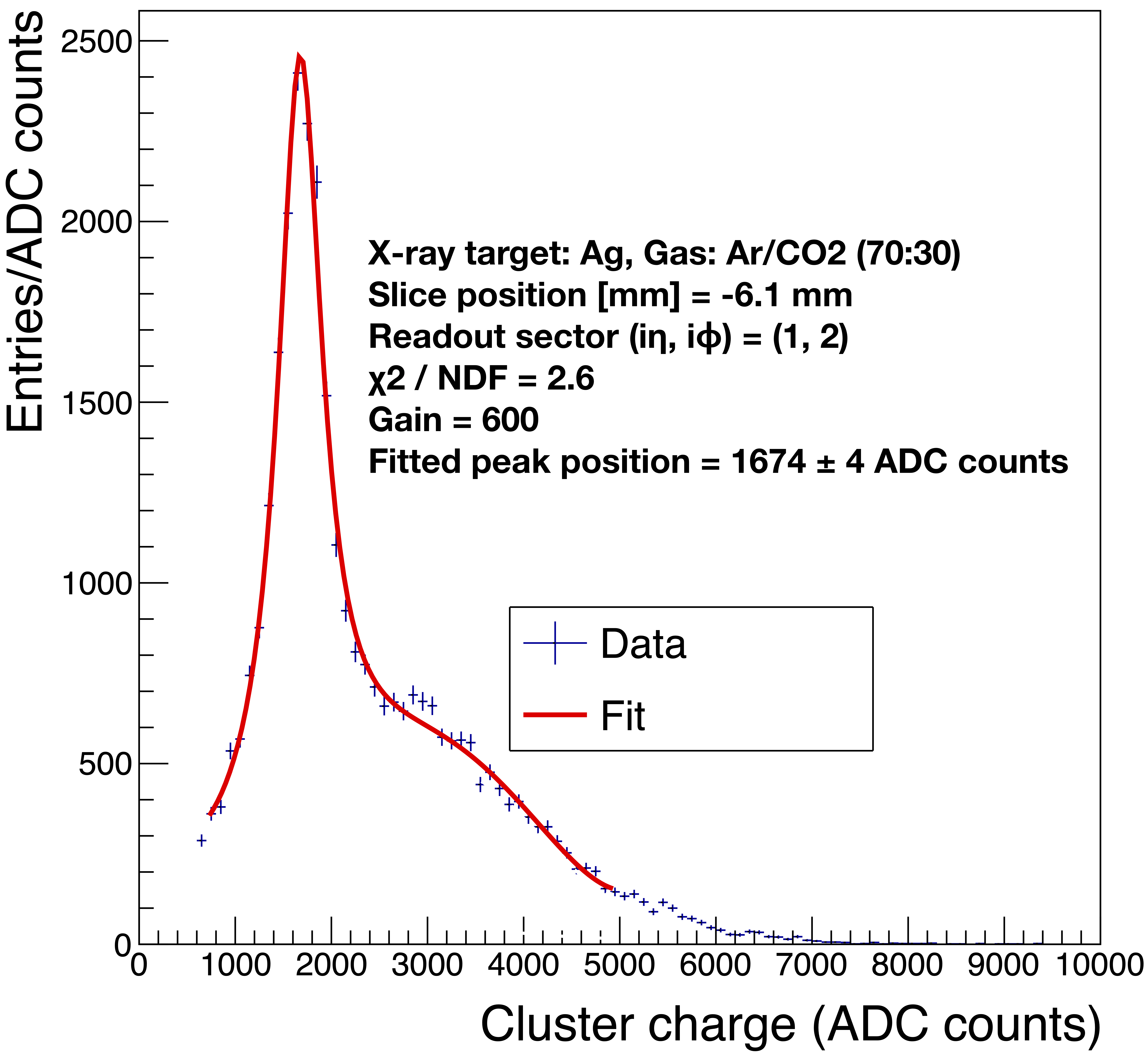}
\caption{\label{fig:RUspectrum}
X-ray spectrum measured within one 4-strip slice of a GE1/1 production detector (serial number GE1/1-X-S-BARI-0007) in ADC units. The spectrum is dominated by the copper fluorescence peak that occurs when irradiating the GEM detector and its various Cu components with X-rays from an X-ray tube with Ag or Au anode. The data are fit to a Cauchy function (peak) plus a fifth-order polynomial for the underlying continuum.}
\end{figure}

As the APV25 ASICs were initially designed for detection of minimum-ionizing particles with silicon-based detectors~\cite{French:2001xb}, the amplifiers tend to saturate when operating with X-rays in gaseous detectors at full nominal gain. This saturation typically prevents the proper reconstruction of the collected charge and diminishes the data quality. To overcome this issue, the GEM chambers under test are operated at a reduced gas gain (typically between 500 and 600), below the saturation level. Figure~\ref{fig:QC5UniformityVsHV} demonstrates that at lower operating voltages, i.e.\ at lower total effective gains, the response uniformity measured across all readout sectors of the chamber is very similar to that measured at higher voltages and gains. This is an important result as it allows flexibility when setting the operating point of the chamber. A broad range of drift voltages can be used without compromising the uniform response of the detector.
\newpage
In the data analysis, the readout of the GE1/1 detector is divided into 768 slices with each slice comprising four readout strips that are adjacent in the $i\phi$ coordinate. For each detector slice, the position of the main peak in the spectrum---the copper fluorescence photopeak---is determined from fitting a Cauchy distribution to the ADC spectrum of the total measured strip-cluster charges. Here, the total charge in each X-ray conversion is obtained by summing the ADC counts over the strips in the strip cluster. To model the background, a fifth-order polynomial is fit to the data. Figure~\ref{fig:RUspectrum} shows a typical example of this strip-cluster charge ADC spectrum for one slice of a production GE1/1 detector.

\subsubsection{Selection Criteria}\label{sect:RespUni}
\begin{figure}[!ht]
\centering
\includegraphics[width=0.6\columnwidth]{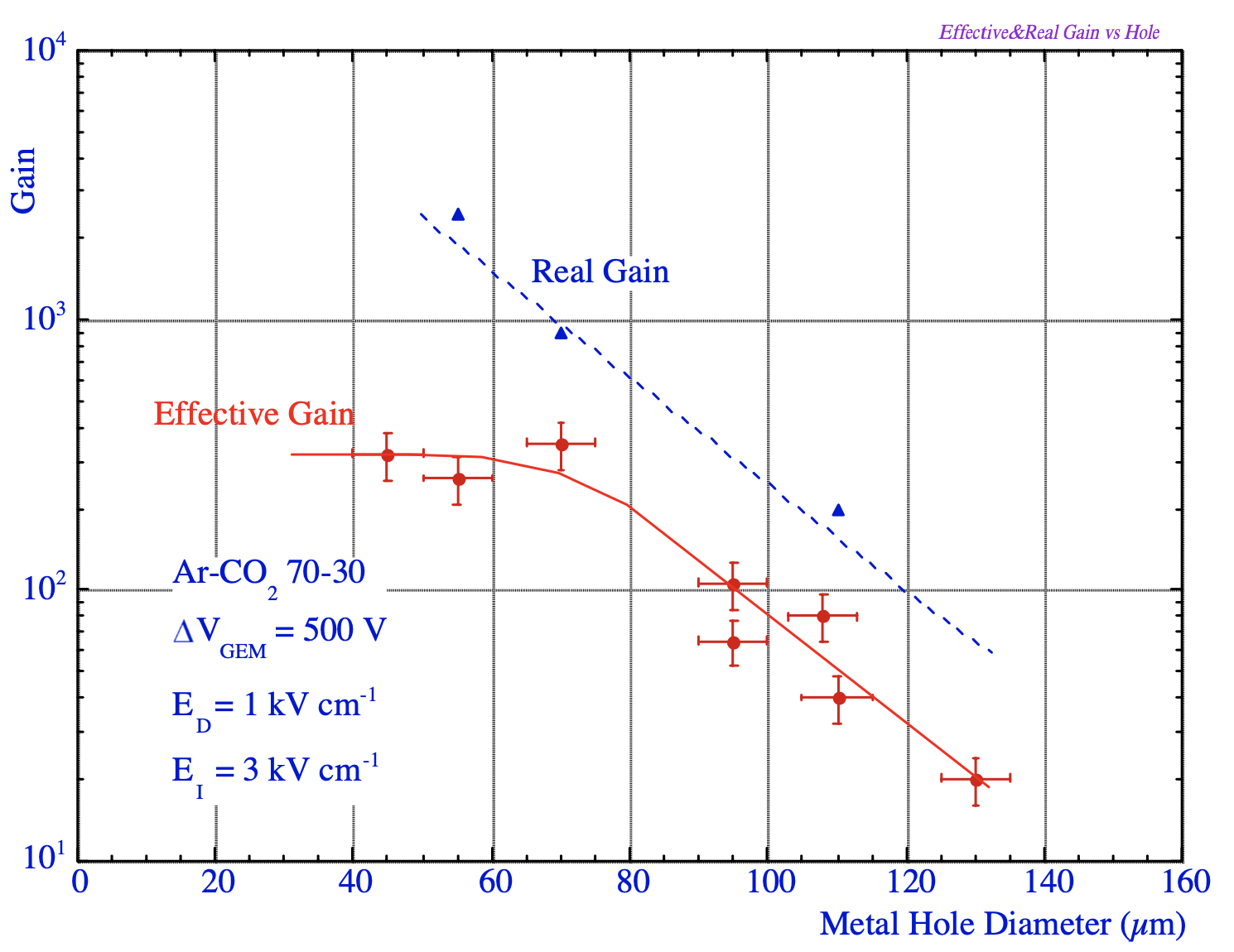}
\caption{\label{fig:ChargeAmpli}
Real and effective gas gains of a single-GEM detector as a function of the outer hole diameter (reproduced from \cite{Bachmann}). The real gain is the absolute multiplication factor, which neglects losses of the secondary electrons to the bottom of the GEM foils and electron-ion recombination. The effective gain, which considers these losses, is the more relevant measure of gain.}
\end{figure}

As discussed in Sec.~\ref{effgasgain}, the effective gas gain is a combination of the amplification factor given by each GEM foil and the transparency factor. The uniformity of the amplification factor is mostly driven by the consistency of the GEM hole geometry across the foil since the applied voltage is usually constant. Past measurements~\cite{Bachmann} show that the main variations in the GEM geometry are the thickness of the polyimide layer and the inner and outer diameters of the biconical holes. From the manufacturer’s specifications, the polyimide thickness is typically $(50.0 \pm 0.1)\,\mu$m. The corresponding uncertainty on the electric field propagates into a gain variation of 2.4\% for a single GEM and 4.2\% for a triple-GEM stack.
Similarly, the effect of the uncertainty on the hole diameter can be translated in terms of gain variations. Based on the measurements shown in Fig.~\ref{fig:ChargeAmpli}~\cite{Bachmann}, we estimate that with a typical outer GEM hole diameter of $(70.0\pm5.0)\,\mu$m, the corresponding gain uncertainty is on the order of 15\% for a single foil, and 26\% for a triple-GEM stack.

\begin{figure}[!ht]
\centering
\includegraphics[width=0.55\columnwidth]{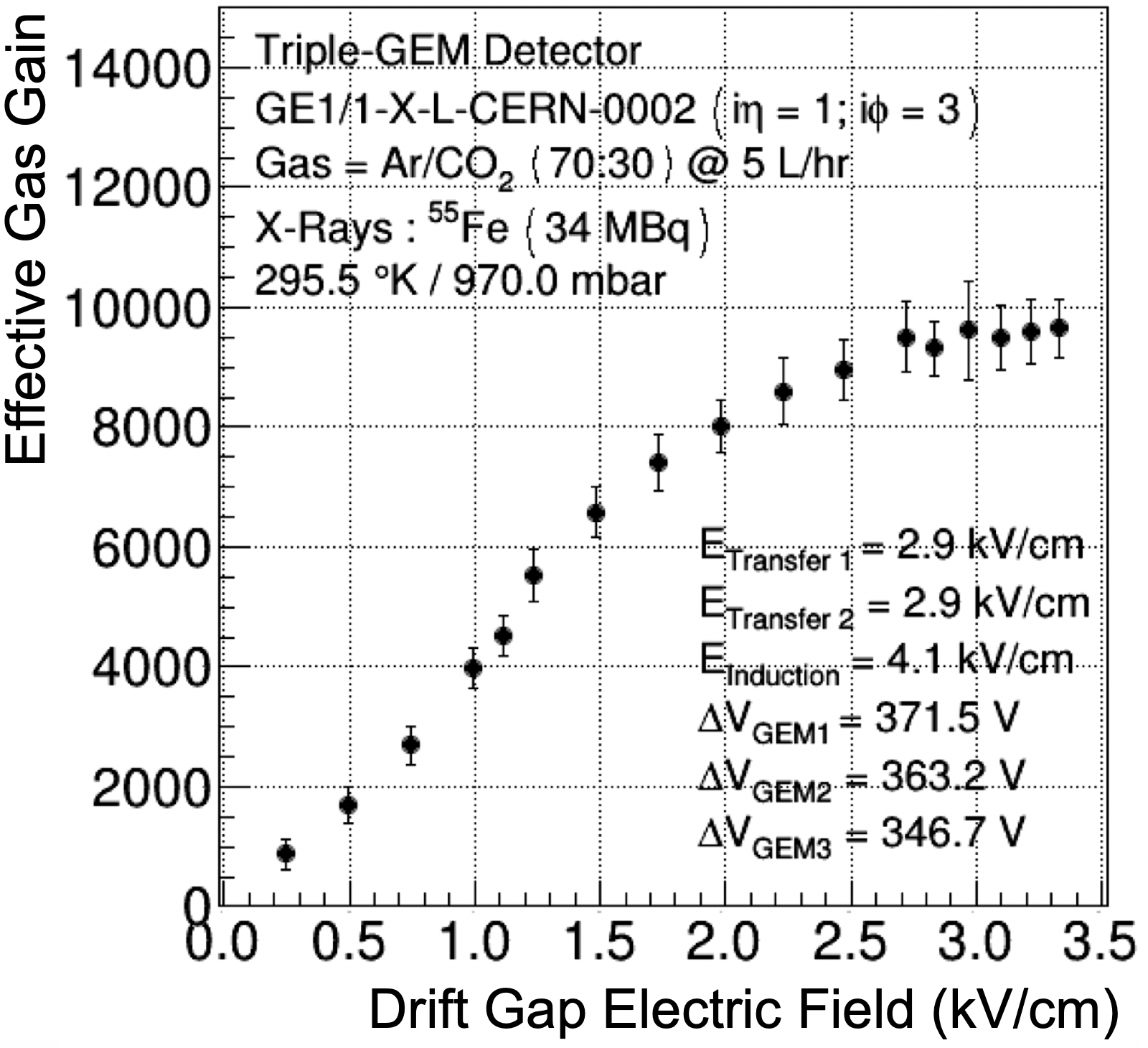}
\caption{\label{fig:QC5DriftField}Effective gas gain as a function of the electric field in the drift gap. The results are shown for detector GE1/1-X-L-CERN-0002.}
\end{figure}

\begin{figure}[!hb]
\centering
\includegraphics[width=0.55\columnwidth]{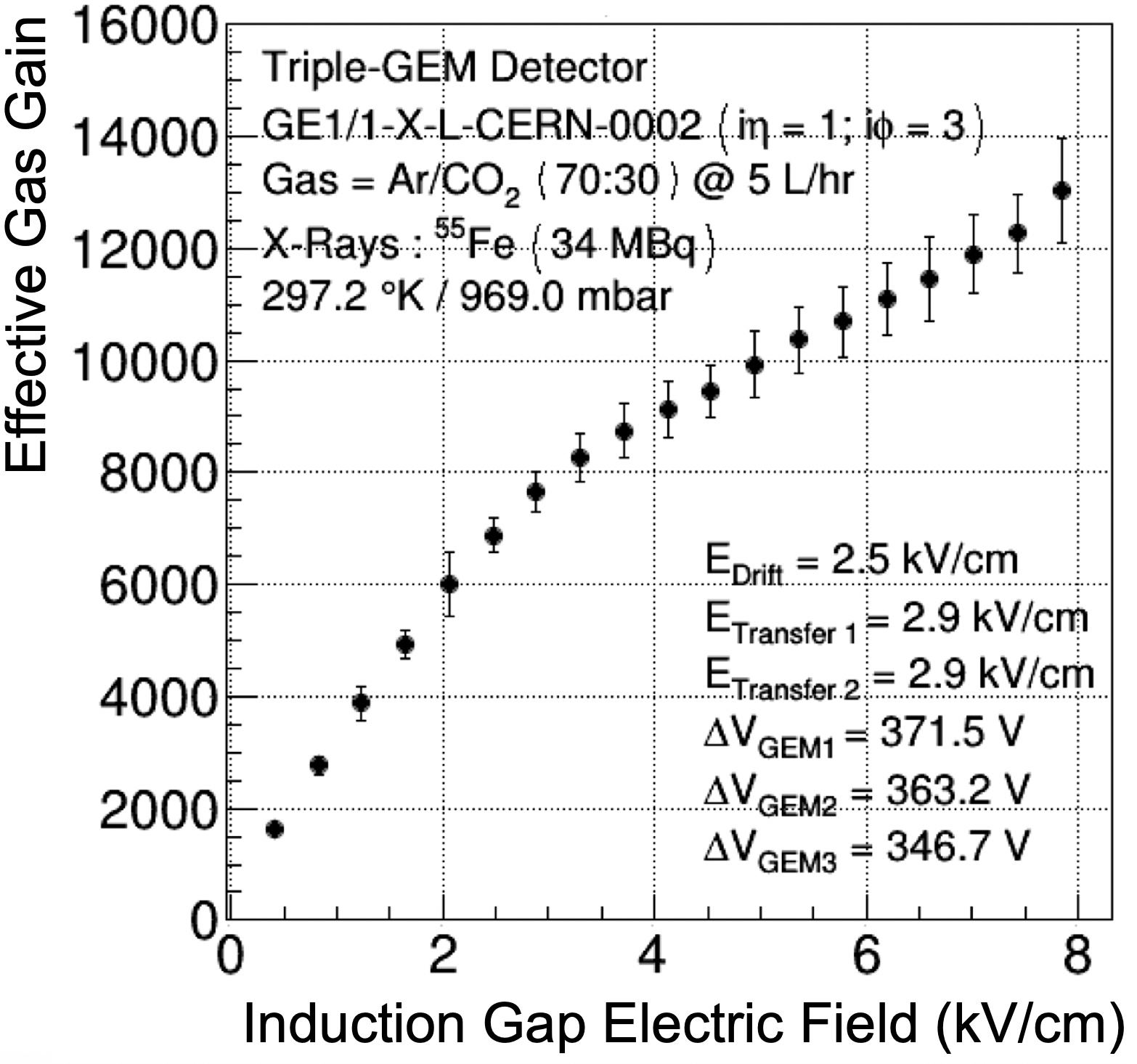}
\caption{\label{fig:QC5InductionField}Effective gas gain as a function of the electric field in the induction gap. The results are shown for detector GE1/1-X-L-CERN-0002.}
\end{figure}

The transparency factor is mostly affected by the variations of the electric field across the detector volume. While the input voltage is unique for each entire electrode, the variation of the gaps between the electrodes impacts the electric field uniformity. In the particular case of the GE1/1 detector, the gaps between the foils are defined by the thickness of the internal frame pieces and the foil stretching applied from the periphery of the GEM stack. The corresponding variation of the gap sizes is below 1\% and introduces only negligible gain variations. However, the drift and readout boards are large PCBs with uneven copper deposits on one side, and both are subject to significant mechanical tensions when they are fixed to the foil pull-outs. As a result, those boards can slightly bend, causing variations in gap thickness, especially near the center of the trapezoid. In addition, the boards can bulge in the center due to the applied gas pressure. Consequently, the typical dimension and variation of the drift gap is $(3.0\pm0.6)$~mm and $(1.0 \pm 0.4)$~mm for the readout gap. The corresponding relative errors on the electric fields are 20\% for the drift field and 40\% for the induction field in the readout gap. Figures~\ref{fig:QC5DriftField} and \ref{fig:QC5InductionField} show the influence of these electric field variations on the effective gain. From these values, we estimate the gain variations arising from the deformations of the PCBs to be 7.5\% on the drift side and 25\% on the induction side. The impact of the various effects are summarized in table~\ref{tab:gainParams}.

\begin{table}[!hbt] 
\centering
\caption{\label{tab:gainParams}Parameters contributing to the uncertainty in uniformity response.}
\begin{tabular}{|c|c|c|}\hline
Uncertainty                       & Description                 & Contribution (\%) \\
\hline
$\sigma_{\mathrm{thickness}}$   & Gap Size between the GEM foils    & 1\\
\hline
$\sigma_{\mathrm{diameter}}$    & Diameter of holes in the GEM foil & 4.2\\
\hline
$\sigma_{\mathrm{drift-bend}}$  & Bending of the drift PCB      & 25\\
\hline
$\sigma_{\mathrm{RO-bend}}$     & Bending of the RO PCB         & 7.5\\
\hline
\end{tabular}
\end{table}

Combining the uncertainties for each parameter discussed above in quadrature yields the total expected relative uncertainty on the effective gain within a GE1/1 chamber:
\begin{equation}
\sigma _{\textrm{tot}}=\sqrt{\sigma _{\textrm{thickness}}^2 + \sigma _{\textrm{diameter}}^2 + \sigma _{\textrm{drift-bend}}^2 + \sigma _{\textrm{RO-bend}}^2} = 37.1\%
\end{equation}
\noindent Thus, a maximum gain variation of 37\% is taken as the upper threshold on the allowed response variation across a GE1/1 detector.

\newpage
\subsubsection{Results}

\begin{figure}[!ht]
\centering
\includegraphics[width=0.52\columnwidth]{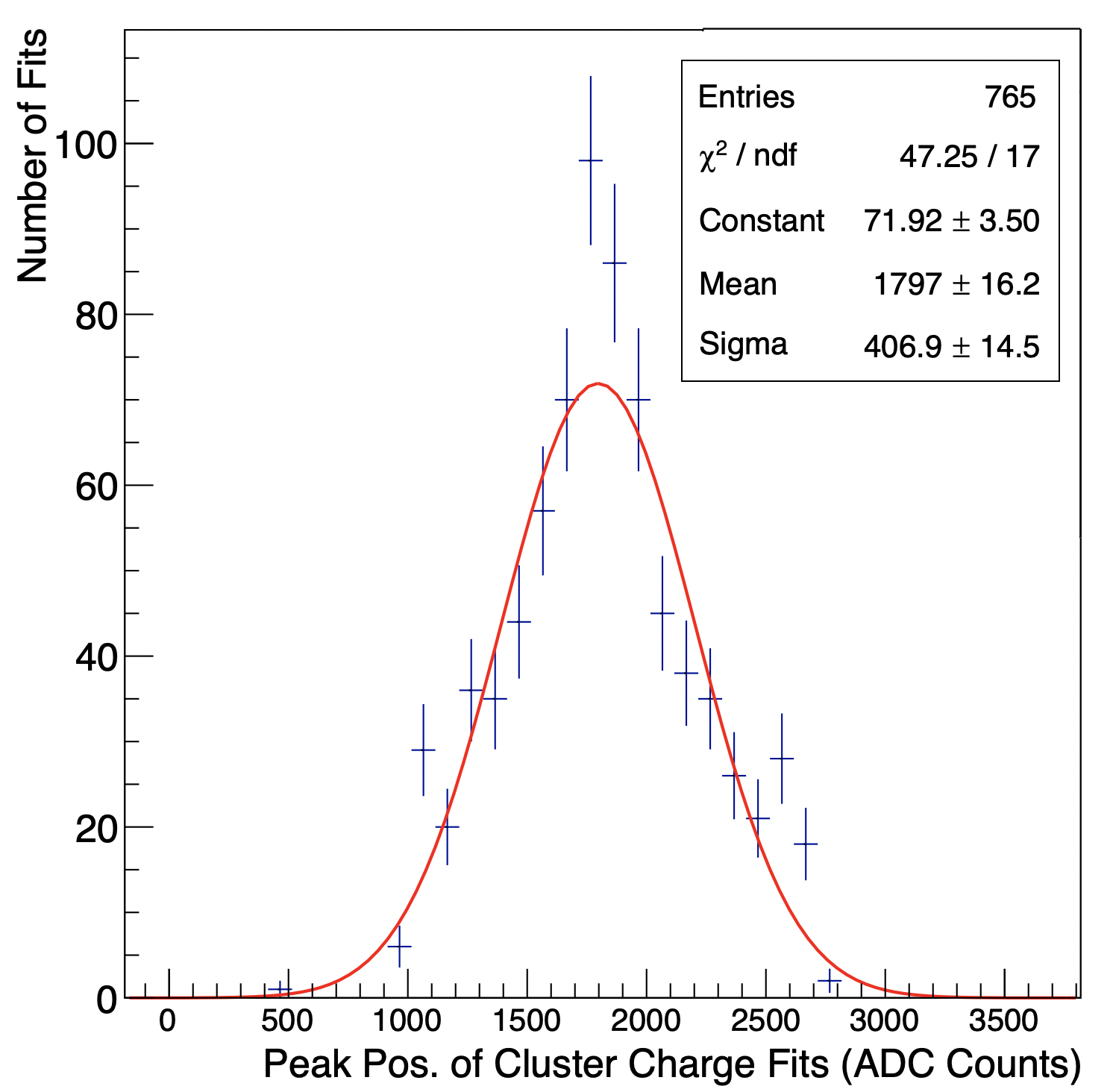}
\caption{\label{fig:pkPosdataset}Example of the aggregated ADC spectra of strip-cluster charge across one entire long chamber (serial number GE11-X-L-PAKISTAN-0005). The peak positions of strip-cluster charges obtained from the fits as shown in Fig.~\ref{fig:RUspectrum} are histogrammed for all slices here, where one slice corresponds to four strips. The standard deviation obtained from the Gaussian fit, shown relative to the mean of that Gaussian, is taken as an overall measure of the gain uniformity for the chamber; for this particular chamber \mbox{$\sigma/\mu$ = 22.6\%}.}
\end{figure}

The position of the copper fluorescence photopeak in the ADC spectrum of the total strip-cluster charges is a good measure of the local gain across the chamber. A typical distribution of the fitted positions of these photopeaks across all 768 slices in a chamber is plotted in Fig.~\ref{fig:pkPosdataset}. This distribution is fitted with a Gaussian to extract the mean ($\mu$) and the standard deviation ($\sigma$) of the distribution. 

\begin{figure}[!ht]
\centering
\includegraphics[width=0.55\columnwidth]{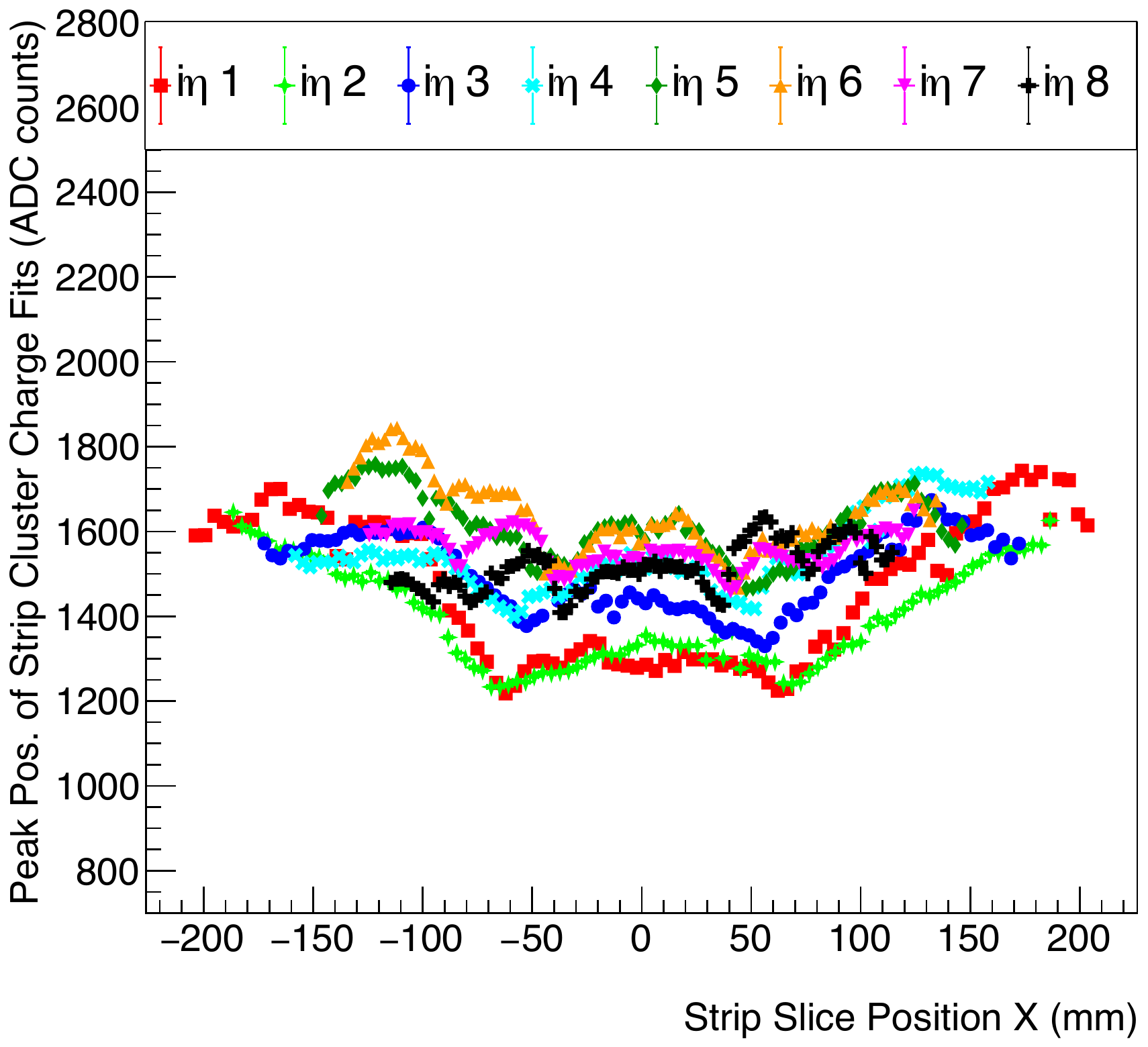}
\includegraphics[width=0.55\columnwidth]{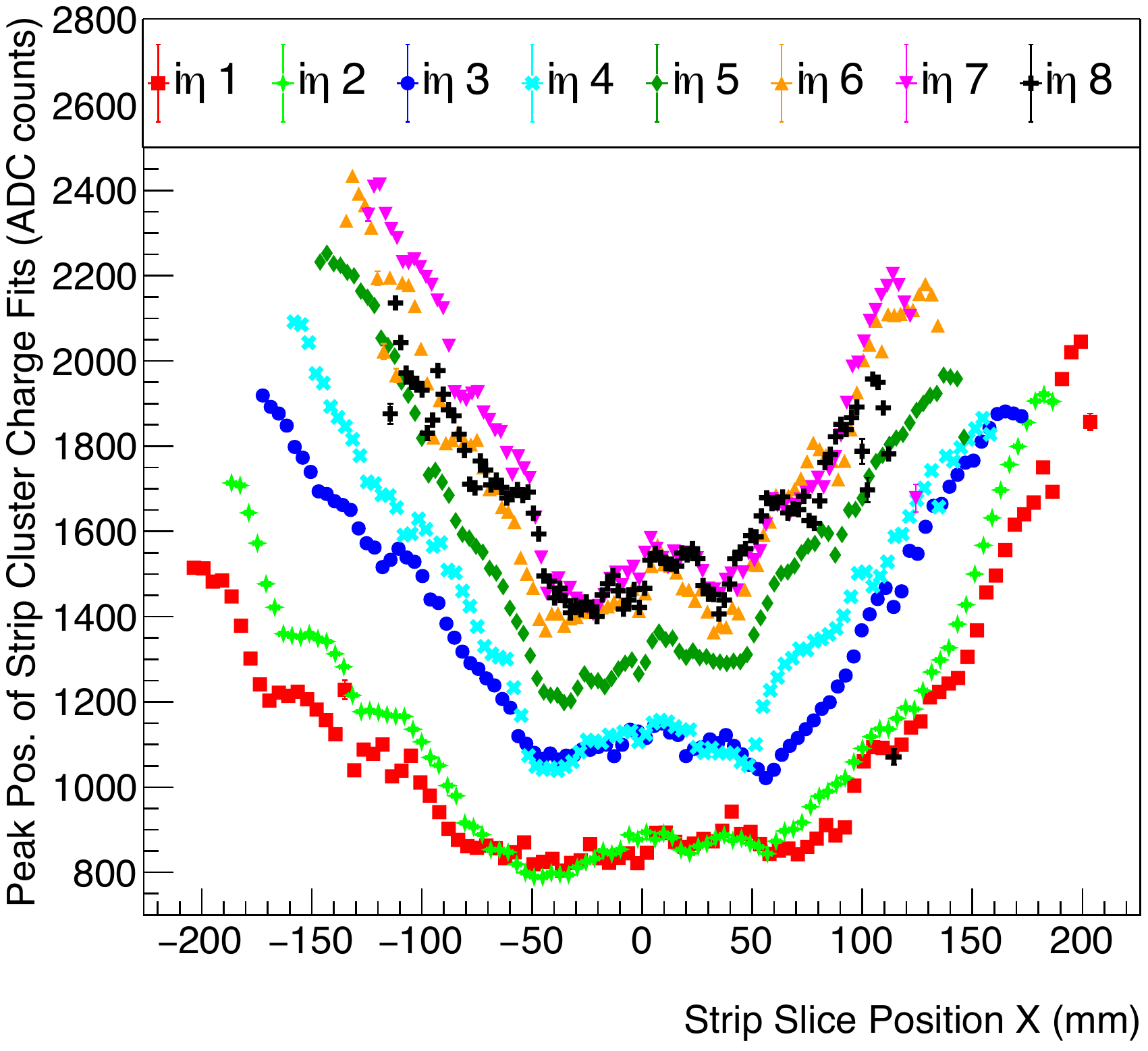}
\caption{\label{fig:plotPkPos}Example of the gain variation across two GE1/1 chamber (serial numbers: GE11-X-L-GHENT-0003 [top] and GE11-X-L-GHENT-0023 [bottom]). The fitted peak position of the distribution of all strip cluster charges (Fig.~\ref{fig:RUspectrum}) within slices of four strips is plotted as a function of the center of the strip slice position across the eight $i\eta$ sectors. The response uniformity, $\sigma/\mu\cdot100\%$, for GE11-X-L-GHENT-0003 is $(6.6\pm0.3)\%$, and the response uniformity for GE11-X-L-GHENT-0023 is $(24.9\pm0.9)\%$. The systematic difference in gain variation between these two detectors is primarily due to a difference in the planarity of the drift and readout boards.} 
\end{figure}

\begin{figure}[!ht]
\centering
\includegraphics[width=0.42\columnwidth]{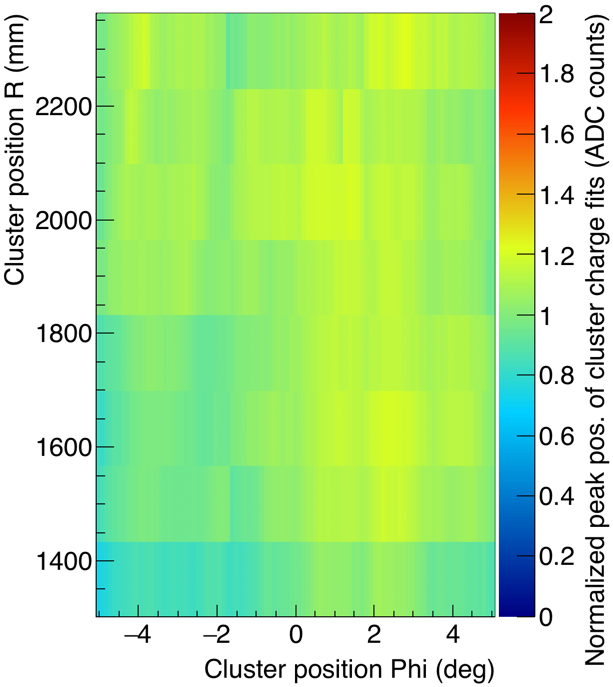}
\caption{\label{fig:2DHist}Example of the gain variation across a GE1/1 chamber (serial number GE11-X-S-INDIA-0009) normalized to the chamber average. Note that the coordinates are plotted in radial distance from the beamline in the CMS experiment (y-axis) and angular distance from the centerline of the chamber ($i\phi$). The horizontal binning corresponds to slices of four strips while the vertical binning corresponds to the eight $i\eta$ sectors.}
\end{figure}

\begin{figure}[!hb]
\centering
\includegraphics[width=0.55\columnwidth, trim={0.4cm 0.4cm 1.2cm 0.2cm}, clip]{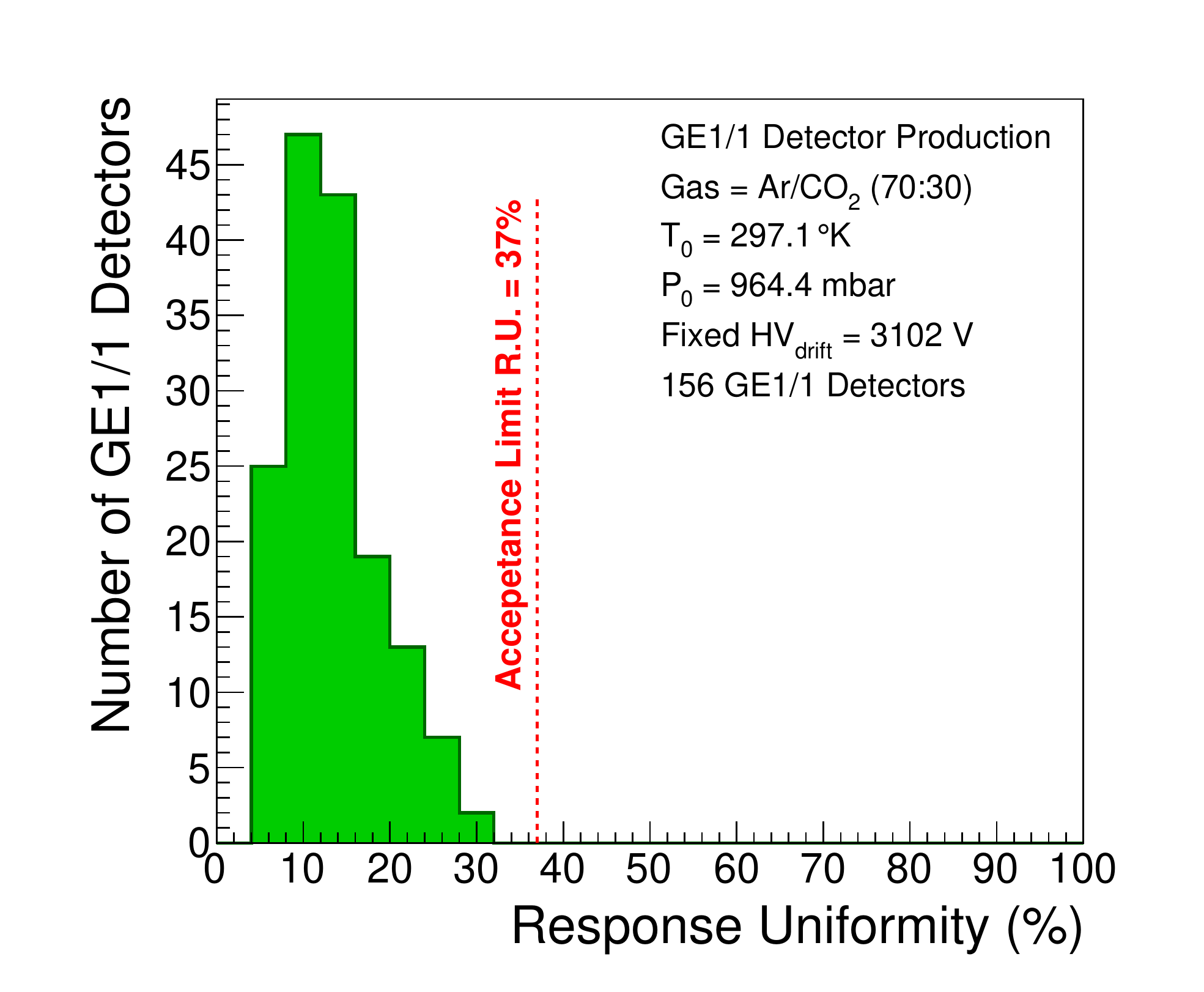}
\caption{\label{fig:RUallchambers}
Relative response uniformity of all 156 GE1/1 detectors as determined by $\sigma$/mean obtained from the respective distributions of the type shown in Fig.~\ref{fig:pkPosdataset}.}
\end{figure}

The standard deviation is a measure of the response uniformity of the detector. Specifically, the relative response uniformity of a GE1/1 detector, quoted as a percent, is defined as $\frac{\sigma}{\mu}\cdot 100\%$, which characterizes the gain variation across the whole detector area.
Figure~\ref{fig:plotPkPos} shows two examples for the absolute azimuthal gain variations across a chamber for each $i\eta$ sector, and Fig.~\ref{fig:2DHist} gives an example of the relative gain variation across a chamber. 
In general, we observe a continuous, ``u"-shaped curve that peaks at the edges of the detector and attains a minimum at the center of the detector (Fig.~\ref{fig:plotPkPos}, bottom). This phenomenon can be attributed to the bending of the readout and drift electrodes previously discussed. We also observe chambers with less variation along the azimuthal direction ($i\phi$), e.g.\  Fig.~\ref{fig:plotPkPos} (top), which results from more planar drift and readout PCBs. As Fig.~\ref{fig:RUallchambers} shows, all tested GE1/1 detectors exhibit gain variations below 30\%. This value is well below the threshold (37\%) that was described in the previous section and consequently all detectors pass this final QC step.
\newpage

\newpage

\newpage
\section{Choice of the operational working point}\label{sec:workingPoint}
\begin{figure}[!ht]
\centering
\includegraphics[width=0.65\columnwidth]{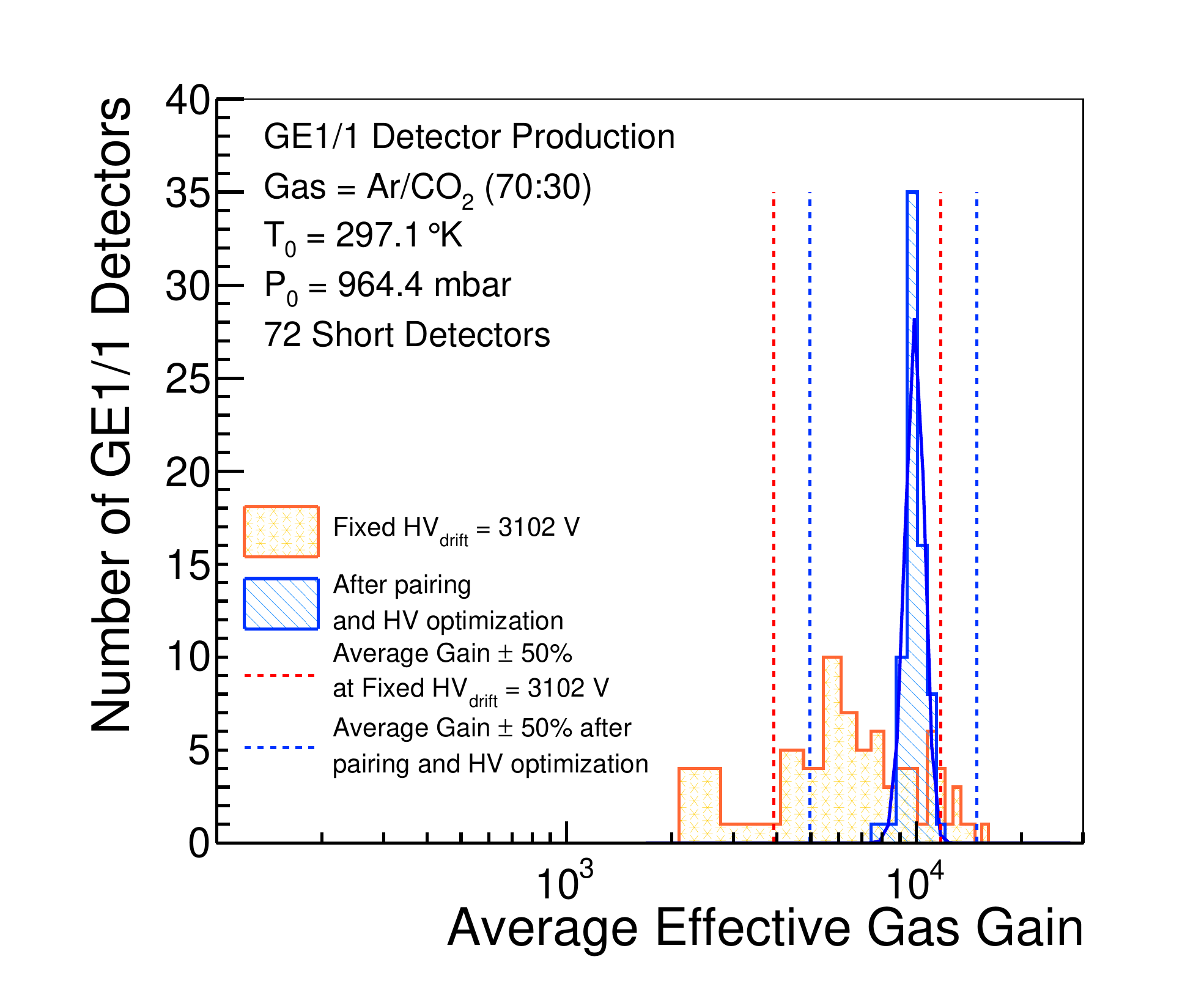}
\caption{\label{fig:GainSCshort}
Average effective gas gain distribution for short GE1/1 detectors before (yellow) and after (blue) the procedure for pairing detectors into superchambers is applied.
The dashed lines represent the $\pm 50\%$ range of the gain around the average before (orange) and after (blue) the pairing procedure.
The gain is normalized to the pressure and temperature expected at the GE1/1 station in the CMS experiment.
Results are shown for all 72 short GE1/1 detectors initially installed in CMS.}
\end{figure}

\begin{figure}[!ht]
\centering
\includegraphics[width=0.65\columnwidth]{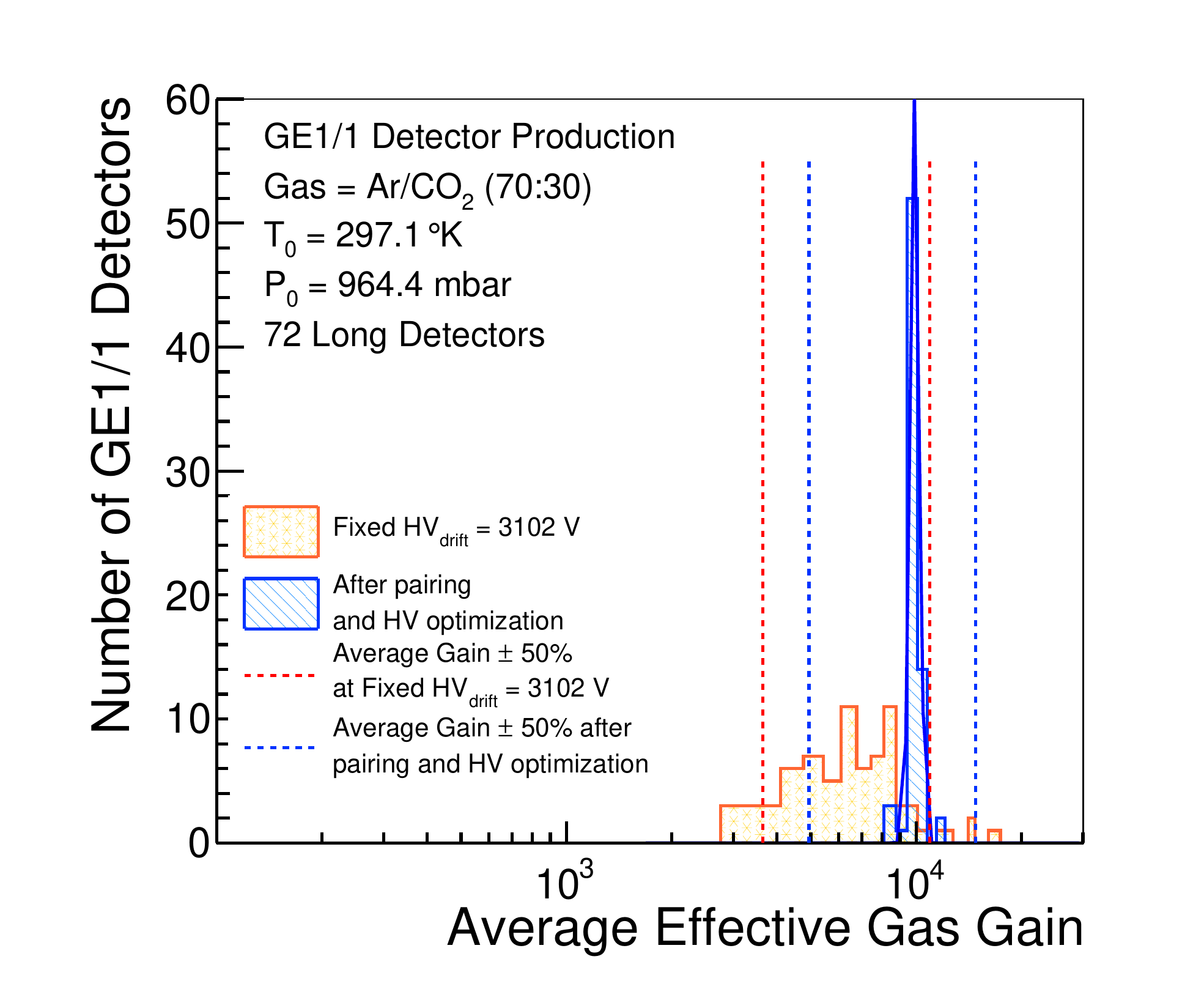}
\caption{\label{fig:GainSClong}
Average effective gas gain distribution for long GE1/1 detectors  before (yellow) and after (blue) the procedure for pairing detectors into superchambers is applied.
The dashed lines represent the $\pm 50\%$ range of the gain around the average before (orange) and after (blue) the pairing procedure.
The gain is normalized to the pressure and temperature expected at the GE1/1 station in the CMS experiment.
Results are shown for all 72 long GE1/1 detectors initially installed in CMS.}
\end{figure}

By combining the detector response uniformity with the gain measurement in the reference readout sector, we can infer the gas gain of each detector slice for each HV operating point.
The average detector gain, i.e.\ the mean of the gain values for all 768 strip slices in the detector, at a fixed drift voltage of \mbox{3102 V}, is plotted in yellow for all short (Fig.~\ref{fig:GainSCshort}) and long (Fig.~\ref{fig:GainSClong}) detectors initially installed in the GE1/1 station. 

It is important to choose HV working points that guarantee uniform behavior of the SCs in terms of gain and, consequently, uniform efficiency and time resolution across the GE1/1 system. During operation in CMS, the two detectors in a SC are powered together by the same power supply, i.e.\ the corresponding electrodes of the two GE1/1 detectors in one SC are connected in parallel and supplied by the same output channel of the HV supply. While gain variations between SCs can be compensated by adjusting the HV working points of the SCs, gain variations between two detectors in the same SC cannot be compensated by individual HV adjustments. To minimize the impact of this constraint, the detectors are sorted by increasing value of the gain and then neighboring chambers in this sorted list are paired into a SC. This ensures that the two detectors in each SC have similar gain performance. 

Finally, for each SC, the HV operation point that ensures a SC gain of $10^{4}$ averaged over the two chambers, is computed from the QC data. The corresponding gas gain distributions, obtained after this optimization procedure, are shown in blue in Figs.~\ref{fig:GainSCshort} and \ref{fig:GainSClong}, for short and long detectors, respectively. As seen in the yellow histograms in Figs.~\ref{fig:GainSCshort} and \ref{fig:GainSClong}, the distribution of the average effective gas gain before the pairing procedure is quite spread out. The width of these distributions, as quantified by the RMS, is found to be 2822 for long SCs and 3122 for short SCs. After the pairing procedure, the width of the distribution of the average effective gas gain is strongly reduced. A Gaussian fit of the blue histograms in Figs.~\ref{fig:GainSClong} and \ref{fig:GainSCshort} is performed and the sigma values obtained from the fits are found to be $290 \pm 29$ for long SCs and $619\pm83$ for short SCs.

\section{Summary, Conclusions, and Perspectives}\label{sec:summary}
In total, 161 GE1/1 detectors were assembled and subjected to stringent quality control tests at several production sites. Of these, three were rejected in the QC2 HV test and two in the QC3 leakage test, while 156 were successfully validated following all the quality control processes described in this paper. Subsequently, 144 detectors have been installed in the CMS muon endcaps with 12 detectors being kept as spares. The five rejected chambers are also kept as one can recuperate components as needed due to the assembly technique without the use of glue.

The QC procedure succeeds in ensuring robust performance and comparable results for detectors assembled and tested at different production sites. 
This is a great success for the distributed-sites production model, which has established a widespread community of GEM technology experts in CMS. The high success rate is indicative of the quality and pre-testing of the components, which is described in \cite{CMS-Muon:2018wsu}.

A few lessons have been learned that now inform the construction and testing of the detectors for the other two stations, GE2/1 and ME0.
For example, the intrinsic noise can be reduced by slightly modifying the shape of the inner frames and the T-nuts.
Also, a central pillar is now introduced into modules that provides a rigid connection between drift and readout PCBs which reduces the deformation of the pressurized detector, which leads to a more uniform response across the detector.
This community is now ready to produce new modules for the two additional muon stations based on GEM technology, i.e.\ GE2/1 and ME0~\cite{bib:MuTDR}, which are currently slated to be installed in CMS in 2023 and 2026, respectively, to further enhance the muon system during Runs 4 and 5 at the HL-LHC.

\section*{Acknowledgments}
We gratefully acknowledge support from FRS-FNRS (Belgium), FWO-Flanders (Belgium), BSF-MES (Bulgaria), MOST and NSFC (China), BMBF (Germany), CSIR (India), DAE (India), DST
(India), UGC (India), INFN (Italy), NRF (Korea), MoSTR (Sri Lanka), DOE (USA), and NSF
(USA).

\section*{References}

\bibliography{QC} 

\end{document}